\documentclass[a4paper,11pt]{article}

\usepackage{jheppub}

\usepackage[T1]{fontenc} 
\usepackage{dsfont}
\usepackage{framed}

\usepackage[makeroom]{cancel}

\allowdisplaybreaks[1]

\newcommand{\pt}[1]{\left( #1 \right)}

\newcommand{\ad}[1]{\pt{a^{\dagger}}^{#1}}

\def\({\left(}
\def\){\right)}
\def\[{\left[}
\def\]{\right]}
\def\<{\left\langle}
\def\>{\right\rangle}

\def\dd{\mathrm{d}}

\def\sp{\hspace{.05em}}

\def\adl{a^{\dagger}}

\def\d{\delta}

\def\e{\epsilon}

\def\t{\tau}

\def\nn{\nonumber\\}
\def\pa{\partial}

\title{Solving anharmonic oscillator with null states: Hamiltonian bootstrap and Dyson-Schwinger equations}

\author{Yongwei Guo}
\author{and Wenliang Li}
\emailAdd{liwliang3@mail.sysu.edu.cn}
\affiliation{School of Physics, Sun Yat-Sen University, Guangzhou 510275, China}

\abstract{
As basic quantum mechanical models, anharmonic oscillators are recently revisited by bootstrap methods. 
An effective approach is to make use of the positivity constraints in Hermitian theories.  
There exists an alternative avenue based on the null state condition, which applies to both Hermitian and non-Hermitian theories. 
In this work, we carry out an analytic bootstrap study of the quartic oscillator 
based on the weak coupling expansion.
In the Hamiltonian formalism, we obtain the anharmonic generalization of Dirac's ladder operators. 
Furthermore, the Schr\"{o}dinger equation can be interpreted as  
a null state condition generated by an anharmonic ladder operator. 
This provides an explicit example in which dynamics is incorporated into the principle of nullness.
In the Lagrangian formalism, we show that the existence of null states can effectively eliminate the indeterminacy of the Dyson-Schwinger equations and systematically determine $n$-point Green's functions.}

\begin{document}
\maketitle 

\section{Introduction}
\label{Introduction}
Two main goals of the bootstrap methods are to achieve 
a deeper understanding of the strong coupling physics and 
to provide concrete computational schemes for extracting precise predictions of strongly coupled theories. 
Before delving into the intricate quantum field theories in physical dimensions, 
a useful strategy is to first study their low-dimensional counterparts, such as zero-dimensional and one-dimensional models, hoping that certain insights may be independent of the spacetime dimension. 
Analogously, the perturbative expansion in a small coupling constant may also elucidate some strong coupling physics if certain general structure is independent of the coupling constant. 
With these motivations in mind, we study the quantum mechanical bootstrap of the quartic oscillator analytically  based on the weak coupling expansion in this work.\footnote{The meaning of the quantum mechanical bootstrap is that the observables, such as energy spectra and matrix elements, are studied using consistency relations, without referring to explicit wave functions. 
This approach can be traced back to Heisenberg's original perspective that led to the establishment of quantum mechanics. }

Recently, matrix theories and quantum mechanical models have been investigated by bootstrap methods 
\cite{Anderson:2016rcw,Lin:2020mme,Han:2020bkb,Han,Hessam:2021byc, Kazakov:2021lel,Berenstein:2021dyf,Bhattacharya:2021btd,Aikawa:2021eai,Berenstein:2021loy,Tchoumakov:2021mnh,Aikawa:2021qbl,Du:2021hfw,Lawrence:2021msm,Bai:2022yfv,Nakayama:2022ahr,Khan:2022uyz,Kazakov:2022xuh,Cho:2022lcj,Morita:2022zuy,Berenstein:2022ygg,Blacker:2022szo,Nancarrow:2022wdr,
Berenstein:2022unr,Lawrence:2022vsb,Lin:2023owt}. 
They are usually implemented with positivity constraints 
associated with the physical assumption of unitarity.\footnote{See however \cite{Khan:2022uyz} for the use of positivity constraints in non-Hermitian models. }
However, the violation of reflection positivity frequently occurs in statistical physics models.
The relevant theory can be related to nonunitary quantum systems, 
where the positivity principle does not apply.
The bootstrap study of such models necessitates alternative principles.
One of the potential candidates is the principle of nullness, i.e. the existence of many null states \cite{Li:2022prn}.\footnote{Here we considered the Hermitian quartic oscillator.
However, the null bootstrap can also be applied to non-Hermitian theories \cite{Li:2022prn}.
The perturbative null bootstrap method presented in the present work had been applied to the study of  $\mathcal{PT}$ symmetric non-Hermitian theories in \cite{John:2023him}.}
In the context of 2D conformal field theory \cite{Belavin:1984vu,DiFrancesco:1997nk}, the existence of null states is closely related to the quantization conditions on the scaling dimensions and the central charges of the minimal models, 
which imply that these physical parameters can only take certain discrete values. 
Only a small subset of the minimal models further obey the unitarity assumption.
A prominent example of the nonunitary case is the $\mathcal M(5,2)$ minimal model \cite{Cardy:1985yy,Cardy:1989fw}, 
which describes the critical behavior of the Yang-Lee edge singularity \cite{Yang:1952be,Lee:1952ig, Kortman:1971zz, Fisher:1978pf}.\footnote{The general $d$ conformal bootstrap program \cite{Ferrara:1973yt,Polyakov:1974gs} was revived by the seminal work \cite{Rattazzi:2008pe}. We refer to \cite{Poland:2018epd} for a comprehensive review. }

The principle of nullness postulates that many states are orthogonal to all states.
From the algebraic perspective, the null states are related to the left ideals in the operator algebra, 
 since the action of any operator on a null state also gives a null state.
For the standard quantum mechanics with a single position operator, the operator algebra is generated by 
the position operator $x$ and the momentum operator $p$.  
They satisfy the canonical commutation relation
\begin{align}
[x,p]\equiv xp-px=i\hbar\,.
\end{align}
Below, we will set $\hbar$ to 1. 
A  representation of the abstract operator algebra can be induced by a state
\begin{align}
\rho:\quad\mathcal A\rightarrow \mathbb C\,,
\label{A-C}
\end{align}
which is a linear functional mapping the elements of the operator algebra to complex numbers. 
Then one may construct the space of states as a representation of $\mathcal A$ on $\mathcal H$
\begin{align}
\pi:\quad\mathcal A\rightarrow \text{End}(\mathcal H)\,,
\end{align}
and show the existence of a vector $\psi_\rho\in \mathcal H$ with
\begin{align}
\rho(A)=\langle \psi_\rho|A|\psi_\rho\rangle:=\langle \psi_\rho,\pi(A)\,\psi_\rho\rangle\,,
\end{align}
for all $A\in \mathcal A$. 
Typically, $\mathcal H$ is a quotient vector space
\begin{align}
\mathcal H:=\mathcal A/N\,,
\end{align}
where $N$ is a left ideal in $\mathcal A$, corresponding to the subspace of null states. 
The null subspace plays a crucial role in the null bootstrap program, 
which aims to classify physical solutions and extract concrete predictions from the null state condition \cite{Li:2022prn}. 
From the algebraic viewpoint, this can be viewed as a classification program based on the ideals in operator algebra. 
Under some conditions, 
the rigorous construction of a Hilbert space $\mathcal H$ with a cyclic vector  $\psi_\rho$ 
is known as the Gelfand-Naimark-Segal construction \cite{GN,S}.

For physicists, the dynamics of a concrete quantum mechanical model is specified by a Hamiltonian, 
whose eigenstates are labeled by the energy $E$.\footnote{We will restrict to the bound states in a discrete and nondegenerate spectrum. }
The measurable information includes the energy spectrum and the matrix elements.
The choice of a Hamiltonian\footnote{A Hermitian theory implies 
$\langle E|\mathcal O(H-E)|E\rangle=\langle E|(H-E)\,\mathcal O|E\rangle=0$, 
corresponding to a family of operator-algebra representations parametrized by the energy $E$. }
and then an energy eigenstate 
leads to a concrete representation of the operator algebra. 
The mapping \eqref{A-C} is realized by the expectation values of different operators in the chosen state. 
One can also reconstruct the space of states. 
Matrix elements can be obtained from ladder operators that connect different eigenstates. 
For example, the off-diagonal matrix elements in the energy representation are given by
$\langle E|A|E'\rangle=\langle E|A\, L_{E'E}|E\rangle$.

For concrete applications of the null bootstrap, let us consider some basic quantum mechanical models.
In the textbook example of the harmonic oscillator
\begin{align}
H=\frac{1}{2}p^{2}+\frac{1}{2}x^2\,,
\end{align}
the eigenstates satisfying $H|n\rangle=E_{n}|n\rangle$ are connected by Dirac's ladder operators
\begin{align}
|n+k_1-k_2\rangle\propto (a)^{k_2}(a^\dagger)^{k_1}|n\rangle\,,
\end{align}
where the lowering and raising operators are
\begin{align}
	\label{a-ad-in-terms-of-xp}
	a=\frac{1}{\sqrt{2}}(x+ip)\,,\qquad a^{\dagger}=\frac{1}{\sqrt{2}}(x-ip)\,.
\end{align}
The energy levels are labeled by $n$.
It is well known that the Hamiltonian $H=a^\dagger a+\frac 1 2$ is linear in the number operator 
\begin{align}
\mathcal N=a^\dagger a\,.
\end{align} 
Together with Dirac's ladder operators, they form a closed algebra
\begin{align}
[a, a^\dagger]=1\,,\quad
[\mathcal N, a]=-a\,,\quad
[\mathcal N, a^\dagger]=a^\dagger\,.
\end{align}
A direct consequence of the commutators is that the energy spectrum has a constant spacing, i.e., $E_{n+1}-E_n=1$. 
Therefore, Dirac's ladder operators furnish a natural set of building blocks for the operator algebra of the harmonic oscillator.
For a spectrum that is bounded from below, the ground state with the lowest energy should be annihilated by the lowering operator
\begin{align}
	\label{annihilation-equation-harmonic}
	a|0\rangle=0\,.
\end{align}
This annihilation equation provides an example of the null state condition generated by the lowering operator.
We can also construct the null states from excited states, such as $a^{k}|n\rangle=0$ with $k=n+1$.
The stationary Schr\"{o}dinger equation also gives rise to null states
\begin{align}
	\label{null-state-trivial-harmonic}
	\(H-E_{n}\)|n\rangle=0\,.
\end{align}
which is called trivial in \cite{Li:2022prn} 
because it is satisfied by definition and does not lead to any constraint on $E$.

As the harmonic oscillator is well understood, 
it is more interesting to study the anharmonic oscillators, which are usually not exactly solvable at finite coupling.\footnote{Some special potentials can also lead to exact solutions, 
such as the Morse potential and the P\"{o}schl-Teller potential. 
We refer to \cite{Dong} for a review of the factorization method and the ladder operators associated with the underlying Lie algebra. 
We would like to emphasize that the existence of ladder operators does not rely on dynamical symmetries. }
Since quantum mechanics can be viewed as a $(0+1)$-dimensional quantum field theory, 
the anharmonic oscillators provide a testing ground for novel field theory methods. 
For example, a potential with a quartic term $x^4$ can be viewed as a $\phi^4$ theory in $(0+1)$ dimension \cite{Bender:1969si}. 
A curious question is 
whether there exists a natural set of building blocks for the anharmonic operator algebra. In this work, we will focus on the quartic case:
\begin{align}
	\label{quartic-anharmonic-hamiltonian}
	H_{\text{AH}}=\frac{1}{2}p^{2}+\frac{1}{2}x^2+g\sp x^4\,,
\end{align}
which is related to the Ising universality class in higher dimensions. 
It is known that the corresponding energy spectrum does not have a constant spacing, 
which in fact depends on the occupation number nonlinearly. 
Energy eigenstates are still expected to be connected by certain ladder operators. 
We would like to know if these ladder operators have a simple algebraic structure. 
If not, we may need completely different operators to connect different pairs of eigenstates.

Recently, the nonperturbative null bootstrap results of the quartic and cubic anharmonic oscillators   
suggest the existence of some underlying algebraic structure in the anharmonic ladder operators \cite{Li:2022prn}. 
However, these properties are only studied numerically and approximately 
due to the nonperturbative truncation scheme. 
In order to obtain analytical and exact results,  
we assume $g$ is small and make use of perturbation theory in this work. 
The null state in the anharmonic oscillators receives perturbative corrections
\begin{align}\label{annihilation-anharmonic}
	\(a+O(g)\)|0\rangle_{\text{AH}}=0\,,
\end{align}
The trivial null state again takes the form $\(H_{\text{AH}}-E_{0,\text{AH}}\)|0\rangle_{\text{AH}}=0$.
We will use the null state condition to formulate the bootstrap constraints for the observables, to determine the energy spectrum and to derive the analytic expressions of the ladder operators. 

More ambitiously, the bootstrap program aims to classify and solve the dynamical information by basic principles and consistency constraints. We have a curious question: 
\begin{itemize}
	\item
	How is the dynamics encoded in the null bootstrap?
\end{itemize}
To address this question to some extent, we will show that dynamical constraints from the Schr\"{o}dinger equation are related to certain null states generated by ladder operators. 

After investigating the quantum mechanical bootstrap in the Hamiltonian formalism, 
it is natural to consider the Lagrangian formalism. 
Therefore, we also apply the null state condition to solving Dyson-Schwinger (DS) equations \cite{Dyson:1949ha,Schwinger:1951ex,Schwinger:1951hq}, 
the self-consistency equations for the $n$-point\footnote{Here the $n$ should not be confused with the label for energy levels.} Green's functions. 
Since the DS equations can serve as an alternative to operator theory, 
we expect to obtain the same results as those in the Hamiltonian formalism. 
When solving the DS equations, one obstacle is that they form an underdetermined system,\footnote{An early reference on the DS equations in $\phi^{4}$ theory is \cite{Bender:1988bp}.
It was shown that the existence of a weak-coupling expansion can uniquely determine the Green's functions, while additional conditions are needed for the strong-coupling expansion.} 
as higher DS equations involve higher-point Green's functions.

In a simple scheme, one can close the system by setting high-point connected Green's functions to zero, but this produces results that do not converge to the exact values, as emphasized recently in \cite{Bender:2022eze,Bender:2023ttu}.
A more sophisticated approach is to replace high-point connected Green's functions by their large-$n$ asymptotic behaviors \cite{Bender:2022eze,Bender:2023ttu}, which gives numerically accurate results.
This approach has been carried out at $d=0$ and seems more challenging at higher dimensions.
A different avenue proposed recently in \cite{Li:2023nip} is to resolve the DS indeterminacy by the null state condition.\footnote{The two approaches can be unified by the principle of minimal singularity \cite{Li:2023ewe}. }
It was shown that the approximate, numerical results converge rapidly to the exact values for both $d=0$ and $d=1$.

To obtain analytic and exact results, we will investigate the null state approach in perturbation theory.
To be more explicit, we want to solve the following set of DS equations in the weak-coupling expansion:
\begin{align}
	\label{DS-equations}
	&\(\pa_{t}^{2}+1\)G_{n}(t,t_{1},t_{2},\ldots)
	+4g\sp G_{n+2}(t,t,t,t_{1},t_{2},\ldots)
	\nn
	=&\;-i\sum_{j=1}^{n}
	\d(t-t_{j})G_{n-2}(t_{1},t_{2},\ldots,t_{j-1},t_{j+1},\ldots)\,,
\end{align}
where the coupling constant $g$ is small. 
The Green's functions are the correlation functions of the Heisenberg picture operators $x(t_{k})$
\begin{align}
	G_{n}(t_{1},t_{2},\ldots)
	=\langle0|T\{x(t_{1})x(t_{2})\ldots\}|0\rangle\,,
\end{align}
where $T$ is the time-ordering operator and $|0\rangle$ denotes the ground state.

Assuming the existence of the weak-coupling expansion, the DS equations can  reproduce the standard perturbation theory results.
The solutions for the Green's functions will generally have many free parameters.
To obtain the physical solutions, one can fix the free parameters by imposing certain boundary conditions on two-point Green's function $G_{2}(t_{1},t_{2})$ at large $|t_1-t_2|$.\footnote{This is similar to the Feynman's $i\e$ prescription in quantum field theory.}
For higher-point Green's functions, the free parameters can be fixed by the cluster decomposition principle.\footnote{The implications of the cluster decomposition are more clear after Wick rotation $t\rightarrow -i\t$.}
In the null state approach, the derivation of the physical solutions does not rely on the boundary conditions at infinity.
Instead, it is crucial that the action of certain linear combinations of $x(t)$ and $\frac{\dd}{\dd t}{x}(t)$ on the ground state amounts to higher-order terms in the perturbative expansion.
In the harmonic limit $g=0$, we have
\begin{align}
	\(x(t)+i\,\frac{\dd}{\dd t}{x}(t)\)|0\rangle=0\,,
\end{align}
which is precisely the annihilation equation \eqref{annihilation-equation-harmonic}. 
According to the Heisenberg equation of motion, we have $\frac{\dd}{\dd t}{x}(t)=-i[x(t),H]=p(t)$ and
this relation remains exact after turning on the quartic perturbation in $x(t)$.
Since the DS equations are formulated in the Lagrangian formalism, we will use the time derivative $\frac{\dd}{\dd t}{x}(t)$ instead of the momentum $p(t)$.
As in \eqref{annihilation-anharmonic}, the annihilation equation receives perturbative corrections in $g$
\begin{align}
	\(x(t)+i\frac {\dd}{\dd t}x(t)+O(g)\)|0\rangle_{\text{AH}}=0\,.
\end{align}
An important consequence is that its inner products with certain states give rise to a set of relations 
for the Green's functions.
We can solve for the Green's functions order by order using the perturbed ladder operators.

In Table \ref{results}, we summarize the main results in the Hamiltonian and Lagrangian formalisms.
Motivated by the nonperturbative approach in \cite{Li:2022prn}, 
we also obtain higher order results for the perturbative low energy levels in \eqref{E0-K=1}, \eqref{E0-K=2}, \eqref{E1-K=2} and \eqref{E0-K=3}-\eqref{K=4-exact}.

\begin{table}[h!]
	\centering
	\begin{tabular}{|c|c|c|c|c|c|} 
		\hline
		& $E_n$ & $L_{\pm1}$ & $G_2$ & $G_4$ & $G_6$ \\ [0.5ex] 
		\hline
		$g^{0}$ & \eqref{En-g0} & \eqref{ladder-nontrivial-g0} & \eqref{G2-g0} & \eqref{G4-g0} & \eqref{G6-g0}  \\ 
		$g^{1}$ & \eqref{En-g1} & \eqref{ladder-nontrivial-g1} & \eqref{G2-g1} & \eqref{G4-g1} &  \\
		$g^{2}$ & \eqref{En-g2} & \eqref{ladder-nontrivial-g2} & \eqref{G2-g2} &  &  \\
		$g^{3}$ & \eqref{En-g3} & \eqref{ladder-nontrivial-g3} &  &  &  \\
		\hline
	\end{tabular}
	\caption{Perturbative results for the quartic anharmonic oscillator \eqref{quartic-anharmonic-hamiltonian}, where the coupling constant $g$ is small.
		We derive the eigenenergies $E_n$, anharmonic ladder operators $L_{\pm1}$ and some $n$-point Green's functions $G_n$ from the null state condition.}
	\label{results}
\end{table}

The rest of the paper is organized as follows.
In Sec. \ref{The null bootstrap in perturbation theory}, we use the null bootstrap to investigate the quartic anharmonic oscillator in perturbation theory.
We present two different procedures in Sec. \ref{Complete procedure 1} and Sec. \ref{Reduced procedure 1}, and discuss the algebraic properties of the anharmonic operator algebra in Sec. \ref{Anharmonic operator algebra}.
We then consider the DS equations in section \ref{Dyson-Schwinger equations}, 
where we make use of the null state condition and solve for the $n$-point Green's functions.
In Sec. \ref{Discussion}, we summarize the results and discuss future directions.
For comparison, the results from the traditional perturbation method are summarized in the Appendix.

\section{The null bootstrap}
\label{The null bootstrap in perturbation theory}
In the null bootstrap, physical solutions are derived from the null state condition. 
By definition, an exact null state $\psi_\text{null}$ is orthogonal to arbitrary test states, so we have
\begin{align}
\label{null-condition}
\langle\psi_\text{test}|\psi_\text{null}\rangle=0\,,
\end{align}
where $\psi_\text{test}$ can be any state. 
By considering more general types of test states, one can deduce stronger constraints, 
then the properties of the physical states annihilated by the ladder operators are determined more precisely. 
At finite coupling, this can be performed numerically, then approximate results of high precision 
are obtained by truncating the search spaces of the null and test states \cite{Li:2022prn}. 
In perturbation theory, we can carry out the null bootstrap analytically  
and derive the exact perturbative series using truncated search spaces of finite dimensions.

We will focus on the quartic anharmonic oscillator with $H=\frac{1}{2}p^{2}+\frac{1}{2}x^2+g\sp x^4$.
To simplify the notation, we will not write ``AH'' explicitly.
In the small $g$ expansion, 
the null state condition $(a^k+O(g))|\psi\rangle=0$ determines the $k$ low-lying states, where $a$ is Dirac's lowering operator.
We will focus on the expectation values associated with an energy eigenstate labeled by $E$
\begin{align}
	\langle E|\mathcal O| E\rangle=\langle\mathcal O\rangle_E\,.
\end{align}
We carry out the analysis of the quartic anharmonic oscillator based on three assumptions:
\begin{enumerate}
\item
The Hamiltonian is Hermitian and the eigenvalue $E$ is real.\footnote{In general, the eigenvalues of a bounded self-adjoint operator are known to be real. }
We have the Schr\"{o}dinger-like equations
\begin{align}
	\label{consistency-relation}
	\left\langle H \mathcal{O} \right\rangle_{E} = E \left\langle \mathcal{O} \right\rangle_{E} = \left\langle \mathcal{O} H \right\rangle_{E}\,,
\end{align}
which can be derived from the stationary Schr\"{o}dinger equation $H| E\rangle=E| E\rangle$ and the Hermitian inner product $\langle\,\cdot\,|\,\cdot\,\rangle$.
It turns out that the expectation value $\langle x^{m_{1}} p^{m_{2}} \rangle_E$ can be expressed in terms of the coupling constant $g$, the energy $E$, the expectation values $\langle x^{2} \rangle_E$ and $\langle 1 \rangle_E$.
We choose the normalization convention $\langle 1 \rangle_E=1$.
\item
The independent parameters $E$ and $\langle x^{2} \rangle_{E}$ have the perturbative expansions
\begin{align}\label{E-x2-g-expansion}
	E = \sum_{i=0,1,2,\dots}g^i E^{(i)}\,,\quad 
	\langle x^{2} \rangle_{E} =\sum_{i=0,1,2,\dots} g^{i} \langle x^{2} \rangle^{(i)}_{E}\,,
\end{align}
which are formal power series in the coupling constant $g$. 
\item
The expectation value $\langle x^{m_{1}} p^{m_{2}} \rangle_{E}$ is regular in the $g \rightarrow 0$ limit, 
where $m_1,m_2$ are non-negative integers.
This implies that $\langle x^{2} \rangle_{E}$ can be expressed in terms of $g$ and $E$
\begin{align}
	\label{x2-E}
	\langle x^{2} \rangle_{E}
	=\;&E
	-\frac{9}{8}g\pt{
		1+4E^{2}
	}
	+\frac{3}{2}g^{2}\pt{
		19E+26E^{3}
	}
	\nn
	&-\frac{9}{256}g^{3}\pt{
		1825+17048E^{2}+11792E^{4}
	}+\ldots\,,
\end{align}
so $\langle x^{m_{1}} p^{m_{2}} \rangle_{E}$ is a function of the coupling constant $g$ and the energy $E$.
\end{enumerate}

The third assumption is particularly interesting. 
It is not immediately clear why we have \eqref{x2-E}.
Some numerical signatures about the subtlety of the $g\rightarrow 0$ limit were noticed in \cite{Bhattacharya:2021btd}.
In fact, this is similar to the additional constraints from the existence of the weak-coupling expansion when solving the DS equations.
The crucial point is that the consistency relations, whether they are the Schr\"{o}dinger-like equations \eqref{consistency-relation} or the DS equations \eqref{DS-equations}, allow singular behavior of the expectation values or the Green's functions in the limit $g\rightarrow0$.
The assumption that the $g\rightarrow0$ limit is regular or the weak-coupling expansion exists 
implies the absence of singularities and leads to additional constraints on the free parameters.\footnote{This is also similar to the multiplet recombination approach to the $d=4-\e$ Wilson-Fisher conformal field theory, where one assumes that the $\e\rightarrow0$ limit is smooth; see \cite{Rychkov:2015naa} for more details.}
The basic idea behind this type of constraints is that the physical data of a weakly interacting theory should allow continuous deformation into the free theory limit.

To be more explicit, let us examine some concrete expectation values.
Under the first assumption, $\langle x^{m_{1}} p^{m_{2}} \rangle_{E}$ is expressed as
\begin{align}
	\label{xmpn-expectation-value}
	\langle x^{m_{1}} p^{m_{2}} \rangle_{E}
	&=\frac{1}{g^{j}}P_{1}(E,\langle x^2 \rangle_{E})
	+\frac{1}{g^{j-1}}P_{2}(E,\langle x^2 \rangle_{E})
	+\ldots\,,
\end{align}
where $P_{i}(E,\langle x^2 \rangle_{E})$ are polynomials in $E$ and $\langle x^2 \rangle_{E}$.
The number $j$ is the highest order of $1/g$, which depends on $m_1, m_2$:
\begin{align}
	q=\begin{cases}
		\text{max}\(\frac{m_{1}+m_{2}}{2}-1,0\)&\quad\text{if $m_{1},m_{2}$ are both even} \\
		\text{max}\(\frac{m_{1}+m_{2}}{2}-2,0\)&\quad\text{if $m_{1},m_{2}$ are both odd.}
	\end{cases}
\end{align}
We assume that the parity symmetry is unbroken, 
so the expectation value vanishes if $m_{1}+m_{2}$ is odd.

The important point is that the expression \eqref{xmpn-expectation-value} contains terms with negative powers of $g$, which are singular in the $g\rightarrow0$ limit.
To eliminate these singularities, we impose additional constraints on the small $g$ expansion of $E$ and $\langle x^2 \rangle_{E}$. 
This will lead to a set of relations among $\langle x^2 \rangle_{E}^{(j)}$ and $E^{(j)}$.
For example, the case of $m_1=m_2=2$ reads
\begin{align}
	\label{expectation-value-x2p2}
	\langle x^{2}p^{2} \rangle_{E}
	=\frac{1}{15g}\pt{\langle x^{2}\rangle_{E}-E}
	+\frac{1}{10}\pt{-3+8E\langle x^{2}\rangle_{E}}\,,
\end{align}
whose $g\rightarrow0$ limit is singular if $E$ and $\langle x^{2}\rangle_{E}$ are completely independent.
To avoid the singular behavior, the perturbative series of $E$ and $\langle x^{2} \rangle_{E}$ should satisfy 
the constraint $\langle x^{2} \rangle^{(0)}_{E}=E^{(0)}$. 
At higher powers, the expectation value
$\langle x^{4}p^{2} \rangle_{E}$ reads
\begin{align}
	\langle x^{4}p^{2} \rangle_{E}=\frac{2}{105g^{2}}\(E-\langle x^{2} \rangle_{E}\)+\frac{1}{420g}\(-9+80E^{2}-116E\langle x^{2} \rangle_{E}\)+\text{regular}\,.
\end{align}
Note that this is consistent with \eqref{expectation-value-x2p2} 
because  the $1/g^{2}$ singularity is removed by $\langle x^{2} \rangle^{(0)}_{E}=E^{(0)}$.
In addition, the $1/g$ singularity is eliminated by 
\begin{align}
	\langle x^{2} \rangle^{(1)}_{E}
	&=\frac{1}{8}\pt{
		-9-36\pt{E^{(0)}}^{2}+8E^{(1)}
	}\,.
\end{align}
By considering higher powers in $x$ and $p$, we obtain stronger constraints and determine higher-order terms in the perturbative series, such as
\begin{align}
	\langle x^{2} \rangle^{(2)}_{E}
	&=\frac{1}{2}\pt{
		57E^{(0)}+78\pt{E^{(0)}}^{3}-18E^{(0)}E^{(1)}+2E^{(2)}
	}\,,
	\\
	\langle x^{2} \rangle^{(3)}_{E}
	&=\frac{1}{256}\left(
		-16425-153432\pt{E^{(0)}}^{2}-106128\pt{E^{(0)}}^{4}+7296E^{(1)}
	\right.
	\nn
	&\phantom{=}\left.
		+29952\pt{E^{(0)}}^{2}E^{(1)}-1152\pt{E^{(1)}}^{2}-2304E^{(0)}E^{(2)}+256E^{(3)}
	\right)\,.
\end{align}
The regularity of $\langle x^{6}p^{2} \rangle_{E}$ determines $\langle x^{2} \rangle^{(2)}_{E}$, while that of $\langle x^{8}p^{2} \rangle_{E}$ fixes both $\langle x^{2} \rangle^{(2)}_{E}$ and $\langle x^{2} \rangle^{(3)}_{E}$.
If a high power expectation value has a regular limit, then the expectation values with low powers are automatically regular. 
The regularity constraints of different expectation values are consistent with each other. 
In fact, it suffices to impose that $\langle x^{m} \rangle$ is regular in the $g\rightarrow0$ limit. 
Furthermore, we can repackage these relations by expressing $\langle x^{2} \rangle_{E}$ in terms of $E$, which is shown in \eqref{x2-E}. 

In principle, the variable $\langle x^{2} \rangle_{E}$ is completely determined by $E$ to all orders in perturbation theory. In practice, we only need to know $\langle x^{2} \rangle_{E}$ in terms of $E$ to a certain order in $g$. This is because we will deal with a finite set of expectation values. Suppose that the strongest singularity among them is $1/g^{j_{0}}$ and we are interested in perturbative corrections up to order $g^{i}$. 
We should examine an expectation value with a singularity $1/g^{j_{0}+i+1}$, which is not in this set of expectation values. 
The regularity assumption then determines $\langle x^{2} \rangle_{E}$ in terms of $E$ to order $g^{j_{0}+i}$.  
As a result, all the expectation values under consideration are regular and expressed only in terms of $E$ to order $g^{i}$. 

We are now in the position to carry out the complete procedure of the null bootstrap, which consists of two steps:
\begin{enumerate}
	\item
	Determine the full spectrum and exact ladder operators by Schr\"{o}dinger-like equations \eqref{schrodinger-like-equation}.  
	\item
	Impose the null state condition \eqref{null-condition-explicit} on the unboundedly low energy states and solve the remaining parameters.
\end{enumerate}
In general, it is challenging to derive the complete, nonperturbative spectrum in the first step.\footnote{We notice that the solutions for the full energy spectrum and exact (level-$1$) ladder operators are always derived at the same time. We are led to the question whether they contain equivalent information. It is clear that we can deduce the full spectrum from the exact (level-$1$) ladder operators. On the other hand, 
the equation $[H,\mathcal{O}]=\mathcal{O}(\ldots)$ is sufficient for deducing an exact ladder operator $\mathcal{O}$ if the full energy spectrum is known, 
where the ellipsis becomes the energy spacing. 
Therefore, the full energy spectrum and exact (level-$1$) ladder operators do contain equivalent information.}
In Sec. \ref{Reduced procedure 1}, this obstacle is circumvented by the reduced procedure, 
in which we directly solve for the low-lying data using the null state condition. 
In perturbation theory, we can show that the low-lying energy spectrum and matrix elements obtained in the reduced procedure match with those from the complete procedure. 
Therefore, the reduced procedure is at least as strong as the complete procedure concerning the low-lying information. 
In fact, we can derive higher order results more easily, which will be explained in Sec. \ref{Reduced procedure 1}. 
These perturbative results help us to better understand the convergence pattern of the nonperturbative results in \cite{Li:2022prn}. 

\subsection{Complete procedure}
\label{Complete procedure 1}
In the complete procedure, we first determine the energy spectrum by 
the Schr\"{o}dinger-like null state condition 
\begin{align}
	\label{schrodinger-like-equation}
	\left\langle \mathcal{O}_{\text{test}} (H - E') \, L_{E'E} \right\rangle_{E} = 0\,,
\end{align}
which corresponds to the stationary Schr\"{o}dinger equation $(H - E')\, |{E'}\rangle= 0$. 
The two eigenstates $|\psi_E\rangle$ and $|\psi_{E'}\rangle$ is connected by the ladder operator $L_{E'E}$. 
We require that \eqref{schrodinger-like-equation} is valid for arbitrary $\mathcal{O}_{\text{test}}$ 
in the form of polynomials in $x$ and $ip$. 
We first focus on the level-$1$ ladder operators and solve the null state condition to order $g^{3}$.  Then we derive the level-$k$ ladder operators using the level-$1$ ladder operators.
In the end, we use these ladder operators to compute matrix elements.

Before presenting the details, it is useful to note some general features of the solutions to \eqref{schrodinger-like-equation}.
In the $g \rightarrow 0$ limit, the reference energy $E$ is labeled by an integer $n$. 
The solutions to the Schr\"{o}dinger-like equation with energy $E'$ are labeled by $k$, 
which denotes the number of energy levels shifted by the ladder operator $L_{E'E}$.
We adopt the following notations for convenience:
\begin{align}
	E\rightarrow E_{n}\,,\quad
	E'\rightarrow E_{n\pm k}\,,\quad
	L_{E'E}\rightarrow L_{\pm k}\,,\quad
\end{align}
where $L_{\pm k}$ are the level-$k$ ladder operators. 
The discrete energy spectrum and ladder operators can be continuously deformed to the anharmonic case and we assume the existence of the following perturbative series
\begin{align}
	E_{n} &= E^{(0)}_{n}+g E^{(1)}_{n} + g^{2} E^{(2)}_{n} + g^{3} E^{(3)}_{n} +\ldots\,,
	\\
	E_{n \pm k} &= E^{(0)}_{n \pm k} + g E^{(1)}_{n \pm k} + g^{2} E^{(2)}_{n \pm k} + g^{3} E^{(3)}_{n \pm k} + \ldots\,,
	\\
	\label{ladder-g-expansion}
	L_{\pm k} &= L^{(0)}_{\pm k} + g L^{(1)}_{\pm k} + g^{2} L^{(2)}_{\pm k} + g^{3} L^{(3)}_{\pm k} + \ldots\,.
\end{align}
In general, there exists a family of trivial solutions for $L_{\pm k}$ 
due to the Schr\"{o}dinger equation
\begin{align}
	\label{trivial-general}
	L_{\pm k}\big|_{\text{trivial}}=(\text{arbitrary operator})\times(H-E_{n})\,,
\end{align}
where $L_{\pm k}\big|_{\text{trivial}}|{E_n}\rangle=0$ automatically. 
A nontrivial solution for $L_{\pm k}$ should connect two energy eigenstates
\begin{align}
L_{\pm k}\big|_{\text{n.t.}}|{E_n}\rangle\propto |{E_{n\pm k}}\rangle\,,
\end{align}
where we have used ``n.t.'' to indicate the nontrivial part. 
From explicit calculations, we find that the Schr\"{o}dinger-like equation \eqref{schrodinger-like-equation} leads to recursion relations for $E_n$, so one can deduce the complete energy spectrum from one energy level. 
For a stable system, the energy spectrum should be bounded from below. 
The states lower than the ground state should satisfy the null state condition, such as
\begin{align}
	\label{null-condition-explicit}
	\left\langle
	\mathcal{O}_{\text{test}} \, L_{-1}
	\right\rangle_{E_{0}} 
	=0\,.
\end{align}
In this way, the ground-state energy $E_0$ is determined order by order in $g$.
Using this boundary condition, we can further deduce the energy spectrum from the energy recursion relations. 
The null state condition \eqref{null-condition-explicit} can be generalized to the expectation values of the excited states, which reads 
$\left\langle\mathcal{O}_{\text{test}} \, L_{-k}\right	\rangle_{E_{n}} =0$ with $n=k-1$.

\subsubsection*{The order $g^{0}$}
First, we consider \eqref{schrodinger-like-equation} at order $g^{0}$.
The explicit expressions of $L_{\pm k}$ are formulated in terms of polynomials in $x$ and $ip$.
We truncate the search space of null states by using the ansatz
\begin{align}\label{ladder-ansatz-g0}
	L^{(0)}_{\pm k} = \sum_{m_{2} = 0}^{K-m_{1}} \sum_{m_{1} = 0}^{K} A_{\pm k,m_{1},m_{2}}^{(0)} x^{m_{1}} (i p)^{m_{2}}\,,
\end{align}
where $K$ denotes the degree of the polynomial and the coefficients $A_{\pm k,m_{1},m_{2}}^{(0)}$ are assumed to be real.
To remind the reader, the integer $k$ is the number of energy levels shifted by the ladder operator.
For notational simplicity, we suppress the $K$ index in $A_{\pm k,m_{1},m_{2}}^{(0)}$. 

At the lowest truncation order $K=1$, we find two sets of solutions labeled by $\pm1$. 
They correspond to the raising and lowering operators
\begin{align}
	L^{(0)}_{\pm 1}=A_{\pm 1,1,0}^{(0)}\(x \mp ip\)\,,
\end{align}
where $A_{\pm 1,1,0}^{(0)}$ are free real parameters related to the normalization.
The trivial terms are absent because they are at least quadratic in $x$ and $ip$.
For higher $K$,\footnote{The $k>1$ solutions appear when $K>1$, but we focus on the $k=1$ solutions at the moment.} the Schr\"{o}dinger-like equation \eqref{schrodinger-like-equation} determines  $A_{\pm 1,m_{1},m_{2}}^{(0)}$ up to some free real parameters, which will be $E_{n}^{(0)}$ dependent and denoted as $B_{\pm 1}^{(0)}$, $C_{\pm 1,j_{1},j_{2}}^{(0)}$. 
Again the $K$ index is suppressed for simplicity. 
The general solutions\footnote{We assume that $n$ is sufficiently large such that $|n\rangle$ cannot be annihilated by ladder operators in any truncation $K$.} for the level-$1$ ladder operators take the form
\begin{align}
\label{general-g0}
	L_{\pm 1}^{(0)}
	=B_{\pm 1}^{(0)}\(x \mp ip\)
	+\(
		\sum_{j_{1},j_{2}=0}^{\infty}
		C_{\pm 1,j_{1},j_{2}}^{(0)}\,
		x^{j_{1}}(ip)^{j_{2}}
	\)
	\(\frac{1}{2}x^{2}+\frac{1}{2}p^{2}-E_{n}^{(0)}\)\,,
\end{align}
where $B_{\pm 1}^{(0)}$ are related to the normalization and $C_{\pm 1,j_{1},j_{2}}^{(0)}$ are related to the trivial terms. 
Since the sum of a nontrivial solution and a trivial solution is also nontrivial, 
there are some ambiguities in the explicit expressions of the nontrivial part. 
Below, we will remove these ambiguities  
by fixing the normalization and requiring the nontrivial part is $E_{n}$ independent.\footnote{The $E_{n}^{(0)}$ dependence of $B_{\pm 1}^{(0)}$ is completely fixed by the normalization condition. 
To completely fix the $E_{n}^{(0)}$ dependence of $C_{\pm 1,j_{1},j_{2}}^{(0)}$, 
we further impose the gauge-fixing condition.}

Since the expectation values are expressed as functions of $E_{n}$, it is useful to find the explicit expression of $E_{n}$ first.
At order $g^{0}$, the Schr\"{o}dinger-like equation \eqref{schrodinger-like-equation} gives the energy recursion relation\footnote{There is also an equivalent relation $E_{n+1}^{(0)}=E_{n}^{(0)}+1$.
}
\begin{align}
	E_{n-1}^{(0)}=E_{n}^{(0)}-1\,.
\end{align}
The null state condition \eqref{null-condition-explicit} yields $E_{0}^{(0)}=\frac{1}{2}$, so we have
\begin{align}\label{En-g0}
	E_{n}^{(0)}=n+\frac{1}{2}\,.
\end{align}
We will choose a specific normalization for $L_{\pm1}|n\rangle$ based on this result.

To determine the free parameters, we impose the following conditions:
\begin{itemize}
	\item 
	Normalization condition:  $|L_{+1}|n\rangle|^2=n+1$ and $|L_{-1}|n\rangle|^2=n$.\footnote{The difference in the norms is consistent with the commutation relation of $L_{+1}$ and $L_{-1}$.}	
	\item 
	Gauge-fixing condition: the nontrivial part is independent of $E_n$.
\end{itemize}
The trivial terms are not constrained by the first condition since they do not contribute to the norm.
To completely fix the expressions of the ladder operators, we impose the second condition 
to quotient out the trivial part. 
From the operator algebra perspective, 
it is natural that the expressions of the ladder operators 
do not depend on the choice of the eigenstates, i.e., $E_n$. 

Since the parameters are assumed to be real, there is an ambiguity in the sign of the ladder operators.
Our choice is that the zeroth-order nontrivial parts are the same as Dirac's ladder operators
\begin{align}
	\label{ladder-nontrivial-g0}
	L_{\pm 1}^{(0)}\big|_{\text{n.t.}}=\frac{1}{\sqrt{2}}(x \mp ip)\,.
\end{align}
where ``n.t.'' means the nontrivial part.
The general solution \eqref{general-g0} can be recovered by adding trivial solutions and overall normalization factors. 
Furthermore, we fix the relative phases of the energy eigenstates by
\begin{align}
	L_{+1}\big|_{\text{n.t.}}|n\rangle=\sqrt{n+1}\,|n+1\rangle\,,\quad
	L_{-1}\big|_{\text{n.t.}}|n\rangle=\sqrt{n}\,|n-1\rangle\,.
\end{align}
We will also use the same normalization and gauge-fixing conditions to determine the explicit expressions of the ladder operators at higher orders. 

\subsubsection*{The order $g^{1}$}
We extend the analysis to order $g^{1}$.
Since the zeroth-order ladder operators have been solved,  
we set the zeroth-order part to \eqref{ladder-nontrivial-g0}
\begin{align}
	L_{\pm k}=L_{\pm k}^{(0)}\big|_{\text{n.t.}}+gL^{(1)}_{\pm k}\,+\ldots,
\end{align}
and use $K$ to denote the truncation order of the $g^{1}$ ansatz
\begin{align}
\label{L1-g1}
	L^{(1)}_{\pm k} = \sum_{m_{2} = 0}^{K-m_{1}} \sum_{m_{1} = 0}^{K} A_{\pm1,m_{1},m_{2}}^{(1)} x^{m_{1}} (i p)^{m_{2}}\,.
\end{align}
As the free parameters start at first order, the trivial part at order $g^{1}$ takes a simple form 
\begin{align}
	\(\text{arbitrary operator}\)\times(H_{\text{harmonic}}-E_{n}^{(0)})\,,
\end{align}
and the anharmonic corrections to $H$ and $E_{n}$ are of higher order. 

To solve the Schr\"{o}dinger-like equation \eqref{schrodinger-like-equation} for arbitrary test operators, 
the truncation order in \eqref{L1-g1} should satisfy $K\geqslant 3$, 
which corresponds to the minimal shift in the energy level, i.e., $k=1$.
We gradually increase the value of $K$ and extract the general solution
\begin{align}
	\label{lowering-g1}
	L^{(1)}_{\pm 1}=&\;B_{\pm 1}^{(1)}(x \mp ip)+\frac{1}{\sqrt{2}}x^{3}\pm \frac{3}{2\sqrt{2}}x^{2}(ip)	-\frac{3}{2\sqrt{2}}x(ip)^{2}
	\nn
	&+\pt{
		\sum_{j_{1},j_{2}=0}^{\infty}
		C_{\pm 1,j_{1},j_{2}}^{(1)}\,
		x^{j_{1}}(ip)^{j_{2}}}
		\pt{\frac{1}{2}x^{2}+\frac{1}{2}p^{2}-E_{n}^{(0)}
	}\,.
\end{align}
The remaining freedom resides in the choice of the normalization and the trivial terms.
Here the sign of $L_{\pm1}^{(1)}$ is determined because we require that the parameters are real and the normalization does not depend on $g$.

As above, we first solve for $E_n$ before choosing a normalization.
At order $g^{1}$, the energy recursion relation reads
\begin{align}
	E_{n-1}^{(1)}=\frac{1}{2}
	\pt{
		3-6E_{n}^{(0)}+2E_{n}^{(1)}
	}\,.
\end{align}
The null state condition \eqref{null-condition-explicit} leads to
the boundary condition $E_{0}^{(1)}=\frac{3}{4}$,
so the first-order energy corrections are given by
\begin{align}\label{En-g1}
	E_{n}^{(1)}=\frac{3}{4}\pt{1+2n+2n^{2}}\,.
\end{align}
Then we impose the normalization and gauge-fixing conditions to
fix the expression of the first-order nontrivial part
\begin{align}
	\label{ladder-nontrivial-g1}
	L_{\pm 1}^{(1)}\big|_{\text{n.t.}}
	=\frac{1}{8\sqrt{2}}\(\pm 15x- 9ip+5x^{3}\pm 15x^{2}(ip)-9x(ip)^{2} \mp 3(ip)^{3}\)  \,,
\end{align}
which is different from simply setting $C_{\pm 1,j_{1},j_{2}}^{(1)}$ to zero due to our choice of the normalization condition. 
One can check that the lowering and raising operators are Hermitian conjugate to each other
\begin{align}\label{Hermitian-conjugate-ladder}
	L_{-1}^{(1)}\big|_{\text{n.t.}}
	=\pt{
		L_{+1}^{(1)}\big|_{\text{n.t.}}
	}^{\dagger}\,.
\end{align}
In the Appendix, we summarize the results from the traditional perturbation method and they agree exactly with the above results.

\subsubsection*{Higher orders}
At order $g^{2}$, we use the known expressions of the nontrivial parts at order $g^{0}$ and $g^{1}$
\begin{align}
	L_{\pm k}=L_{\pm k}^{(0)}\big|_{\text{n.t.}}+gL_{\pm k}^{(1)}\big|_{\text{n.t.}}+g^{2}L^{(2)}_{\pm k}\,+\ldots,
\end{align}
and the $g^2$ order terms are
\begin{align}
	L^{(2)}_{\pm k} = \sum_{m_{2} = 0}^{K-m_{1}} \sum_{m_{1} = 0}^{K} A_{\pm1,m_{1},m_{2}}^{(2)} x^{m_{1}} (i p)^{m_{2}}\,.
\end{align}
When solving the null state condition \eqref{schrodinger-like-equation}, the lowest truncation order for $L^{(2)}_{\pm k}$ is $K_\text{min}=5$, corresponding to the $k=1$ case.
By increasing $K$, we obtain the general solution
\begin{align}
	L_{\pm 1}^{(2)}
	=\;&B_{\pm 1}^{(2)}(x \mp ip)+\frac{27}{8\sqrt{2}}x\mp \frac{5}{4\sqrt{2}}x^{3}+9\sqrt{2}x^{2}(ip) \pm \frac{3}{2\sqrt{2}}x(ip)^{2}-\frac{13}{8\sqrt{2}}x^{5}
	\nn
	&+\frac{11}{2\sqrt{2}}x^{4}(ip)+\frac{37}{4\sqrt{2}}x^{3}(ip)^{2}\pm \frac{15}{4\sqrt{2}}x^{2}(ip)^{3}-\frac{39}{8\sqrt{2}}x(ip)^{4} 
	\nn
	&+\pt{
		\sum_{j_{1},j_{2}=0}^{\infty}
		C_{\pm 1,j_{1},j_{2}}^{(2)}\,
		x^{j_{1}}(ip)^{j_{2}}}
	\pt{\frac{1}{2}x^{2}+\frac{1}{2}p^{2}-E_{n}^{(0)}
	}\,.
\end{align}
The recursion relation from the null state condition \eqref{schrodinger-like-equation} is
\begin{align}
	E_{n-1}^{(2)}=\frac{1}{16}\pt{153-204E_{n}^{(0)}+276\pt{E_{n}^{(0)}}^{2}-48E_{n}^{(1)}+16E_{n}^{(2)}}\,.
\end{align}
Together with the boundary condition $E_{0}^{(2)}=-\frac{21}{8}$ from \eqref{null-condition-explicit}, we obtain
\begin{align}\label{En-g2}
	E_{n}^{(2)}=-\frac{1}{8}
	\pt{
		21+59n+51n^{2}+34n^{3}
	}\,.
\end{align}
Imposing the normalization and the gauge-fixing conditions, we have a unique expression for the nontrivial part
\begin{align}
	\label{ladder-nontrivial-g2}
	L_{\pm 1}^{(2)}\big|_{\text{n.t.}}=\;
	\frac{1}{128\sqrt{2}}\Big(&1509x\pm1419ip\mp 1670x^{3}+2766x^{2}(ip)\pm 2226x(ip)^{2}-986(ip)^{3}-77x^{5}
	\nn
	&\mp 835x^{4}(ip)+922x^{3}(ip)^{2}\pm 742x^{2}(ip)^{3}-493x(ip)^{4}  \mp 131(ip)^{5}  \Big)\,.
\end{align}
It is straightforward to repeat the procedure at order $g^{3}$ and the results are
\begin{align}\label{En-g3}
	E_{n}^{(3)}=\frac{3}{16}
	\pt{
		111+347n+472n^{2}+250n^{3}+125n^{4}
	}\,,
\end{align}
and
\begin{align}
	\label{ladder-nontrivial-g3}
	L_{\pm 1}^{(3)}\big|_{\text{n.t.}}
	=\frac{1}{1024\sqrt{2}}\Big(&\mp 223275x+234621ip-190455x^{3}\mp 541425x^{2}(ip)+573159x(ip)^{2}  
	\nn
	&\pm 184017(ip)^{3}\pm 74829x^{5}-188655x^{4}(ip)\mp 318150x^{3}(ip)^{2}
	\nn
	&+338538x^{2}(ip)^{3}\pm 145545x(ip)^{4} -57771(ip)^{5}+949x^{7}  \pm 24943x^{6}(ip)
	\nn
	&-37731x^{5}(ip)^{2}\mp 53025x^{4}(ip)^{3} +56423x^{3}(ip)^{4}  \pm 29109x^{2}(ip)^{5}
	\nn
	&-19257x(ip)^{6} \mp 4227(ip)^{7} \Big)\,.
\end{align}
The lowering and raising operators at higher orders are also related by Hermitian conjugation
\begin{align}\label{Hermitian-conjugate-ladder-2}
	L_{-1}^{(2)}\big|_{\text{n.t.}}
	=\(
	L_{+1}^{(2)}\big|_{\text{n.t.}}
	\)^{\dagger}\,,\qquad
	L_{-1}^{(3)}\big|_{\text{n.t.}}
	=\(
	L_{+1}^{(3)}\big|_{\text{n.t.}}
	\)^{\dagger}\,.
\end{align}
All these results agree with those from the traditional method in the Appendix.
In principle, one can perform  the complete procedure of the null bootstrap and determine the ladder operators and the energy spectrum to arbitrarily high order in $g$.

\subsection*{Ladder operators with higher level $k>1$}

So far we have not considered the $k>1$ solutions.
In fact, we can construct the nontrivial part of the level $k>1$ ladder operators from multiple level-1 ladder operators
\begin{align}
	\label{ladder-k}
	L_{\pm k}\big|_{\text{n.t.}}=\pt{L_{\pm 1}\big|_{\text{n.t.}}}^{k}\,.
\end{align}
They can be obtained from the normalization and gauge-fixing conditions:
\begin{itemize}
	\item 
	The norms are set by $|L_{+ k}|n\rangle|^{2}=(n+1)_{k}$ and $|L_{- k}|n\rangle|^{2}=(n+1-k)_{k}$, where $(x)_{y}={\Gamma(x+y)}/{\Gamma(x)}$ is the Pochhammer symbol.
	\item 
	The nontrivial part is independent of $E_{n}$.
\end{itemize}
The second condition is the same as that for the level-1 ladder operators, while the first one can be understood as the consequences of the repeated action of $L_{\pm 1}$. 

Although the general solutions are much more involved, 
we verify that they can be written in terms of $L_{\pm 1}\big|_{\text{n.t.}}$
\begin{align}\label{level-k-L}
L_{\pm k}=(\text{normalization})\big(\pt{L_{\pm 1}\big|_{\text{n.t.}}}^{k}+(\text{arbitrary operator})\times(H-E)\big)\,.
\end{align}
Therefore, there is no new independent solution at $k>1$. 
All the nontrivial algebraic information is encoded in the level-1 ladder operators 
$L_{\pm 1}\big|_{\text{n.t.}}$. 

Below are some general comments from the algebraic perspective. 
In contrast to the ideals in commutative algebra, 
the noncommutative nature of operator algebra leads to nontrivial constraints on the eigenenergies $E_k$ and the fundamental ladder operators $L_{\pm 1}$ that are compatible with the Hamiltonian $H$. 
The null bootstrap is a program about the systematic classification of the set of consistent data $\{\{H^{(i)}, E^{(i)}\}, \{L^{(j)}\}\}$.\footnote{Note that the explicit form of the Hamiltonian is not necessarily known in a general investigation.}
We can generalize the Hamiltonian and energy eigenvalues to a set of mutually commuting operators $\{H^{(i)}\}$, i.e., conserved quantities, 
and the corresponding ``good'' quantum numbers $\{E^{(i)}\}$. 
For example, we could include the $\mathcal Z_2$ parity operator 
that commutes with the Hamiltonian of the quartic anharmonic oscillator. 
The meaning of ``consistent data'' is that the dynamical constraints associated with the Hamiltonian 
are encoded in the null state conditions associated with the ladder operators.  
This will be discussed in more detail in the Sec. \ref{Anharmonic operator algebra}.

\subsection*{Matrix elements}

Besides the energy spectrum, there are other physical observables, such as the matrix elements of an operator $\mathcal{O}$.
The diagonal elements are the expectation values discussed above, 
which can be expressed in terms of the energy $E_{n}$ and the coupling constant $g$.
The off-diagonal elements can be computed using the ladder operators
\begin{align}
	\label{matrix-element}
	\langle n | \mathcal{O} | n' \rangle
	=\begin{cases}
		\frac{1}{\sqrt{(n+1)_{n'-n}}}\langle \mathcal{O} L_{n'-n} \rangle_n&\quad\text{if $n'>n$} \\
		\frac{1}{\sqrt{(n'+1)_{n-n'}}}\langle \mathcal{O} L_{n'-n} \rangle_n&\quad\text{if $n'<n$,}
	\end{cases}
\end{align}
where we have written $\<\ldots\>_{n}\equiv\<\ldots\>_{E_{n}}$.
Let us consider $\mathcal{O}=x$ as a simple example.
To order $g^1$, we use the expressions of the nontrivial parts of $L_{+1}$ in \eqref{ladder-nontrivial-g0} and \eqref{ladder-nontrivial-g1}. We obtain
\begin{align}
	\langle n|x|n+1 \rangle
	&=\frac{1}{\sqrt{n+1}}\langle xL_{+1} \rangle_{n}
	=\sqrt{\frac{n+1}{2}}\(1-\frac{3}{2}(n+1)g+O(g^{2})\)\,,
	\\
	\langle n|x|n+3 \rangle
	&=\frac{1}{\sqrt{(n+1)_{3}}}\langle x\(L_{+1}\)^3 \rangle_{n}
	=\frac{1}{4}\sqrt{\frac{(n+1)(n+2)(n+3)}{2}}\,g+O(g^{2})\,.
\end{align}
The nonvanishing matrix elements, to order $g^{1}$, are $\langle n|x|n\pm1 \rangle$ and $\langle n|x|n\pm3 \rangle$ because the position operator $x$ can be written as
\begin{align}
	x=\frac{1}{\sqrt{2}}\pt{L_{+1}+L_{-1}}+\frac{g}{4\sqrt{2}}\pt{
		L_{-3}-6L_{+1}L_{-2}-6L_{+2}L_{-1}+L_{+3}-6L_{-1}-6L_{+1}
	}+O(g^{2})\,,
\end{align}
and the eigenstates with different energies are orthogonal\footnote{This follows from the assumption that the Hamiltonian is Hermitian.}
\begin{align}\label{orthogonality-condition}
	\<m|n\>=0\,,\qquad (m\neq n)\,.
\end{align}
The other two cases can be obtained by a change of variable and taking the complex conjugate $\langle n|x|n-j \rangle=\(\langle n|x|n+j \rangle|_{n\rightarrow n-j}\)^\ast$ with $j=1,3$.

\subsection{Reduced procedure}
\label{Reduced procedure 1}
The reduced procedure is motivated by the difficulties in obtaining the complete energy spectrum and the exact ladder operators in a nonperturbative scheme. 
Alternatively, we can study the low-lying states directly and obtain their energies and matrix elements, which is the main idea of the reduced procedure.
It requires fewer steps to carry out the null bootstrap in this reduced approach. 
We can also obtain the higher order perturbative series of energies using the same spaces of test states.
Although we only obtain the low-lying data in the reduced procedure, 
it is well known that the weak-coupling-expansion results cease to be good approximations at high energies.\footnote{At high energies, it is more reasonable to use the Wentzel–
Kramers–Brillouin (WKB) method. It would be interesting to figure out the algebraic counterpart of the WKB method from the null bootstrap perspective.}

We have three versions of the reduced procedure.
In the first version, the null state condition holds for arbitrary test states.
However, as mentioned above, the discussion here is motivated by the nonperturbative applications, where the test operators are not arbitrary.
We will modify the first version in accordance with the nonperturbative method, which leads to the second version of the reduced procedure. 
In the third version, we further reduce the number of null state constraints and obtain higher order results in $g$. 
Below, we will study the low energy levels in all the three versions, but we will only examine the matrix elements in the first version of the perturbative null bootstrap. 

Let us begin with the first version of the perturbative null bootstrap, where the test operators are arbitrary.
The null state condition for a low-lying state reads
\begin{align}
	\label{null-condition-reduced}
	\langle \mathcal{O}_{\text{test}}\, \tilde{L} \rangle_{E}=0\,. 
\end{align}
We have put a tilde on the null operator to emphasize that $\tilde{L}$ is not exactly a ladder operator, which will be explained later.
In this naive version, the test operators $\mathcal{O}_{\text{test}}$ are arbitrary polynomials in $x$ and $ip$.
There will be two types of nontrivial perturbative solutions, corresponding to two types of quantization conditions.
In the first case, the low energy state is annihilated by the variation of a lowering operator and the energy spectrum is bounded from below.  
In the second case, the situation is opposite and the energy spectrum is bounded from above. 
\footnote{There may be other issues in this perturbative solution, as one finds a complex conjugate pair of solutions at finite coupling. }
Usually, a physical solution should have a bounded-from-below energy spectrum, so we will focus on the first case.
We assume that $\tilde{L}$ is of the form
\begin{align}\label{reduced-null-operator}
	\tilde{L}=\sum_{m_{2} = 0}^{K-m_{1}} \sum_{m_{1} = 0}^{K} a_{m_{1},m_{2}} x^{m_{1}} (i p)^{m_{2}}\,.
\end{align}
The truncation order is denoted by $K$, i.e. the null operators $\tilde{L}$ are degree-$K$ polynomials in $x$ and $ip$. 
The coefficients have the small $g$ expansion
\begin{align}
	a_{m_{1},m_{2}}=\sum_{i=0,1,2,\ldots} g^{i} a_{m_{1},m_{2}}^{(i)}\,.
\end{align}
where $a_{m_{1},m_{2}}^{(i)}$ are real numbers.

For finite $K$, the null state condition can hold to a certain order in $g$. 
We will consider the null state condition order by order in $g$, and search for the correct values of the energies.
They are extracted based on the following phenomenon: when the variable $E$ takes certain correct values, the expectation value $\langle\mathcal{O}_{\text{test}}\tilde{L}\rangle_{E}$ can be exactly zero to higher orders in $g$ than when $E$ is arbitrary.
Suppose that the expectation value $\langle\mathcal{O}_{\text{test}}\tilde{L}\rangle_{E}$ is exactly zero to order $g^{j}$.
As we consider higher energy levels, $j$ becomes smaller, and eventually we cannot distinguish the correct values of the energies from the arbitrary ones. 
In this way, we obtain a finite number of low energy levels for a fixed $K$.
The results for $K=1,2,3,4$ are\footnote{To avoid the solutions where $\tilde{L}$ vanish at order $g^{0}$, we set $a_{1,0}^{(0)}=1$ when $K$ is odd, and we set $a_{1,1}^{(0)}=1$ when $K$ is even.
At $K=4$, this also avoids the trivial solution $\tilde{L}=H-E$.
We do not consider the case of $K>4$.}
\begin{align}
	K=1:\qquad \langle \mathcal{O}_{\text{test}} \tilde{L}_{- 1} \rangle_{E_0}=O(g)\;\, \quad \Rightarrow \quad E_{0}&=\frac{1}{2}+O(g)\,, \label{naive-K=1}\\
	K=2:\qquad \langle \mathcal{O}_{\text{test}} \tilde{L}_{- 2} \rangle_{E_0}=O(g^{2}) \quad \Rightarrow \quad E_{0}&=\frac{1}{2}+\frac{3}{4}g+O(g^{2})\,, \label{naive-E0-K=2}\\ \langle \mathcal{O}_{\text{test}} \tilde{L}_{- 2} \rangle_{E_1}=O(g^{2}) \quad \Rightarrow \quad E_{1}&=\frac{3}{2}+\frac{15}{4}g+O(g^{2})\,, \label{naive-E1-K=2}\\
	K=3:\qquad \langle \mathcal{O}_{\text{test}} \tilde{L}_{- 1} \rangle_{E_0}=O(g^{5}) \quad \Rightarrow \quad  E_{0}&=\frac{1}{2}+\frac{3}{4}g-\frac{21}{8}g^2+\frac{333}{16}g^3-\frac{30885}{128}g^4\nn
	&\phantom{=}+O(g^{5})\,, \label{naive-K3-n1}\\
	\langle \mathcal{O}_{\text{test}} \tilde{L}_{- 3} \rangle_{E_1}=O(g^{3}) \quad \Rightarrow \quad E_{1}&=\frac{3}{2}+\frac{15}{4} g-\frac{165}{8} g^2+O(g^{3})\,,\\
	\langle \mathcal{O}_{\text{test}} \tilde{L}_{- 3} \rangle_{E_2}=O(g^{3}) \quad \Rightarrow \quad E_{2}&=\frac{5}{2}+\frac{39}{4} g-\frac{615}{8} g^2+O(g^{3})\,,\\
	\langle \mathcal{O}_{\text{test}} \tilde{L}_{- 5} \rangle_{E_3}=O(g^{2}) \quad \Rightarrow \quad E_{3}&=\frac{7}{2}+O(g)\,, \label{naive-E3-K=3}\\
	\langle \mathcal{O}_{\text{test}} \tilde{L}_{- 5} \rangle_{E_4}=O(g^{2}) \quad \Rightarrow \quad E_{4}&=\frac{9}{2}+O(g)\,, \label{naive-E4-K=3}\\
	K=4:\qquad \langle \mathcal{O}_{\text{test}} \tilde{L}_{-2} \rangle_{E_0}=O(g^{6}) \quad \Rightarrow \quad E_{0}&=\frac{1}{2}+\frac{3}{4} g-\frac{21}{8} g^2+\frac{333}{16} g^3-\frac{30885}{128} g^4\nn
	&\phantom{=}+\frac{916731}{256} g^5 +O(g^{6}) \,, \\
	\langle \mathcal{O}_{\text{test}} \tilde{L}_{- 2} \rangle_{E_1}=O(g^{6}) \quad \Rightarrow \quad E_{1}&=\frac{3}{2}+\frac{15}{4} g-\frac{165}{8} g^2+\frac{3915}{16} g^3-\frac{520485}{128} g^4\nn
	&\phantom{=}+\frac{21304485}{256} g^5+O(g^{6})\,, \\
	\langle \mathcal{O}_{\text{test}} \tilde{L}_{- 4} \rangle_{E_2}=O(g^{4}) \quad \Rightarrow \quad E_{2}&=\frac{5}{2}+\frac{39}{4} g-\frac{615}{8} g^2+\frac{20079}{16} g^3+O(g^{4}) \,, \\
	\langle \mathcal{O}_{\text{test}} \tilde{L}_{- 4} \rangle_{E_3}=O(g^{4}) \quad \Rightarrow \quad E_{3}&=\frac{7}{2}+\frac{75}{4} g-\frac{1575}{8} g^2+\frac{66825}{16} g^3+O(g^{4})\,, \\
	\langle \mathcal{O}_{\text{test}} \tilde{L}_{- 6} \rangle_{E_4}=O(g^{3}) \quad \Rightarrow \quad E_{4}&=\frac{9}{2}+\frac{123}{4}g+O(g^{2})\,, \label{naive-E4-K=4}\\
	\langle \mathcal{O}_{\text{test}} \tilde{L}_{- 6} \rangle_{E_5}=O(g^{3}) \quad \Rightarrow \quad E_{5}&=\frac{11}{2}+\frac{183}{4}g+O(g^{2})\,. \label{naive-E5-K=4}
\end{align}
The solutions for the energies are labeled by the $n=0,1,2,\ldots$, 
where a larger $n$ corresponds to a higher level and the zeroth order terms of $E_n$ are the same as the harmonic cases. 
The solutions for $\tilde L$ are labeled by $-k$, indicating the relation with the level-$k$ lowering operators.
The value of $k$ is fixed by the explicit expression of the solution [see the discussion below \eqref{L-reduced}].
In most cases with $n<K$, we can solve the null state conditions and obtain $E_n$ to the same orders in $g$.
The special cases are \eqref{naive-E3-K=3}, \eqref{naive-E4-K=3}, \eqref{naive-E4-K=4}, and \eqref{naive-E5-K=4}, where we need to solve the null state conditions to one order higher in $g$.
All results to order $g^{3}$ agree with those from the complete procedure.
Here and below, we verify the higher-order coefficients by comparing them to the results from the Bender-Wu method \cite{Bender:1973rz,Sulejmanpasic:2016fwr}.

It is surprising that the null state condition holds to higher orders in $g$, and the low energy levels can be determined to high orders in $g$.
If the state $|n\rangle$ is annihilated by a lowering operator $L_{-k}$, 
the null state $L_{-k}|n\rangle=O(g^{j+1})$ is constructed using $L_{-k}$ to order $g^{j}$.
The lowering operator $L_{-k}$ to order $g^{j}$ has a minimal degree in $x$ and $ip$.
Naively, the truncation order $K$ should be higher than or equal to the minimal degree, but this is not true.
For example, to order $g^{1}$, the minimal degree of $L_{-2}$ in $x$ and $ip$ is 4,\footnote{\label{level-k-minimal}The level-$k$ lowering operator with minimal degree in $x$ and $ip$ is constructed using $L_{-k}=(\text{normalization})L_{-k}|_{\text{n.t.}}+(\ldots)(H-E_{n})+O(g^{j+1})$,
which is the minus case of \eqref{level-k-L} to order $g^{j}$.
The trivial terms can cancel out the higher-degree terms in $x$ and $ip$ from $(\text{normalization})L_{-k}|_{\text{n.t.}}$.
}
but the null state condition with $K=2$ holds to order $g^{1}$  in \eqref{naive-E1-K=2}.
Therefore, $\tilde{L}_{- 2}$ cannot be the same as $L_{-2}$.
For instance, the solution for $\tilde{L}_{-2}$ at $K=2$ and $n=1$ is 
\begin{align}
	\tilde{L}_{-2}=\frac{1}{2} \(x+ i p\)^{2}+\Bigg[&\frac{25}{4} x^2+\frac{15}{4} x(ip)+ \(a_{0,2}^{(1)}-a_{0,0}^{(1)}\)(x+i p)^{2}
	\nn&-\(3a_{0,2}^{(1)}-a_{0,0}^{(1)}\)\(\frac{x^2}{2}+\frac{p^2}{2}-\frac{3}{2}\)\Bigg]g+O(g^{2}) \,,
\end{align}
where the first order terms are different from those of $L_{-2}$. 
Why does the degree-two polynomial in $\tilde{L}_{-2}$ annihilate $|1\rangle$?
We find that $\tilde{L}_{-2}$ can be written as
\begin{align}\label{L-reduced-K=2}
	\tilde{L}_{-2}=&\( 1 + g\( 6+3a_{0,2}^{(1)}-a_{0,0}^{(1)}+2x^{2} \) \) L_{-2}\big|_{\text{n.t.}} 
	\nn&+ g\( \frac{3}{2}+a_{0,2}^{(1)}-a_{0,0}^{(1)}-\frac{13}{4}x^{2}-\frac{3}{2}x(ip)-\frac{3}{4}(ip)^{2} \) \(H-E_{1}\)+O(g^{2})\,,
\end{align}
where we have used \eqref{ladder-k} and the nontrivial part of the lowering operator \eqref{ladder-nontrivial-g0} and \eqref{ladder-nontrivial-g1}.
The operator $\tilde{L}_{-2}$ annihilates $|1\rangle$ because the level-2 lowering operator annihilates the first excited state $L_{-2}|_{\text{n.t.}}|1\rangle=0$ and $(H-E_{1})|1\rangle$ is a trivial null state.
The factor in front of $L_{-2}|_{\text{n.t.}}$ in \eqref{L-reduced-K=2} is not a normalization factor, as opposed to the case of $L_{-2}$ (see footnote \ref{level-k-minimal}).
Although the two parts in \eqref{L-reduced-K=2} contain terms of degree higher than 2 in $x$ and $ip$,  
the higher-degree terms cancel out, 
so the final expression of $\tilde{L}_{-2}$ is given by a degree-2 polynomial.

More generally, we only impose that $\tilde{L}_{-k}$ annihilates a specific state, which is weaker than the requirement that $L_{-k}$ is a ladder operator for all energy eigenstates.
So the solution space of $\tilde{L}_{-k}$ is larger than that of $L_{-k}$.
Suppose that the null state condition holds to order $g^{j}$, i.e., $\langle \mathcal{O}_{\text{test}}\tilde{L}_{-k} \rangle_{E_n}=O(g^{j+1})$.
The generalization of \eqref{L-reduced-K=2} reads\footnote{There are different ways to write a solution for the null operator as the right-hand side of \eqref{L-reduced}.
We consider the case where $k$ is the highest possible.} 
\begin{align}\label{L-reduced}
	\tilde{L}_{- k} \propto \(\ldots \) L_{- k}\big|_{\text{n.t.}}  + \(\ldots\) \(H-E_{n}\) +O(g^{j+1})\,,
\end{align}
where the proportionality factor is a $g$-independent constant and the right-hand side annihilates $|n\rangle$ to order $g^{j}$.
The ellipses represent certain power series in $g$. 
In \eqref{L-reduced}, there are terms of degree higher than $K$ in $x$ and $ip$ in the two parts, but they cancel out to order $g^{j}$.
Then \eqref{L-reduced} becomes a polynomial of degree $K$, which is lower than the minimal degree of $L_{- k}$.
So we can construct the null state $\tilde{L}_{- k}|n\rangle=O(g^{j})$, despite that $K$ does not reach the minimal degree. 
In most cases with $n<K$, the first $(\ldots)$ in \eqref{L-reduced}  is given by $1+O(g)$, which indicates that the solutions for the null operators are level-$k$ lowering operators at order $g^{0}$.
In the special cases \eqref{naive-E3-K=3}, \eqref{naive-E4-K=3}, \eqref{naive-E4-K=4}, and \eqref{naive-E5-K=4}, the first $(\ldots)$ in \eqref{L-reduced} starts at order $g^{1}$.\footnote{The null state $\tilde{L}_{- k}|n\rangle$ is then a trivial null state at order $g^{0}$, but it is nontrivial at higher orders in $g$, so it constrains the value of $E$ as well.}

As $K$ increases, we obtain the energies to higher orders in $g$, and the orders increase sometimes rapidly and sometimes slowly. 
In other words, there are pairs of results with similar orders in $g$ as $K$ increases. 
This pattern is due to the parity constraints. 
For example, the ground-state energy is calculated to order $g^{0}$, $g^{1}$, $g^{4}$, and $g^{5}$ at $K=1,2,3$, and $4$.
The results show greater improvement from $K=2$ to $K=3$.
At $K=1$, the ground state is annihilated by $\tilde{L}_{-1}=x+ip+O(g)$.
At $K=2$, if we still use $\tilde{L}_{-1}$ to annihilate the ground state, the null state condition will hold to the same order as that in the case of $K=1$, and the result for $E_{0}$ will not improve.
This is due to the fact that $L_{-1}|_{\text{n.t.}}$ is parity odd.
More specifically, $L_{-1}|_{\text{n.t.}}$ has degree 1 and 3 terms in $x,p$ at order $g^{0}$ and $g^{1}$.
If the degree-3 terms in $\tilde{L}_{-1}$ cancel out to order $g^1$, we should have
\begin{align}\label{K=2-if}
	&\( 1 + g\(\text{constant $\,+\,$ (degree-2 terms in $x$ and $ip$)}\) \)L_{-1}|_{\text{n.t.}} \nn&+ g \text{(degree-1 terms in $x$ and $ip$)} \( H-E_{1} \) +O(g^{2}) \,.
\end{align}
We do not consider degree-1 terms in front of $L_{-1}|_{\text{n.t.}}$ because they will not help to cancel the third-degree terms from $L_{-1}|_{\text{n.t.}}$ at order $g^{1}$.
Since the parity is odd, the result of \eqref{K=2-if} can only be a degree-1 polynomial in $x$ and $ip$, 
then it should have been found already at $K=1$ if this solution does exist.
Therefore, the ground-state results cannot improve at $K=2$ if we stick to $L_{-1}$. 
It turns out that the null state condition can hold to order $g^{1}$ if we use $\tilde{L}_{-2}$ to annihilate the ground state, and the result for $E_{0}$ can be slightly improved.
At $K=3$, the cancellation mechanism for $\tilde{L}_{-1}$ is not restricted by parity,  
so the result from the $\tilde{L}_{-1}$ annihilation improves more rapidly.
At $K=4$, we again need to use $\tilde{L}_{-2}$ to improve the result, which is less significant.
There are similar patterns for other low energy levels.
In the optimal solutions, the state $|n\rangle$ is annihilated by $\tilde{L}_{-(n+1)}$ when $K+n$ is odd, 
but by $\tilde{L}_{-(n+2)}$ when $K+n$ is even.
As $K$ increases, the results for the energies show greater improvements when $K+n$ is odd.

For a fixed $K$, there are pairs of results with the same order in $g$.
For example, both $E_{0}$ and $E_{1}$ are determined to order $g^{1}$ at $K=2$. 
This is because they are associated with the null operator $\tilde{L}_{-2}$.
In general, $(E_{2m},E_{2m+1})$ are calculated to the same order in $g$ when $K$ is even, and $(E_{2m+1},E_{2m+2})$ are solved to the same order in $g$ when $K$ is odd.
Here $m$ is a non-negative integer. 
The results in the same pair are given by the null operators with the same level $k$, 
which is also due to the parity constraints discussed above.

We have described a naive way to carry out the reduced procedure perturbatively.
If we compare the perturbative results with those from the nonperturbative approach 
in \cite{Li:2022prn}, we find that the nonperturbative results are more precise and do not match those from the above naive procedure. 
Below we will summarize the nonperturbative approach and the corresponding results at small $g$. 

Let us give a brief review of the nonperturbative null bootstrap proposed in \cite{Li:2022prn}.
The idea is the same as that of \eqref{null-condition}, but it is usually difficult to obtain the exact nontrivial null states nonperturbatively.
In practice, we consider approximate null states $\widehat{L}|n\rangle$, where $\widehat{L}$ are finite-degree polynomials in $x$ and $ip$. The hat indicates that $\widehat{L}$ is different from $\tilde{L}$ in the naive approach.
The null state condition here is an approximate equation and does not need to hold for arbitrary test operators.
As shown in \cite{Li:2022prn}, 
accurate results can be obtained from low-degree test operators in $x$ and $ip$. 
We should modify the naive perturbative approach accordingly.

The approximate null state condition is
\begin{align}\label{null-state-condition-nonpert}
	\langle\mathcal{O}_{\text{test}}\widehat{L}\rangle_{E} \approx 0\,.
\end{align}
The operator $\widehat{L}$ and test operators $\mathcal{O}_{\text{test}}$ are degree-$K$ and degree-$M$ polynomials
\begin{align}
	\widehat{L} &= \sum_{m_{2} = 0}^{K-m_{1}} \sum_{m_{1} = 0}^{K} a_{m_{1},m_{2}} x^{m_{1}} (i p)^{m_{2}}\,, \label{L-nonpert} \\
	\mathcal{O}_{\text{test}} &= \sum_{m_{4} = 0}^{M-m_{3}} \sum_{m_{3} = 0}^{M} b_{m_{3},m_{4}} x^{m_{3}}(ip)^{m_{4}}\,, \label{test-nonpert}
\end{align}
where $a_{m_{1},m_{2}}$ and $b_{m_{3},m_{4}}$ are real numbers.
Following \cite{Li:2022prn}, we consider the test operators with $M= K+2$.\footnote{The system is underconstrained if $M\leqslant K+1$.}
Let us explain the meaning of ``$\approx$'' in \eqref{null-state-condition-nonpert}.
The left-hand side $\langle \mathcal{O}_{\text{test}}\widehat{L} \rangle_{E}$ cannot be exactly zero for a finite $K$ unless $\widehat{L}$ is a trivial solution.\footnote{We impose $\sum_{m_{2} = 0}^{K-m_{1}} \sum_{m_{1} = 0}^{K} a_{m_{1},m_{2}}=1$ to avoid the solution $a_{m_{1},m_{2}}=0$.}
The meaning of ``trivial solution'' is the same as that in \eqref{trivial-general}, i.e. the trivial solution satisfies the null state condition automatically and does not yield any constraint on $E$.
For a nontrivial solution $\widehat{L}$ with a finite $K$, the null state is approximated by $\widehat{L}|E\rangle$.
The approximate null state condition \eqref{null-state-condition-nonpert} means that $\langle\mathcal{O}_{\text{test}}\widehat{L}\rangle_{E}$ almost vanishes in the following sense.
Each summand in \eqref{test-nonpert} is associated with an expectation value $\langle x^{m_{3}}(ip)^{m_{4}} \widehat{L} \rangle_{E}$, and all of them should be close to zero.
Therefore, we use the $\eta$ function to measure the violation of the exact null state condition
\begin{align}\label{eta-function}
	\eta=\sum_{m_{4} = 0}^{M-m_{3}} \sum_{m_{3} = 0}^{M} \left| \frac{1}{m_{3}!m_{4}!} \frac{ \pa \langle \mathcal{O}_{\text{test}} \widehat{L} \rangle_{E} }{ \pa \, b_{m_{3},m_{4}} } \right|^{2}\,,
\end{align}
which is a weighted sum of the squared expectation values.
The higher-degree terms in $\mathcal{O}_{\text{test}}$ lead to larger errors, so they are suppressed by $(m_3! m_4!)^{-2}$.
We will obtain the low energy levels $E_n$ and expectation values $\langle x^{2} \rangle_{E_n}$ by finding the local minima of $\eta$ at $K=1,2$.
For small coupling constant $g\leqslant 10^{-2}$, we should be more careful as $K$ increases.
Although there is no obstruction in principle, we do not present the results with $K\geqslant 3$ for practical reasons, 
as it takes more effort to deal with high-precision numerical computations. 

To minimize the $\eta$ function, we need to know the explicit expressions of the expectation values in \eqref{eta-function}.
Since  the consistency relation \eqref{consistency-relation} holds nonperturbatively, 
we can express the expectation values $\langle x^{m_{1}} p^{m_{2}} \rangle_{E}$ in terms of the coupling constant $g$, the energy $E$ and the expectation value $\langle x^{2} \rangle_E$.\footnote{The normalization is given by $\langle 1 \rangle_E=1$.}
As opposed to \eqref{x2-E}, $E$ and $\langle x^{2} \rangle_{E}$ are independent variables in the nonperturbative approach.
Therefore, $\eta$ is a function of  $E$, $\langle x^{2}\rangle_{E}$ and $a_{m_{1},m_{2}}$.
The solutions for $E_{n}$, $\langle x^{2}\rangle_{E_n}$ and $a_{m_{1},m_{2}}$ are determined by minimizing the $\eta$ function locally. 
We will discuss the asymptotic behaviors of the errors as the coupling constant approaches zero, i.e., $g\rightarrow 0$.

As $K$ increases, we obtain more precise results and more energy levels $E_n$, together with the expectation values $\langle x^{2} \rangle_{E_n}$.
We will focus on $E_n$.\footnote{The errors in $E_n$ and $\langle x^{2} \rangle_{E_n}$ have similar orders of magnitudes.}
We denote the null bootstrap results for the energies by $E_{n,K}$, where $K$ indicates the truncation degree of $\widehat{L}$ in \eqref{L-nonpert}.
The errors in the null bootstrap results are $E_{n,K}-E_{n}^{\star}$, where the reference energies $E_{n}^{\star}$ are computed from diagonalizing the Hamiltonian of size $30\times 30$ in the basis of harmonic oscillator eigenfunctions.
We evaluate the errors at $g=10^{-2},10^{-3},\ldots,10^{-7}$.
As $g\rightarrow 0$, the $K=1$ results exhibit the following asymptotic behavior
\begin{align}\label{asymptotic-K=1}
	E_{0,K=1}-E_{0}^{\star} \approx -0.94 \times 10^{-2} g\,,
\end{align}
and the $K=2$ results give
\begin{align}
	E_{0,K=2}-E_{0}^{\star} &\approx -2.5 \times 10^{-1} g^{3}\,, \label{asymptotic-K=2} \\
	E_{1,K=2}-E_{0}^{\star} &\approx 0.70 \, g^{3}\,. \label{asymptotic-K=2-first-excited}
\end{align}
These results are more precise than those in the naive reduced procedure with the same $K$.
At $K=1$, the error in \eqref{naive-K=1} from the naive approach is $-\frac{3}{4}g$, where the absolute value of the coefficient $\frac{3}{4}$ is much larger than $0.94 \times 10^{-2}$.
At $K=2$, the errors in \eqref{naive-E0-K=2} and \eqref{naive-E1-K=2} from the naive approach are $\frac{21}{8}g^{2}$ and $\frac{165}{8}g^{2}$, but the errors in the nonperturbative $K=2$ results are of order $g^3$.

To be consistent with the nonperturbative method, 
we introduce the second version of the perturbative reduced procedure. 
As explained above, the null state condition does not have to hold for arbitrary test operators.
We can restrict the test operators to polynomials of low degree in $x$ and $ip$. In accordance with the nonperturbative approach, we consider the test operators that are degree-$(K+2)$ polynomials in $x$ and $ip$.
The results then match those from the nonperturbative method.
Below we will derive the asymptotic behaviors \eqref{asymptotic-K=1}, \eqref{asymptotic-K=2}, and \eqref{asymptotic-K=2-first-excited} analytically.\footnote{
This approach seems more efficient than the traditional perturbation method.
It is easier to compute the  perturbative low energies to high order in $g$.}

We need to make sense of \eqref{null-state-condition-nonpert} in perturbation theory.
What is $\langle\mathcal{O}_{\text{test}}\widehat{L}\rangle_{E}$ on the left-hand side of \eqref{null-state-condition-nonpert}?
As in \eqref{xmpn-expectation-value}, the expectation values $\langle x^{m_{1}} p^{m_{2}} \rangle_E$ are expressed in terms of the coupling constant $g$, the energy $E$ and the expectation value $\langle x^{2} \rangle_E$.
Besides \eqref{E-x2-g-expansion}, we also have the small $g$ expansion of the coefficients in $\widehat{L}$
\begin{align}
	a_{m_{1},m_{2}}=\sum_{i=0,1,2,\ldots} g^{i} a_{m_{1},m_{2}}^{(i)}\,.
\end{align}
As opposed to the third assumption at the beginning of Sec. \ref{The null bootstrap in perturbation theory} (see the discussion near \eqref{x2-E}), we do not assume the regularity of all expectation values in the limit $g\rightarrow 0$.
So $\langle\mathcal{O}_{\text{test}}\widehat{L}\rangle_{E}$ is a Laurent series in $g$, which has terms proportional to negative powers of $g$.
We require that $\langle\mathcal{O}_{\text{test}}\widehat{L}\rangle_{E}$ is exactly zero for any $\{b_{m_{3},m_{4}}\}$ to the highest order possible in $g$.
We extract the energies according to a phenomenon similar to that in the first version.
The expectation value $\langle\mathcal{O}_{\text{test}}\widehat{L}\rangle_{E}$ can be exactly zero to higher order in $g$ for certain correct values of energies than for arbitrary $E$.\footnote{As in the first version, not all the correct values of the energies satisfy this condition. }
As before, the solutions for the null operators are labeled by $k$, 
and the solutions for the energies are labeled by $n$. 
We write $\langle\mathcal{O}_{\text{test}}\widehat{L}_{-k}\rangle_{E_{n}}=O(g^{j+1})$ and emphasize that $\mathcal{O}_{\text{test}}$ is a degree-$(K+2)$ polynomial in $x$ and $ip$.
In general, the terms proportional to negative powers of $g$ are exactly zero, implying the regularity of $\langle\mathcal{O}_{\text{test}}\widehat{L}_{-k}\rangle_{E_{n}}$ in the $g\rightarrow0$ limit.
However, since other expectation values are still allowed to be singular in the $g\rightarrow 0$ limit, $\langle x^{2}\rangle_{E}$ is not completely determined by $E$, and we do not have the full relation \eqref{x2-E}. 
Suppose that we have found a correct energy $E_n$ to a certain order, and $\langle\mathcal{O}_{\text{test}}\widehat{L}_{-k}\rangle_{E_{n}}$ vanishes to order $g^{j}$. Then the $\eta$ function \eqref{eta-function} vanishes to order $g^{2j+1}$. 
In the perturbative procedure, we require that the $\eta$ function is minimized at order $g^{2j+2}$ 
and discard the higher order terms.\footnote{At $K=1,2,3$, we also consider the $\eta$ minimization when $E$ is not determined.
In that case, we assume $\langle\mathcal{O}_{\text{test}}\widehat{L}\rangle_{E}$ vanishes to order $g^{j'}$ for arbitrary $E$, i.e. $\langle\mathcal{O}_{\text{test}}\widehat{L}\rangle_{E}=O(g^{j'})$.
So the $\eta$ function is zero to order $g^{2j'+1}$.
We search for the local minima of the $\eta$ function at order $g^{2j'+2}$.
They correspond to the correct values of the energies that are extracted above, and no other local minimum is found.
In fact, these local minima are the zeros of the $\eta$ function, since $\langle\mathcal{O}_{\text{test}}\widehat{L}\rangle_{E}$ can be exactly zero at order $g^{j'+1}$, if $E$ takes the correct values from the analysis above. } 

Let us consider $K=1$ for example.
The test operators are of the form \eqref{test-nonpert} with $M=3$. 
For $n=0$, the left-hand side of \eqref{null-state-condition-nonpert} starts at order $g^{-1}$.
The null state condition is exact at this order, i.e., $\langle \mathcal{O}_{\text{test}}\widehat{L}_{-1} \rangle_{E_0}=O(1)$.
We obtain the relation $\langle x^{2}\rangle_{E_0}^{(0)} = E_0^{(0)}$, which agrees with the result from the regularity assumption.
At order $g^{0}$, the null state condition can be solved exactly, and we find $E_0^{(0)}=\frac{1}{2}$.\footnote{We restrict to the solution where the energy spectrum is bounded from below.}
At order $g^{1}$, the null state condition cannot be exact.
We minimize the $\eta$ function \eqref{eta-function}, and obtain $E_{0}^{(1)}\approx\frac{165805}{223884}$.
We use the subscript $K$ to indicate the degree of $\widehat{L}_{-k}$ in the approximate null state condition \eqref{null-state-condition-nonpert}.
We have
\begin{align}\label{E0-K=1}
	E_{0,K=1} \approx \frac{1}{2} + \frac{165805}{223884} g + O(g^{2})\,,
\end{align}
where the zeroth-order coefficient is exact and the first-order coefficient is approximate.
For the coefficients, by ``exact'' we mean that they are the same as those from the Bender-Wu method \cite{Bender:1973rz,Sulejmanpasic:2016fwr}.
In the $g\rightarrow 0$ limit, the error is
\begin{align}
	\frac{165805}{223884} g - \frac{3}{4} g \approx -0.94 \times 10^{-2} g \,,
\end{align}
which agrees with the nonperturbative result \eqref{asymptotic-K=1}.
For other values of $E$, the null state condition can be exact to order $g^{-1}$, but not to order $g^0$. 
Therefore, we do not find other energy levels.

At $K=2$, we consider $\langle\mathcal{O}_{\text{test}}\widehat{L}_{-2}\rangle_{E_0}=O(g^{3})$ as well as the $\eta$ minimization, and obtain
\begin{align}\label{E0-K=2}
	E_{0,K=2} \approx \frac{1}{2} + \frac{3}{4} g - \frac{21}{8} g^{2} + \frac{24044253411}{1169246192} g^{3} + O(g^{4})\,.
\end{align}
where the coefficients are exact at order $g^{0}$, $g^{1}$, and $g^{2}$, but the third-order coefficient is an approximate result from the $\eta$ minimization.
The leading error is
\begin{align}
	\frac{24044253411}{1169246192} g^{3} - \frac{333}{16} g^{3} \approx -2.5 \times 10^{-1} g^{3} \,,
\end{align}
which corresponds to \eqref{asymptotic-K=2} in the nonperturbative approach.
We also obtain the first-excited-state energy from $\langle\mathcal{O}_{\text{test}}\widehat{L}_{-k}\rangle_{E_1}=O(g^{3})$ and the $\eta$ minimization
\begin{align}\label{E1-K=2}
	E_{1,K=2} \approx \frac{3}{2}+\frac{15}{4}g-\frac{165}{8}g^2+\frac{16650808149015}{67854323024}g^3+O(g^{4})\,,
\end{align}
where we have the approximate result for the third-order coefficient and the rest of the coefficients are exact.
As $g\rightarrow 0$, the error is
\begin{align}
	\frac{16650808149015}{67854323024}g^3-\frac{3915}{16}g^{3} \approx 0.70 \, g^{3}\,,
\end{align}
which agrees with the asymptotic behavior of the nonperturbative results \eqref{asymptotic-K=2-first-excited}.
At $K=2$, we have obtained two energy levels.
For other values of $E$, the null state condition can hold to order $g^{-1}$, but not to order $g^0$, so no more energy level can be detected.

At $K=3$, we find five energy levels.
Considering $\langle\mathcal{O}_{\text{test}}\widehat{L}_{-1}\rangle_{E_0}=O(g^{8})$ and the $\eta$ minimization, we have the ground-state energy
\begin{align}\label{E0-K=3}
	E_{0,K=3} \approx &\;\frac{1}{2}+\frac{3}{4}g-\frac{21}{8}g^2+\frac{333}{16}g^3-\frac{30885}{128}g^4+\frac{916731}{256}g^5-\frac{65518401}{1024}g^6 \nn
	&+\frac{2723294673}{2048}g^7-\frac{6648684051586933741623}{211416904204288}g^{8}+O(g^9)\,,
\end{align}
where the eighth-order coefficient is the approximate result from the $\eta$ minimization and the rest are exact.
We do not have the nonperturbative null bootstrap results at $K=3$ for sufficiently small $g$, so we do not compare the asymptotic behaviors here.
Using the exact coefficients in the small $g$ expansion, we estimate that the error of \eqref{E0-K=3} at order $g^{8}$ is about $10^{-1}g^{8}$, and the order $g^{9}$ correction is about $10^{8}g^{9}$.\footnote{The exact coefficient at order $g^{9}$ is obtained in \eqref{E0-K=4}.}
The coupling constant should be smaller than $10^{-9}$ in order for the error at order $g^{8}$ to be dominant.
We examine the cases of $g=10^{-2},10^{-3},\ldots,10^{-7}$, and verify that the error of \eqref{E0-K=3} is indeed about $10^{8}g^{9}$ using the same reference values $E_{n}^{\star}$ as those in the paragraph containing \eqref{asymptotic-K=1}.
We also obtain the first- and second-excited-state energies from $\langle\mathcal{O}_{\text{test}}\widehat{L}_{-3}\rangle_{E_{n=1,2}}=O(g^{5})$ and the $\eta$ minimization.
The results are
\begin{align}
	E_{1,K=3} &\approx \frac{3}{2}+\frac{15}{4} g-\frac{165}{8} g^2+\frac{3915}{16} g^3-\frac{520485}{128} g^4+\frac{1548031419879073965}{18608822642944} g^5+O(g^6) \,, \label{E1-K=3} \\
	E_{2,K=3} &\approx \frac{5}{2}+\frac{39}{4} g-\frac{615}{8} g^2\!+\frac{20079}{16} g^3\!-\frac{3576255}{128} g^4\!+\frac{690676244524833539787}{920874095938304}
	g^5\!+O(g^6) \,,
\end{align}
where the $\eta$ minimization gives approximate coefficients at order $g^{5}$, and the other coefficients are exact.
The third- and fourth-excited-state energies are obtained from $\langle\mathcal{O}_{\text{test}}\widehat{L}_{-5}\rangle_{E_{n=3,4}}=O(g^{3})$ and the $\eta$ minimization.
Their results are
\begin{align}
	E_{3,K=3} &\approx \frac{7}{2}+\frac{75}{4} g-\frac{22212548478304275}{25061340698792}g^{2} + O(g^{3})\,, \label{E3-K=3} \\
	E_{4,K=3} &\approx \frac{9}{2}+\frac{123}{4} g+\frac{579546280175463}{8798256371704}g^{2} + O(g^{3}) \label{E4-K=3} \,,
\end{align}
where the coefficients at order $g^{2}$ are approximate results from the $\eta$ minimization and the rest are exact.
However, the $\eta$ minimization does not give reasonable approximations for the order $g^2$ coefficients of $E_{3,K=3}$ and $E_{4,K=3}$.
The errors are larger if we consider the second-order corrections in \eqref{E3-K=3} and \eqref{E4-K=3}.
For other values of $E$, the null state condition can be solved exactly to order $g^{0}$, but not to order $g^1$. So we have not detected other energy levels.

At $K=4$, we obtain six energy levels from 
\begin{align}
	\langle\mathcal{O}_{\text{test}}\widehat{L}_{-2}\rangle_{E_{n=0,1}}&=O(g^{10})\,, \\
	\langle\mathcal{O}_{\text{test}}\widehat{L}_{-4}\rangle_{E_{n=2,3}}&=O(g^{7})\,,\\
	\langle\mathcal{O}_{\text{test}}\widehat{L}_{-6}\rangle_{E_{n=4,5}}&=O(g^{5})\,,
\end{align}
and the $\eta$ minimization. The explicit results are
\begin{align}
	E_{0,K=4}\approx\;&\frac{1}{2}+\frac{3}{4} g-\frac{21}{8} g^2\!+\frac{333}{16} g^3\!-\frac{30885}{128} g^4\!+\frac{916731}{256} g^5\!-\frac{65518401}{1024}
	g^6\!+\frac{2723294673}{2048} g^7 \nn
	&-\frac{1030495099053}{32768} g^8+\frac{54626982511455}{65536} g^9
	\nn&-\frac{2311268895269303699774094514905}{94418664607097552896}g^{10}+O(g^{11})\,, \label{E0-K=4}
	\\
	E_{1,K=4}\approx\;&\frac{3}{2}+\frac{15}{4} g-\frac{165}{8} g^2+\frac{3915}{16} g^3-\frac{520485}{128} g^4+\frac{21304485}{256} g^5-\frac{2026946145}{1024}
	g^6 \nn&+\frac{108603230895}{2048} g^7 -\frac{51448922163885}{32768} g^8+\frac{3325989183831585}{65536}
	g^9
	\nn&-\frac{175501317536123439084638122629745305}{98834461924794527645696} g^{10}+O(g^{11})\,, \label{E1-K=4}
	\\
	E_{2,K=4}\approx\;&\frac{5}{2}+\frac{39}{4} g-\frac{615}{8} g^2+\frac{20079}{16} g^3-\frac{3576255}{128} g^4+\frac{191998593}{256} g^5 \nn
	&-\frac{23513776995}{1024}
	g^6+\frac{35135739024227564792793164073}{45159147304513673216} g^7+O(g^{8})\,, \label{E2-K=4}
	\\
	E_{3,K=4}\approx\;&\frac{7}{2}+\frac{75}{4} g-\frac{1575}{8} g^2+\frac{66825}{16} g^3-\frac{15184575}{128} g^4+\frac{1024977375}{256} g^5 \nn
	&-\frac{155898295875}{1024}
	g^6 +\frac{36203582427161186496240292235625}{5713467480989300844544}
	g^7+O(g^{8})\,, \label{E3-K=4}
	\\
	E_{4,K=4}\approx\;&\frac{9}{2}+\frac{123}{4} g-\frac{3249}{8} g^2\!+\frac{171153}{16}g^{3}\! -\frac{3637623070300184675417824875}{9790734694343084724608}g^{4}\!+O(g^{5}) \,, \label{E4-K=4}
	\\
	E_{5,K=4}\approx\;&\frac{11}{2}+\frac{183}{4} g-\frac{5841}{8} g^2+\frac{369063}{16} g^3
	\nn&-\frac{124940937770350373790920800845}{130510054543768316521088}g^{4}+O(g^{5})\,, \label{E5-K=4}
\end{align}
where the highest-order coefficients are approximate results from the $\eta$ minimization, and all other coefficients are exact.
For other values of $E$, the null state condition can be satisfied exactly to order $g^{2}$, but not to order $g^3$. So no other energy levels are found.

Why are the results of the second version better than those from the first perturbative procedure?
As we restrict the test operators to be of low degree in $x$ and $ip$, there are fewer constraints on $\widehat{L}_{-k}$ than those on $\tilde{L}_{-k}$.
The space of solutions becomes larger.
Suppose that the null state condition holds to order $g^{j}$ in the naive approach.
We can add to the solution $\tilde{L}_{-k}$ a term $h(x,ip)$
\begin{align}\label{Lhat-Ltilde}
	\widehat{L}_{-k}=\tilde{L}_{-k}+h(x,ip)\,,
\end{align}
where $h(x,ip)$ is a polynomial in $x$ and $ip$, satisfying
\begin{align}
	\langle \mathcal{O}_{\text{test}}\, h(x,ip) \rangle_{E_n} = O(g^{j+1})\,.
\end{align}
The test operator $\mathcal{O}_{\text{test}}$ takes the form \eqref{test-nonpert} with $M=K+2$.
Then, the null state condition still holds to order $g^{j}$ for test operators with $M=K+2$.
For certain $h(x, ip)$, the high degree terms on the right-hand side of \eqref{Lhat-Ltilde} cancel out and 
we obtain a polynomial of considerably lower degree in $x$ and $ip$ than $\tilde{L}_{-k}$.
In this way, a low degree $\widehat{L}_{-k}$ is equivalent to a high degree $\tilde{L}_{-k}$ in the null state condition with $M=K+2$. 
As a result, we can obtain significantly higher-order results for the energy levels in the second version of the perturbative null bootstrap.

The pair patterns in the first version also exist in the second version.
As $K$ increases, the results show greater improvements when $K+n$ is odd than when $K+n$ is even, 
so we obtain pairs of results with similar orders in $g$.
We also have the following pairs of results determined to the same order in $g$: $(E_{2m,K},E_{2m+1,K})$ when $K$ is even, and $(E_{2m+1,K},E_{2m+2,K})$ when $K$ is odd.
Here $m$ denotes a non-negative integer. 
These patterns can also be traced back to the parity constraints. 

In \cite{Li:2022prn}, the results are obtained at finite coupling.
We find that the convergence of the nonperturbative approach at finite coupling is similar to that at small $g$. 
The latter has been studied analytically using the above small $g$ expansion.
In the finite coupling case, the precision increases significantly from $K=2$ to $K=3$ for the ground-state energy, but more slowly from $K=1$ to $K=2$ and from $K=3$ to $K=4$ (see Table I in \cite{Li:2022prn}).
This pattern is consistent with that of the perturbative results, where the ground-state energy is calculated to order $g^{1}$, $g^{3}$, $g^{8}$, and $g^{10}$ at $K=1,2,3$, and $4$.
Moreover, the finite coupling results for the first-excited-state energy improve slowly from $K=2$ to $K=3$, but rapidly from $K=3$ to $K=4$.
In the small $g$ expansion, the first-excited-state energy is calculated to order $g^{3}$, $g^{5}$, and $g^{10}$ at $K=2,3$, and $4$ [see \eqref{E1-K=2}, \eqref{E1-K=3} and \eqref{E1-K=4}], showing a similar pattern to the finite coupling results.
For a fixed $K$, the finite coupling results behave similarly to those in the $g$ expansion.
There are pairs of finite coupling results with similar precision (see Table I in \cite{Li:2022prn}).
The pattern is the same as that discussed above in the $g$ expansion.
In addition, we can compare the results for different energy levels and at different $K$.
The finite coupling and small $g$ expansion results match qualitatively.
The higher-precision results in the finite coupling case correspond to higher-order results in the small $g$ expansion.
For example, the $K=3$ result for $E_2$ is more precise than the $K=2$ result for $E_1$ in the finite coupling case (see Table I in \cite{Li:2022prn}), and in the small $g$ expansion we obtain $E_{2,K=3}$ to order $g^{5}$, while we determine $E_{1,K=2}$ to order $g^{3}$.
In conclusion, for the low energy levels, the convergence of the finite coupling results roughly resembles the behavior at small $g$, which can be explained by the small $g$ expansion results discussed above.

A difference is that we obtain more energy levels in the small $g$ expansion at $K=3$. 
We find five energy levels $E_0$, $E_1$, $E_2$, $E_3$, and $E_4$ in the small $g$ expansion, while the finite coupling results only have three energy levels $E_0$, $E_1$, and $E_2$. 
As mentioned above, the $\eta$ minimization results of the additional solutions 
do not give good approximations for the highest order coefficients. 
This may be the reason for their absence in the minimization results at finite coupling.\footnote{At $K=4$, we obtain six energy levels in the small $g$ expansion.
In the finite coupling case, one can also find six local minima of the $\eta$ function, in accordance with the perturbative results.  
In \cite{Li:2022prn}, the $n=4,5$ results at $K=4$ were not presented for the single well potential.   
However, the six local minima in the double well potential case were discussed in the footnote 19 of \cite{Li:2022prn}.  } 

We have introduced the second version of reduced procedure in perturbation theory, which is consistent with the nonperturbative method when the coupling constant $g$ is small.
Interestingly, we can further improve the perturbative results by reducing the number of null state constraints. 
For test operators with $M=K+2$, the null state condition cannot hold to arbitrarily high order in $g$ and the system is overdetermined at higher orders in $g$.
If we remove some of these constraints, 
then more exact coefficients in the perturbative energies can be determined by the null state condition.
Consider the expectation value $\langle\mathcal{O}_{\text{test}}\widehat{L}_{-k}\rangle_{E_n}$.
The test operators are still the lower-degree polynomials in $x$ and $ip$, i.e., $M=K+2$, and we require that $\langle\mathcal{O}_{\text{test}}\widehat{L}_{-k}\rangle_{E_n}$ is exactly zero to the highest order possible in $g$. 
We write $\langle\mathcal{O}_{\text{test}}\widehat{L}_{-k}\rangle_{E_n}=O(g^{j+1})$.
However, we do not minimize the $\eta$ function in the third version. 
Instead, we set $M=K+1$ and require that $\langle\mathcal{O}_{\text{test}}\widehat{L}_{-k}\rangle_{E_n}$ is exactly zero at order $g^{j+1}$ again.\footnote{At order $g^{j+1}$, we also consider $M=K$ and obtain the exact coefficients.
In most cases, the coefficients cannot be determined if $M<K$.
The ground-state energy at $K=3,4$ and the first-excited-state energy at $K=4$ are special.
To obtain the exact energies at order $g^{j+1}$, the minimal value of $M$ is $K-2$ in the special cases.
Furthermore, we obtain two exact coefficients in these special cases below.} 
When $n<K$, this gives the exact coefficients, instead of the approximated ones from the $\eta$ minimization.
Moreover, we can impose that $\langle\mathcal{O}_{\text{test}}\widehat{L}_{-k}\rangle_{E_n}$ is exactly zero for $M=K+1$ at even one order higher in $g$, i.e., at order $g^{j+2}$.
Curiously, we obtain the exact coefficients at order $g^{j+2}$ in the ground-state energy at $K=3,4$ and in the first-excited-state energies at $K=4$.\footnote{In these cases, we can obtain the exact coefficients at order $g^{j+2}$ as long as $M$ satisfies $K-2 \leqslant M \leqslant K+3$ at this order.
There is no solution for higher $M$ and the energies at order $g^{j+2}$ are not fixed for lower $M$.} 
In other cases, the energies at order $g^{j+2}$ are not fixed by the null state condition.\footnote{Since the energies at order $g^{j+2}$ are not fixed, we increase the number of constraints by setting $M=K+2$ at order $g^{j+2}$.
Then, there will be no solution if we impose that $\langle\mathcal{O}_{\text{test}}\widehat{L}_{-k}\rangle_{E_n}$ vanishes at this order, but we can use the $\eta$ minimization to obtain good approximations of the energies at order $g^{j+2}$.
For simplicity, the $\eta$ minimization here is carried out under the constraint that $\langle\mathcal{O}_{\text{test}}\widehat{L}_{-k}\rangle_{E_n}$ is zero for $M=K+1$ at order $g^{j+2}$.} 
In more detail, at $K=1,2,3$ we obtain the exact coefficients $(E_{0,K=1}^{(1)})$, $(E_{0,K=2}^{(3)}, E_{1,K=2}^{(3)})$,  and $(E_{0,K=3}^{(8)}, E_{0,K=3}^{(9)}, E_{1,K=3}^{(5)})$, whose explicit values are contained in \eqref{E0-K=4}--\eqref{E3-K=4}. 
Let us present the additional results at $K=4$
\begin{align}\label{K=4-exact}
	E_{0,K=4}^{(10)}&=-\frac{6417007431590595}{262144} \,, \qquad
	E_{0,K=4}^{(11)}=\frac{413837985580636167}{524288} \,,  \nn
	E_{1,K=4}^{(10)}&=-\frac{465491656557283395}{262144} \,, \qquad
	E_{1,K=4}^{(11)}=\frac{35043703273186461945}{524288} \,, \nn
	E_{2,K=4}^{(7)}&=\frac{1593440096499}{2048} \,, \nn
	E_{3,K=4}^{(7)}&=\frac{12977225578125}{2048} \,.
\end{align}
At order $g^{j+3}$, we do not obtain more exact coefficients by setting $M=K+1$ and requiring that $\langle\mathcal{O}_{\text{test}}\widehat{L}_{-k}\rangle_{E_n}$ vanishes.\footnote{In the special cases where two exact coefficients are obtained, one can consider $M=K+2$ or $M=K+3$ at order $g^{j+2}$ and $M=K-2$ or $M=K-1$ at $g^{j+3}$.
The coefficients at order $g^{j+2}$ and $g^{j+3}$ can be fixed by imposing that $\langle\mathcal{O}_{\text{test}}\widehat{L}_{-k}\rangle_{E_n}$ vanishes at these orders.
The exact coefficients at order $g^{j+2}$ are obtained, but the solutions at order $g^{j+3}$ are different from those in the Bender-Wu method.
Nonetheless, they are good approximations of the exact values. }
This concludes our discussion of the low energy levels.

Now we discuss the matrix elements.
We will only consider arbitrary test operators here. 
Suppose that we have obtained the eigenenergies of two states $|n\rangle$ and $|n'\rangle$.
We can also compute the matrix element using a slightly modified version of \eqref{matrix-element}
\begin{align}
	\label{matrix-element-reduced}
	\langle n | \mathcal{O} | n' \rangle
	=\begin{cases}
		\frac{1}{\sqrt{(n+1)_{n'-n}}}
		\langle \mathcal{O} L_{n',n} \rangle_n&\quad\text{if $n'>n$} \\
		\frac{1}{\sqrt{(n'+1)_{n-n'}}}
		\langle \mathcal{O} L_{n',n} \rangle_n&\quad\text{if $n'<n$,}
	\end{cases}
\end{align}
where $L_{n',n}$ is the operator that connects the two states $|n\rangle$ and $|n'\rangle$ and $\langle\ldots\rangle_{n}\equiv\langle\ldots\rangle_{E_{n}}$.
Note that we have specified the two energy eigenstates connected by the ladder operator.
The reason will be explained shortly.
The ladder operator in \eqref{matrix-element-reduced} is obtained by considering
\begin{align}
	\label{ladder-matrix-element-reduced}
	\langle\mathcal{O}_{\text{test}}(H-E_{n'})L_{n',n}\rangle_{n}=0\,,
\end{align}
which holds for arbitrary test operators. 
We use the ansatz
\begin{align}
	L_{n',n}=\;&\text{degree-$|n'-n|$ polynomial}+g\pt{\text{degree-$\(|n'-n|+2\)$ polynomial}}
	\nn
	&+g^{2}\pt{\text{degree-$\(|n'-n|+4\)$ polynomial}}
	\nn&
	+g^{3}\pt{\text{degree-$\(|n'-n|+6\)$ polynomial}}
	+O(g^{4})\,.
\end{align}
These degrees are the minimal degrees for constructing the nontrivial part of the level-$|n'-n|$ ladder operator. 
For some low-lying states, there are differences between ladder operators obtained in this way and those from the complete procedure.
For example, the zeroth-order ladder operator $L_{1,0}^{(0)}$ is
\begin{align}
	L_{1,0}^{(0)}=\text{(normalization)}\times L_{+1}^{(0)}+\text{($g$-independent operator)}\times L_{-1}^{(0)}\,,
\end{align}
where the last term is associated with a null state at zeroth order when acting on $|0\rangle$ at order $g^{0}$.
Despite the differences in the ladder operators, we obtain the same results for the matrix elements, 
since the null states are orthogonal to all states and do not contribute to the matrix elements.
For example, after choosing the normalization and fixing the sign of $L_{n',n}$, although $L_{1,0}$ is not exactly $L_{+1}$, the results for the matrix elements are identical
\begin{align}
	\langle 0| \mathcal{O} |1 \rangle=\langle 0| \mathcal{O} L_{1,0} |0 \rangle=\langle 0| \mathcal{O} L_{+1} |0\rangle\,.
\end{align}
To be consistent with \eqref{matrix-element-reduced}, we impose the normalization condition:
\begin{itemize}
	\item
	For $n'\neq n$, we have $|L_{n',n} |n\rangle|^{2}=(\text{min}(n',n)+1)_{|n'-n|}$.
\end{itemize}
As in \eqref{ladder-nontrivial-g0}, the sign of $L_{n',n}$ is fixed using Dirac's ladder operators.
As the explicit expression of $L_{n',n}$ is complicated, we will consider the expectation value to simplify the discussion.
As in \eqref{ladder-k}, $L^{(0)}_{n',n}$ should be equivalent to the $|n'-n|$-th power of Dirac's ladder operators when they act on $|n\rangle$
\begin{align}\label{phase-reduced}
	L_{n',n}|n\rangle=\(\frac{1}{\sqrt{2}}\(x-\text{sgn}(n'-n)\, ip\)\)^{|n'-n|}|n\rangle+O(g)\,.
\end{align}
The main point is that the relative phase factor between the left- and right-hand sides should be $+1$.
Since the relative phase factor is encoded in their inner product, the condition \eqref{phase-reduced} implies the constraint
\begin{align}
	\<\(x+\text{sgn}(n'-n)\, ip\)^{|n'-n|}L_{n',n}\>_{n}^{(0)}>0\,.
\end{align}
In this way, the sign of $L_{n',n}$ is consistent with that in the complete procedure.

In general, the matrix elements obtained in the reduced procedure are the same as those in the complete procedure.
Therefore, the reduced-procedure results for the low-lying matrix elements are at least as complete as those from the complete procedure.

\subsection{Anharmonic operator algebra}
\label{Anharmonic operator algebra}
In the complete procedure, we have derived the explicit expressions of the anharmonic ladder operators. 
We can further study their algebraic properties.
It is natural to construct the anharmonic number operator from the ladder operators,
which form a closed algebra as that in the harmonic case.
However, the Hamiltonian $H$ is a nonlinear function of the number operator and the commutators involving $H$ are more complicated.

As shown in \eqref{Hermitian-conjugate-ladder} and \eqref{Hermitian-conjugate-ladder-2}, the level-1 raising operator $L_{+1}$ is precisely the Hermitian conjugate of the level-1 lowering operator $L_{-1}$
\begin{align}
	L_{+1}=\pt{L_{-1}}^{\dagger}\,.
\end{align}
Here and below, we drop ``n.t.'' for simplicity.
The ladder operators in this subsection are always the nontrivial part.
Using the explicit expression of $L_{\pm 1}$ in \eqref{ladder-nontrivial-g0}, \eqref{ladder-nontrivial-g1}, \eqref{ladder-nontrivial-g2} and \eqref{ladder-nontrivial-g3}, we verify that their commutator takes a simple form as in the harmonic case
\begin{align}
\label{L-1L+1}
	[L_{-1},L_{+1}]=1+O(g^4)\,,
\end{align}
so they provide a natural set of building blocks of the anharmonic operator algebra. 
We can further introduce the anharmonic number operator 
\begin{align}
	\mathcal N=L_{+1}L_{-1}\,,
\end{align}
whose commutators with the level-1 ladder operators are
\begin{align}
	\label{commutators-N-L}
	[\mathcal N, L_{-1}]=-L_{-1}\,,\quad
	[\mathcal N, L_{+1}]=L_{+1}\,.
\end{align}
The anharmonic number operator can be viewed as a conserved quantity,  
as the action of a raising and then a lowering operator should leave a nondegenerate eigenstate invariant up to some factor.
In fact, the eigenvalue of $\mathcal N$ is precisely $n$ in our convention
\begin{align}
	\mathcal{N}|n\rangle=n|n\rangle+O(g^{4})\,.
\end{align}
We verify that the number operator commutes with the Hamiltonian as expected
\begin{align}
	[H, \mathcal N]=O(g^4)\,.
\end{align}
However, the commutators of the Hamiltonian and the anharmonic ladder operators 
cannot be linear in the ladder operators $L_{\pm 1}$, since the energy levels have nonconstant spacing. 
In the small $g$ expansion, they are given by
\begin{align}
	[H,L_{-1}]=L_{-1}G_-(H)+O(g^4)\,,\quad
	[H,L_{+1}]=L_{+1}G_+(H)+O(g^4)\,,
\end{align}
where $G_-$ and $G_+$ are functions of $H$. 
As a result, the commutators of $G_{\pm}$ and $H$ vanish 
and the higher-order commutators are associated with higher powers of $G_\pm$.  
For example, the second order case is $[H,[H,L_{\pm 1}]]=[H,L_{\pm 1}G_{\pm }]=[H,L_{\pm 1}]G_{\pm }=L_{\pm 1}(G_\pm )^2$.
The explicit expressions of $G_+$ and $G_-$ are\footnote{Notice that $G_+=-G_-|_{H\rightarrow -H, g\rightarrow -g}+O(g^4)$.
	For the harmonic oscillator, the transformations $x\rightarrow ix, p\rightarrow ip$, i.e. $H\rightarrow-H$, lead to a different quantization condition, where the wave function vanishes at infinity on the imaginary axis, instead of the usual real infinity.
	In the anharmonic case, we pick up an additional transformation $g\rightarrow-g$ in the perturbative treatment.
}
\begin{align}
	G_+(H)=\;&1
	+g\left(\dfrac{3}{2}+3H\right)
	+g^2\left(-\dfrac{153}{16}-\dfrac{51}{4}H-\dfrac{69}{4}H^2\right)
	\nn
	&+g^3\left(\frac{1305}{16}+\frac{3615}{16}H+\frac{639}{4}H^{2}+\frac{633}{4}H^{3}\right)
	+O(g^4)\,,
	\\
	G_-(H)=\;&-1
	+g\left(\dfrac{3}{2}-3H\right)
	+g^2\left(\dfrac{153}{16}-\dfrac{51}{4}H+\dfrac{69}{4}H^2\right)
	\nn
	&+g^3\left(\frac{1305}{16}-\frac{3615}{16}H+\frac{639}{4}H^{2}-\frac{633}{4}H^{3}\right)
	+O(g^4)\,,
\end{align}
It is precisely the nontrivial dependence on $H$ that leads to the nonlinear energy spacing in the occupation number $n$. 
Furthermore, $G_+$ and $G_-$ are not independent 
as they are closely related to the energy differences.
If the operator $H$ in $G_\pm$ is replaced by a number $E$, 
we have
\begin{align}
	G_+(E)+G_-(E+G_+(E))=O(g^{4})\,,\quad G_-(E)+G_+(E+G_+(E))=O(g^{4})\,,
\end{align}
as the action of a ladder operator and then the opposite one should leave the energy invariant. 
In terms of commutation relations, i.e. $[H,L_{-1}L_{+1}]=O(g^4)$ and $[H,L_{+1}L_{-1}]=O(g^4)$, we have
\begin{align}
	G_-L_{+1}+L_{+1}G_+=O(g^4)\,,\quad
	G_+L_{-1}+L_{-1}G_-=O(g^4)\,.
\end{align}
which can be equivalently written as
\begin{align}
	[G_{+}\,,L_{-1}]=-L_{-1}(G_{+}+G_{-})+O(g^4)\,,\quad
	[G_{-}\,,L_{+1}]=-L_{+1}(G_{+}+G_{-})+O(g^4)\,.
\end{align}
However, the diagonal commutators are given by
\begin{align}
	[G_-\,,L_{-1}]&=L_{-1}\(
		3g
		+\frac{51}{2}\pt{
			1-H
		}g^{2}
		+\frac{9}{8}\pt{
			395-500H+284H^{2}
		}g^{3}+O(g^{4})
	\)\,,
	\\
	[G_+\,,L_{+1}]&=L_{+1}\pt{
		3g
		-\frac{51}{2}\pt{
			1+H
		}g^{2}
		+\frac{9}{8}\pt{
			395+500H+284H^{2}
		}g^{3}+O(g^{4})
	}\,,
\end{align}
so it seems that the Hamiltonian and the ladder operators do not form a simple algebraic structure.
Note that the leading orders of the diagonal commutators $[G_-\,,L_{-1}]$ and $[G_+\,,L_{+1}]$ are both $g^{1}$, 
so we have $[G_-,[G_-\,,L_{-1}]]\sim g^{2}$ and $[G_+,[G_+\,,L_{+1}]]\sim g^{2}$.
By considering higher nested commutators, we can see that the algebra is perturbatively closed to some order in $g$.

Nevertheless, we can express the anharmonic Hamiltonian in terms of the anharmonic number operator $\mathcal N$.
Let us recall that the perturbative series for $E_n$ reads
\begin{align}
	\label{En}
	E_n=\;&\frac{1}{2}+n
	+\frac{3}{4}\pt{1+2n+2n^{2}}g
	-\frac{1}{8}
	\pt{
		21+59n+51n^{2}+34n^{3}
	}g^{2}
	\nn
	&+\frac{3}{16}
	\pt{
		111+347n+472n^{2}+250n^{3}+125n^{4}
	}g^{3}
	+O(g^4),
\end{align}
where $n=0,1,2,\ldots$ labels the energy levels.
This perturbative series gives a precise approximation of the energy levels for sufficiently small $g$ at low $n$.
Accordingly, the Hamiltonian can be expressed in terms of the anharmonic number operator
\begin{align}
	\label{Hamiltonian-in-terms-of-N}
	H=\;&E_0+\mathcal N
	+\frac{3}{4}\pt{2\mathcal N+2\mathcal N^{2}}g
	-\frac{1}{8}
	\pt{
		59\mathcal N+51\mathcal N^{2}+34\mathcal N^{3}
	}g^{2}
	\nn
	&+\frac{3}{16}
	\pt{
		347\mathcal N+472\mathcal N^{2}+250\mathcal N^{3}+125\mathcal N^{4}
	}g^{3}
	+O(g^4), 
\end{align}
which is consistent with replacing the occupation number $n$ in \eqref{En} with $\mathcal N=L_{+1} L_{-1}$. 
From the operator algebraic perspective, 
the left ideal generated by $(H-E_0)$ is a subset of that generated by $L_{-1}$. 
Here $\{H,E_0, L_{-1}\}$ provides a concrete example of a set of consistent data.
For higher states, we can use the generalization of \eqref{Hamiltonian-in-terms-of-N}:
\begin{align}
\label{HLS}
	(L_{-1})^{m}(H-E_{m})=\mathcal SL_{+1}(L_{-1})^{m+1}+O(g^{4})\,,
\end{align}
where $m$ denotes the number of lowering operators and $\mathcal S$ is given by
\begin{align}
	\mathcal S=\;&1
	+\(
		\frac{3}{2}(1+2m)+\frac{3}{2}\mathcal{N}
	\)g
	\nn
	&+\(
		-\frac{1}{8}(59+102m+102m^{ 2})-\frac{51}{8}(1+2m)\mathcal{N}-\frac{17}{4}\mathcal{N}^{2}
	\)g^{2}
	\nn
	&+\(
		\frac{3}{16}(347+944m+750 m^{2}+500 m^{ 3})+\frac{3}{8}(236+375m+375m^{ 2})\mathcal{N}
	\right.
	\nn
	&\qquad\left.
		+\frac{375}{8}(1+2m)\mathcal{N}^{2}+\frac{375}{16}\mathcal{N}^{3}
	\)g^{3}+O(g^{4})\,.
\end{align}
For $m=n$, the stationary Schr\"{o}dinger equation $(H-E_n)|n\rangle=0$ is encoded in 
the level-$n$ null state condition $(L_{-1})^{n+1}|n\rangle=0$, together with the assumption $(L_{-1})^{n}|n\rangle\neq0$. 
Although the statements in this subsection are examined to order $g^3$, 
we believe that some are valid  to arbitrarily high orders and even nonperturbatively. 
 
\section{Dyson-Schwinger equations}
\label{Dyson-Schwinger equations}
In Sec. \ref{The null bootstrap in perturbation theory}, we have solved the anharmonic oscillator in the Hamiltonian formalism.
Before presenting a parallel discussion in the Lagrangian formalism, 
let us first give a brief overview of the DS equations.

In the discussion of the DS equations, we consider the Heisenberg picture.
To be consistent with the results in the Hamiltonian approach, the Lagrangian in the generating functional  
$Z[J]=\int Dx\,\exp\[i\int_{-\infty}^{\infty}\dd t(\mathcal L+J(t)x(t))\]$ is given by
\begin{align}
	\mathcal L=\frac{1}{2}\(\frac{\dd}{\dd t}x(t)\)^{2}-\frac{1}{2}x(t)^{2}-gx(t)^{4}\,,
\end{align}
where $J$ is the classical source.
An infinitesimal change of the integration variable $x(t)$ gives\footnote{This corresponds to the classical equation of motion
\begin{align}
	\frac{\dd^2}{\dd t^2}x(t)+x(t)+4gx(t)^{3}=J(t)\,.
\end{align}}
\begin{align}
	\(\pa_{t}^{2}+1\)\frac{\d}{i\d J(t)}Z[J]+4g\frac{\d^{3}}{i^{3}\d^{3} J(t)}Z[J]=J(t)Z[J]\,.
\end{align}
We can derive the DS equations by taking its functional derivatives with respect to $J$ and then setting $J=0$
\begin{align}
	\label{DS-equations-1}
	&\(\pa_{t}^{2}+1\)G_{n}(t,t_{1},t_{2},\ldots,t_{n-1})
	+4g\sp G_{n+2}(t,t,t,t_{1},t_{2},\ldots,t_{n-1})
	\nn
	=&\;-i\sum_{j=1}^{n}
	\d(t-t_{j})G_{n-2}(t_{1},t_{2},\ldots,t_{j-1},t_{j+1},\ldots,t_{n-1})\,,
\end{align}
where we have introduced the full Green's function
\begin{align}
	G_{n}(t_{1},t_{2},\ldots,t_{n})=\frac{1}{Z[0]}\frac{\d^{n}Z[J]}{i^{n}\d J(t_{1})\d J(t_{2})\ldots\d J(t_{n})}\Big|_{J=0}\,.
\end{align}

To show the equivalence of the Lagrangian and Hamiltonian approaches, 
let us derive the DS equations in the Hamiltonian formalism. 
To remind the reader, the explicit definition of the anharmonic oscillator Hamiltonian is $H=\frac{1}{2}p^{2}+\frac{1}{2}x^{2}+gx^{4}$.
At any time,  the definition of $H$ and the canonical commutation relation $[x,p]=i$ imply an operator identity
\begin{align}
	2HxH-xH^{2}-H^{2}x+x+4gx^{3}=0\,.
\end{align}
In the Heisenberg picture, we should write the $t$ dependence of the operators explicitly. 
Then the matrix element associated with two energy eigenstates $|m_{1}\rangle$ and $|m_{2}\rangle$ reads
\begin{align}
\label{EEx}
	\pt{
		-\pt{E_{m_{1}}-E_{m_{2}}}^{2}+1
	}\langle m_{1}|x(t)|m_{2}\rangle
	+4g\langle m_{1}|x^3(t)|m_{2}\rangle
	=0\,.
\end{align}
which can be written in a differential form\footnote{This can be derived from $x(t)=e^{iHt}x_0e^{-iHt}$ with $x_{0}=x(t=0)$.}
\begin{align}
	\label{DS-1pt}
	(\pa^{2}_{t}+1)
	\langle m_1|x(t)|m_2\rangle
	+4g
	\langle m_1|x^3(t)|m_2\rangle
	=\;0\,,
\end{align} 
To derive the DS equations, we first assume the time order $t_{1}>t_{2}>\ldots> t_{j-2}> t_{j-1}>t>t_{j}>t_{j+1}>\ldots>t_{n-2}>t_{n-1}$.
Then we multiply \eqref{DS-1pt} by the matrix elements $\langle 0|x(t_{1})\ldots x(t_{j-1})|m_{1} \rangle$ and $\langle m_{2}|x(t_{j})\ldots x(t_{n-1})|0 \rangle$ and sum over $m_{1},m_{2}$
\begin{align}
	&(\pa^{2}_{t}+1)\sum_{m_{1},m_{2}}
	\langle 0|x(t_{1})\ldots x(t_{j-1})|m_{1} \rangle
	\langle m_{1}|x(t)|m_{2}\rangle
	\langle m_{2}|x(t_{j})\ldots x(t_{n-1})|0 \rangle
	\nn
	&+4g\sum_{m_{1},m_{2}}
	\langle 0|x(t_{1})\ldots x(t_{j-1}) |m_{1} \rangle
	\langle m_{1}|x^{3}(t)|m_{2}\rangle
	\langle m_{2}|x(t_{j})\ldots x(t_{n-1})|0 \rangle
	=0\,,
\end{align}
Then the completeness relation $\sum_{m}|m\rangle\langle m|=1$ implies
\begin{align}
	\(\pa^{2}_{t}+1\)\langle0| x(t_{1}) \ldots x(t_{j-1})x(t)x(t_{j}) \ldots x(t_{n-1}) |0\rangle
	\nn+4g\langle0| x(t_{1}) \ldots x(t_{j-1})x^{3}(t)x(t_{j}) \ldots x(t_{n-1})|0\rangle
	=0\,.
\end{align}
Similarly, for the slightly different time order $t_{1}>\ldots>t_{j}>t>t_{j+1}\ldots>t_{n-1}$, we have
\begin{align}
	\(\pa^{2}_{t}+1\)\langle0| x(t_{1}) \ldots x(t_{j})x(t) x(t_{j+1})\ldots x(t_{n-1}) |0\rangle
	\nn+4g\langle0| x(t_{1}) \ldots x(t_{j})x^{3}(t) x(t_{j+1})\ldots x(t_{n-1})|0\rangle
	=0\,.
\end{align}
Note that
\begin{align}
	\label{DS-1}
	&\pa_{t}\langle0| x(t_{1}) \ldots x(t)x(t_{j}) \ldots x(t_{n-1}) |0\rangle\big|_{t_{j}\rightarrow t}-\pa_{t}\langle0| x(t_{1}) \ldots x(t_{j})x(t) \ldots x(t_{n-1}) |0\rangle\big|_{t_{j}\rightarrow t}
	\nn
	=\;&\langle0| x(t_{1}) \ldots p(t)x(t) \ldots x(t_{n-1}) |0\rangle-\langle0| x(t_{1}) \ldots x(t)p(t) \ldots x(t_{n-1}) |0\rangle
	\nn
	=\;&-i\langle0| x(t_{1}) \ldots x(t_{j-1})x(t_{j+1})\ldots x(t_{n-1}) |0\rangle\,,
\end{align}
where the dependence on $t$ is removed by the canonical commutation relation $[x(t),p(t)]=i$.
In terms of the Green's function
\begin{align}
	\langle x(t_{1})x(t_{2})\ldots \rangle_{0}
	=\langle 0|T\{x(t_{1})x(t_{2})\ldots\}|0\rangle\,,
\end{align}
Eq. \eqref{DS-1} implies that the first-order $t$-derivative of $\langle x(t)x(t_{1})x(t_{2})\ldots \rangle_{0}$ is discontinuous at the coincident limit $t_{j}\rightarrow t$.
This is realized by introducing the contact term with $\d(t-t_{j})$ to the second-order differential equation.
The complete differential equation reads
\begin{align}
	&(\pa^{2}_{t}+1)\langle x(t)x(t_{1})x(t_{2}) \ldots x(t_{n-1})\rangle_{0}
	+4g\langle x^{3}(t)x(t_{1})x(t_{2})\ldots x(t_{n-1})\rangle_{0}
	\nn
	=\;&-i\sum_{j}\d(t-t_j)\langle x(t_{1})x(t_{2}) \ldots x(t_{j-1})x(t_{j+1}) \ldots x(t_{n-1})\rangle_{0}\,,
\end{align}
which is precisely the DS equation \eqref{DS-equations-1}. 

One can also verify explicitly that the DS equations are satisfied by 
the energy spectrum and the matrix elements in Sec. \ref{The null bootstrap in perturbation theory}.
As an example, we consider
\begin{align}\label{two-point-hamiltonian}
	\langle x(t_{1})x(t_{2}) \rangle_{0}
	=\;&\Theta(t_{1}-t_{2})\sum_{n_{1}}\langle0| e^{iHt_{1}}xe^{-iHt_{1}} |n_{1}\rangle\langle n_{1}| e^{iHt_{2}}xe^{-iHt_{2}} |0\rangle +\(t_{1} \leftrightarrow t_{2}\)
	\nn
	=\;&\Theta(t_{1}-t_{2})\[\(\frac{1}{2}-\frac{3g}{2}+\frac{207g^{2}}{16}\)e^{-i(1+3g-18g^{2})(t_{1}-t_{2})}\right.
	\nn
	&\qquad\qquad\quad\left.+\frac{3g^{2}}{16}e^{-3i(t_{1}-t_{2})}\]+\(t_{1} \leftrightarrow t_{2}\)+O(g^{3})\,,
\end{align}
where $\Theta(x)$ is the Heaviside step function
\begin{align}
	\Theta(x)=\begin{cases}
		\;1&\qquad\text{if $x>0$} \\
		\;\frac{1}{2}&\qquad\text{if $x=0$} \\
		\;0&\qquad\text{if $x<0$.}
	\end{cases}
\end{align}
We have used the half-maximum convention $\Theta(0)=\frac{1}{2}$
so that the formula produces the correct result when $t_{1}=t_{2}$.\footnote{The value at $x=0$ is fixed by time reversal symmetry.}
Furthermore, the function $\langle x^{3}(t_{1})x(t_{2}) \rangle_{0}$ is
\begin{align}
	\langle x^{3}(t_{1})x(t_{2}) \rangle_{0}
	&=\Theta(t_{1}-t_{2})\sum_{n_{1}}\langle 0| e^{iHt_{1}}x^{3}e^{-iHt_{1}} |n_{1}\rangle\langle n_{1}| e^{iHt_{2}}xe^{-iHt_{2}} |0\rangle +\(t_{1} \leftrightarrow t_{2}\)
	\nn
	&=\Theta(t_{1}-t_{2})\[\(\frac{3}{4}-\frac{45g}{8}\)e^{-i(1+3g)(t_{1}-t_{2})}+\frac{3g}{8}e^{-3i(t_{1}-t_{2})}\]
	\nn
	&\phantom{=}\;+\(t_{1} \leftrightarrow t_{2}\)+O(g^{2})\,.
\end{align}
As expected, they satisfy the DS equation \eqref{DS-equations-1} with $n=2$
\begin{align}
	\(\pa^{2}_{t_{1}}+1\)\langle x(t_{1})x(t_{2}) \rangle_{0}+4g\langle x^{3}(t_{1})x(t_{2}) \rangle_{0}+O(g^{3})=-i\d(t_{1}-t_{2})\,.
\end{align}

Below, we will investigate the anharmonic oscillator in the DS approach, without reference to the Hamiltonian.
We make a comparison to the Hamiltonian approach by considering one-point functions of composite operators.
The Dyson-Schwinger equations in the one-point limit give the constraint
\begin{align}
	\<\ldots\(\frac{\dd^{2}}{\dd t^{2}}x(t)+x(t)+4gx^{3}(t)\)\ldots
	\>_{0}=0\,,
\end{align}
where the ellipses represent arbitrary operators at time $t$.
The time-translation invariance implies
\begin{align}
\label{time-translation-DS}
	\pa_t\< x(t)^{m_{1}}\(\frac{\dd}{\dd t}x(t)\)^{m_{2}}\>_{0}=0\,,
\end{align}
which is similar to the equation $\langle[H,\mathcal{O}]\rangle_{0}=0$ in the Hamiltonian approach \cite{Han:2020bkb}.
For example, the constraint from $\langle x(t)\frac{\dd}{\dd t}x(t)\rangle_{0}$ reads
\begin{align}
	0=\pa_t\< x(t)\frac{\dd}{\dd t}x(t)\>_{0}&=\< \(\frac{\dd}{\dd t}x(t)\)^{2}+x(t)\frac{\dd^{2}}{\dd t^{2}}x(t)\>_{0}
	\nn
	&=\< \(\frac{\dd}{\dd t}x(t)\)^{2}-x^{2}(t)-4gx^{4}(t)\>_{0}\,,
\end{align}
which is equivalent to the one from $\langle[H,xp]\rangle_{0}=0$.
However, the constraint \eqref{time-translation-DS} is less stringent than \eqref{consistency-relation}.
To extract more information, we consider higher-point functions $G_{n}(t_{1},t_{2},\ldots)$.
To solve the DS equations for these functions, we will impose the null state condition. 
The solutions for the Green's functions are consistent with the results in Sec. \ref{The null bootstrap in perturbation theory}. 
In parallel to Sec. \ref{The null bootstrap in perturbation theory}, we can use the exact expressions of the ladder operators to construct the null state condition, which corresponds to the complete procedure 
in the Lagrangian formalism.
Alternatively, we can also solve the DS equations by only assuming the existence of some null states, without  knowing the exact expressions of the ladder operators.
This can be seen as the reduced procedure for solving the DS equations.

\subsection{Complete procedure}
In the complete procedure of Sec. \ref{Complete procedure 1}, we obtain the exact expressions of the ladder operators.
We now use the corresponding null state condition to solve the DS equations. 
As discussed in Sec. \ref{Introduction}, we will derive the constraints for $G_{n}$ by considering the null state condition and its inner product with some test states.
We will show that $G_{n}$ can be determined order by order in $g$.

The null state condition in the Heisenberg picture reads
\begin{align}\label{null-state-condition-1}
	L_{-1}(t)|0\rangle=0\,,
\end{align}
where $L_{-1}(t)=e^{iHt}L_{-1}e^{-iHt}$ is the lowering operator in the Heisenberg picture. 
According to $p(t)=\frac{\dd}{\dd t}{x}(t)$ and the small $g$ expansion \eqref{ladder-g-expansion}, the null state condition \eqref{null-state-condition-1} can be rearranged into the form
\begin{align}
	\label{null-state-condition}
	\(x(t)+i\frac{\dd}{\dd t}x(t)\)
	|0\rangle
	=-\sqrt{2}
	\(gL_{-1}^{(1)}(t)+g^{2}L_{-1}^{(2)}(t)+O(g^{3})\)
	|0\rangle\,,
\end{align}
where the right-hand side corresponds to the perturbative corrections to the lowering operator.
Below, we will always assume $t_{1}>t_{2}>\ldots>t_{n}$ and sometimes write simply $G_n$ for $G_{n}(t_{1},t_{2},\ldots,t_{n})$.
Note that $G_{n}$ is symmetric in its arguments, so it suffices to solve for $G_{n}$ in this specific time order. 
To relate the states in \eqref{null-state-condition} to $G_{n}$, we consider the inner product with 
the test state $\langle 0|x(t_{1})x(t_{2})\ldots x(t_{n-1})$.
The null state condition \eqref{null-state-condition} implies the null differential equation
\begin{align}
	\label{null-constraint-tn}
	G_{n}+i\pa_{t_{n}}G_{n}
	=gU_{n}^{(1)}+g^{2}U_{n}^{(2)}+O(g^{3})\,,
\end{align}
where the terms on the right-hand side are associated with the perturbative corrections to the lowering operator\footnote{Note that $U_{n}^{(j)}$ admits a small $g$ expansion.}
\begin{align}
\label{definition_Un}
	U_{n}^{(j)}\equiv\;-\sqrt{2}\langle 0|x(t_{1})x(t_{2})\ldots x(t_{n-1})L_{-1}^{(j)}(t_{n})|0\rangle\,.
\end{align}
The explicit expressions of $U_{n}^{(j)}$ are composed of Green's functions and their derivatives.

For the Green's function $G_n$, the functional form in $t_n$ is determined by the null differential equation \eqref{null-constraint-tn}.
But what about other time variables?
It seems that \eqref{null-constraint-tn} is not useful as it only involves the $t_{n}$ derivative.
However, we notice that in certain limits, the $t_n$ derivative is related to the $t_k$ derivatives with $k<n$.
This will lead to constraints on the $t_{k}$ dependence of $G_{n}$.

To see the constraints more explicitly, we consider the coincident limit of several time variables, i.e., 
$(G_{n}+i\pa_{t_{k}}G_{n})_{t_{k+1},t_{k+2},\ldots,t_{n}\rightarrow t_{k}}$. 
Then the derivative with respect to $t_{k}$ can be associated with $\pa_{t_{n}}$  
and we can use \eqref{null-constraint-tn} to constrain the $t_{k}$ dependence as well. 
For example,  the coincident limit of two time variables is
\begin{align}\label{tk-derivative-example}
	\(G_{n}+i\pa_{t_{n-1}}G_{n}\)_{t_{n}\rightarrow t_{n-1}}=&\;G_{n-2}+\(G_{n}+i\pa_{t_{n}}G_{n}\)_{t_{n}\rightarrow t_{n-1}}
	\nn=&\;G_{n-2}+\(gU_{n}^{(1)}+g^{2}U_{n}^{(2)}+O(g^{3})\)_{t_{n}\rightarrow t_{n-1}}\,,
\end{align}
where we have used \eqref{null-constraint-tn} and the DS equation in the $t_{n}\rightarrow t_{n-1}$ limit
\begin{align}
	\(\pa_{t_{n-1}}G_{n}-\pa_{t_{n}}G_{n}\)_{t_{n}\rightarrow t_{n-1}}=-iG_{n-2}\,.
\end{align}
In the coincident time limit $t_{n}\rightarrow t_{n-1}$, the term $i\pa_{t_{n-1}}G_{n}$ in \eqref{tk-derivative-example} is expressed in terms of $i\pa_{t_{n}}G_{n}$ and $G_{n-2}$, so we can use \eqref{null-constraint-tn} to constrain the $t_{n-1}$ dependence of $G_{n}$.
From the perspective of operator theory, 
we move the momentum operator $p(t_{n-1})$ to the right using the commutation relation, 
so we can take advantage of the null state condition.
More generally, we obtain a set of null differential equations
\begin{align}
	\label{null-constraint-tk}
	\(G_{n}+i\pa_{t_{k}}G_{n}\)_{t_{k+1},t_{k+2},\ldots,t_{n}\rightarrow t_{k}}=\[\(n-k\)G_{n-2}+gU_{n}^{(1)}+g^{2}U_{n}^{(2)}\]_{t_{k+1},t_{k+2},\ldots,t_{n}\rightarrow t_{k}}+O(g^{3})\,,
\end{align}
where $k=1,2,\ldots,n$.
Note that there is a term proportional to $G_{n-2}$ on the right-hand side. 
The factor $(n-k)$ is related to the use of $(n-k)$ different DS equations at coincident times, 
which is equivalent to the number of commutation relations in the operator theory perspective.

The null differential equations \eqref{null-constraint-tk} are the main ingredients to obtain the Green's functions in the complete procedure.
Now, the question is whether we can determine $G_{n}$ completely using these constraints, since the null differential equations \eqref{null-constraint-tk} only constrain $G_{n}$ in certain limits.
But the short answer is yes.
Below, we address this question in more detail.

Let us assume that $G_n$ admits a small $g$ expansion
\begin{align}
	G_{n}=\sum_{i=0,1,2,\dots}g^i G^{(i)}_{n}\,.
\end{align}
It is simpler to consider the zeroth order.
We will show by induction that we can in principle determine all $G_{n}^{(0)}$ using \eqref{null-constraint-tk}.
Suppose that $G_{n-2}^{(0)}$ is known. 
For $G_{n}^{(0)}$, the null differential equation \eqref{null-constraint-tk} with $k=n$ at zeroth order determines its functional form in $t_n$. 
Next, we consider the null differential equation \eqref{null-constraint-tk} with $k=n-1$ at zeroth order, where the limit $t_{n}\rightarrow t_{n-1}$ is taken. 
However, the crucial point is that the limit $t_{n}\rightarrow t_{n-1}$ does not lose any information about the functional form in $t_{n-1}$, since the functional dependence on $t_{n}$ is known.
We proceed similarly for $k=n-2,n-1,\ldots,1$, and the functional dependence on each time variable can be determined.
In the end, $G_{n}^{(0)}$ is determined up to a free parameter, which can be fixed by time-translation invariance.\footnote{This is because the general solutions to \eqref{null-constraint-tk} violate time-translation invariance.
}
Therefore, if $G_{n-2}^{(0)}$ is known, we can determine $G_{n}^{(0)}$ completely.
Since we have $G_{0}^{(0)}=1$ by definition, 
we can determine $G_{n}^{(0)}$ one by one.\footnote{We assume that the parity symmetry is unbroken, so the $n$-point Green's function $G_{n}$ vanishes if $n$ is odd.}
For the perturbative corrections in $g$, we need to take into account the contribution of $U_n$, 
but they can be computed using lower order results for $G_n$.
So the argument extends to the perturbative corrections as well. 
In conclusion, we can in principle determine all $G_{n}$ completely order by order using the null differential equation \eqref{null-constraint-tk} and time-translation invariance.

\subsubsection*{Examples: $G_{2}$, $G_{4}$ and $G_{6}$}
We consider some low-point Green's functions $G_{2}$, $G_{4}$, and $G_{6}$ as explicit examples.
At zeroth order, the $(n,k)=(2,2)$ null differential equation \eqref{null-constraint-tk} reads
\begin{align}
	G_{2}^{(0)}(t_{1},t_{2})+i\pa_{t_{2}}G_{2}^{(0)}(t_{1},t_{2})=0\,,
\end{align}
and the solution is
\begin{align}
	\label{G2-g0-t2}
	G_{2}^{(0)}(t_{1},t_{2})=Q_{0}(t_{1})e^{it_{2}}\,.
\end{align}
Here and below, we use $Q_{i}$ to represent the functional dependence that remains to be determined.
They will be obtained by considering more null differential equations. 
Using \eqref{G2-g0-t2}, we see that the $(n,k)=(2,1)$ null differential equation \eqref{null-constraint-tk} becomes
\begin{align}
	e^{it_{1}}Q_{0}(t_{1})+ie^{it_{1}}\pa_{t_{1}}Q_{0}(t_{1})=G_{0}^{(0)}\,,
\end{align}
where by definition $G_{0}^{(0)}=1$.
The solution to this differential equation is
\begin{align}
	Q_{0}(t_{1})=\frac{1}{2}e^{-it_{1}}+c_{0}e^{it_{1}}\,,
\end{align}
where $c_i$ denotes the free parameter. 
On the right-hand side, the second term violates time-translation invariance. 
Therefore, the time-translation invariant solution for $G_{2}^{(0)}$ reads
\begin{align}\label{G2-g0}
	G_{2}^{(0)}(t_{1},t_{2})=\frac{1}{2}e^{-it_{1}+it_{2}}\,.
\end{align}
This is precisely \eqref{two-point-hamiltonian} at zeroth order with $t_1>t_2$. 
In the calculation of $G_{4}^{(0)}$, there are more intermediate functions $Q_i$, 
and the null differential equations involve the previous result for $G_{2}^{(0)}$.
First we have the $(n,k)=(4,4)$ null differential equation
\begin{align}
	G_{4}^{(0)}(t_{1},t_{2},t_{3},t_{4})+i\pa_{t_{4}}G_{4}^{(0)}(t_{1},t_{2},t_{3},t_{4})=0\,,
\end{align}
and the solution is $G_{4}^{(0)}(t_{1},t_{2},t_{3},t_{4})=e^{it_{4}}Q_{1}(t_{1},t_{2},t_{3})$.
Next, the $(n,k)=(4,3)$ null differential equation reads
\begin{align}
	e^{it_{3}}Q_{1}(t_{1},t_{2},t_{3})+ie^{it_{3}}\pa_{t_{3}}Q_{1}(t_{1},t_{2},t_{3})=G_{2}^{(0)}(t_{1},t_{2})\,,
\end{align}
and the solution is
\begin{align}
	Q_{1}(t_{1},t_{2},t_{3})=\frac{1}{4}e^{-it_{1}+it_{2}-it_{3}}+e^{it_{3}}Q_{2}(t_{1},t_{2})\,.
\end{align}
Then we solve the $(n,k)=(4,2)$ and $(n,k)=(4,1)$ null differential equations and obtain
\begin{align}\label{G4-g0}
	G_{4}^{(0)}(t_{1},t_{2},t_{3},t_{4})=\frac{1}{4}e^{-it_{1}+it_{2}-it_{3}+it_{4}}+\frac{1}{2}e^{-it_{1}-it_{2}+it_{3}+it_{4}}\,,
\end{align}
where again we have used time-translation invariance to fix the free parameter. 
Repeating the procedure above for $G_{6}^{(0)}$, we have the time-translation invariant result
\begin{align}\label{G6-g0} 
	G_{6}^{(0)}(t_{1},t_{2},t_{3},t_{4},t_{5},t_{6})=\;&\frac{1}{8} e^{-i t_1+i t_2-i t_3+i t_4-i t_5+i t_6}
	+\frac{1}{4} e^{-i t_1-i t_2+i t_3+i t_4-i t_5+i t_6}
	\nn
	&+\frac{1}{4} e^{-i t_1+i t_2-i t_3-it_4+i t_5+i t_6}
	+\frac{1}{2} e^{-i t_1-i t_2+i t_3-i t_4+i t_5+i t_6}
	\nn
	&+\frac{3}{4} e^{-i t_1-i t_2-i t_3+i t_4+i t_5+i t_6}\,.
\end{align}

At first order in $g$, we need to take into account the contribution of $U^{(1)}_{n}$.
According to the definition \eqref{definition_Un}, it is related to $G_{n}$, $G_{n+2}$ and their derivatives in a certain limit.
For example, the contribution of the term $xp^{2}(t_{n})$ in $L^{(1)}_{-1}(t_{n})$ can be computed by
\begin{align}
	\label{J-calculation-example}
	\langle0|x(t_{1})x(t_{2})\ldots x(t_{n-1}) xp^{2}(t_{n})|0\rangle=\(\pa_{t_{n+1}}\pa_{t_{n+2}}G_{n+2}\)_{t_{n+1},t_{n+2}\rightarrow t_{n}}.
\end{align}
Using the zeroth-order solutions \eqref{G2-g0}, \eqref{G4-g0} and the exact expression of the nontrivial lowering operator \eqref{ladder-nontrivial-g1}, 
we find
\begin{align}
	\label{U2(1)-g0}
	U_{2}^{(1)}\big|_{g^{0}}=-\frac{3}{2}e^{-it_{1}+it_{2}}\,,
\end{align}
where $U_{n}^{(j)}|_{g^{i}}$ represents $U_{n}^{(j)}$ at order $g^{i}$. 
At first order, the $(n,k)=(2,2)$ null differential equation reads
\begin{align}
	G_{2}^{(1)}(t_{1},t_{2})+i\pa_{t_{2}}G_{2}^{(1)}(t_{1},t_{2})=U_{2}^{(1)}\big|_{g^{0}}\,,
\end{align}
which leads to the solution
\begin{align}
	G_{2}^{(1)}(t_{1},t_{2})=\frac{3}{2} i e^{i t_2-i t_1} t_2+e^{i t_2} Q_{3}(t_1)\,.
\end{align}
Then the $(n,k)=(2,1)$ null differential equation reads
\begin{align}
	3it_{1}+e^{it_{1}}Q_{3}(t_1)+ie^{it_{1}}\pa_{t_1}Q_{3}(t_1)=\(G_{0}^{(1)}+U_{2}^{(1)}\big|_{g^{0}}\)_{t_2\rightarrow t_1}\,,
\end{align}
where by definition $G_{0}^{(1)}=0$.
The solution is
\begin{align}
	Q_{3}(t_1)=-\frac{3}{2} e^{-i t_1}-\frac{3}{2} i e^{-i t_1} t_1+c_1 e^{i t_1}\,.
\end{align}
Together with time-translation invariance, we find that
\begin{align}\label{G2-g1}
	G_{2}^{(1)}(t_{1},t_{2})=\frac{3}{2} e^{-i t_1+i t_2} \( -1-it_{1}+it_{2} \)\,,
\end{align}
which is the first-order term in \eqref{two-point-hamiltonian} with $t_1>t_2$. 
For the calculation of $G_{4}^{(1)}$, we need to first compute the zeroth-order expression of $U_{4}^{(1)}$
\begin{align}\label{U4(1)-g0}
	U_{4}^{(1)}\big|_{g^{0}}=-\frac{3}{4} e^{-i t_1+i t_2-i t_3+i t_4}-\frac{3}{2} e^{-i t_1-i t_2+i t_3+i t_4}-\frac{3}{4} e^{-i t_1-i t_2-i t_3+3 i t_4}\,.
\end{align}
The above procedure then gives the time-translation invariant result
\begin{align}\label{G4-g1}
	G_{4}^{(1)}(t_{1},t_{2},t_{3},t_{4})
	=\;& \frac{3}{8} e^{ -3it_{1}+it_{2}+it_{3}+it_{4} } + \frac{3}{8} e^{ -it_{1}-it_{2}-it_{3}+3it_{4} } \nn
	&+ \frac{3}{4} e^{ -it_{1}+it_{2}-it_{3}+it_{4} } \( -2-it_{1}+it_{2}-it_{3}+it_{4} \) \nn
	&+ \frac{3}{2} e^{ -it_{1}-it_{2}+it_{3}+it_{4} } \( -3-it_{1}-2it_{2}+2it_{3}+it_{4} \)\,.
\end{align}
We will not consider $G^{(1)}_{6}$ because that requires the knowledge of $G^{(0)}_{8}$.

At second order, the contributions from the corrections to the lowering operator are $U_{n}^{(2)}|_{g^{0}}$ and $U_{n}^{(1)}|_{g^{1}}$.
Using the first-order solutions \eqref{G2-g1} and \eqref{G4-g1}, we obtain
\begin{align}
	\label{U2(1)-g1}
	U_{2}^{(1)}\big|_{g^{1}}
	=-\frac{3}{8} e^{ -3it_{1}+3it_{2} } + \frac{9}{8} e^{ -it_{1}+it_{2} } \( 3+4it_{1}-4it_{2} \)\,.
\end{align}
The zeroth-order solutions \eqref{G2-g0}, \eqref{G4-g0} and \eqref{G6-g0} give
\begin{align}
	\label{U2(2)-g0}
	U_{2}^{(2)}\big|_{g^{0}}=\frac{81}{8}e^{-it_{1}+it_{2}}\,.
\end{align}
At order $g^{2}$, the $(n,k)=(2,2)$ null differential equation is
\begin{align}
	G_{2}^{(2)}(t_{1},t_{2})+i\pa_{t_{2}}G_{2}^{(2)}(t_{1},t_{2})=U_{2}^{(1)}\big|_{g^{1}}+U_{2}^{(2)}\big|_{g^{0}}\,,
\end{align}
which has the solution
\begin{align}
	G_{2}^{(2)}(t_{1},t_{2})=-\frac{9}{4} e^{-it_{1}+it_{2}} t_2^2-\frac{27}{2} i e^{-it_{1}+it_{2}} t_2+\frac{9}{2} e^{-it_{1}+it_{2}} t_1
	t_2+\frac{3}{16} e^{-3it_{1}+3it_{2}}+e^{i t_2} Q_{4}(t_1)\,.
\end{align}
In the end, the $(n,k)=(2,1)$ null differential equation reads
\begin{align}
	\frac{9 t_1^2}{2}-\frac{45 i t_1}{2}+\frac{3}{4}+e^{i t_1} Q_{4}(t_1)+i e^{i t_1} \pa_{t_{1}}Q_{4}(t_1)=\(U_{2}^{(1)}\big|_{g^{1}}+U_{2}^{(2)}\big|_{g^{0}}\)_{t_{2}\rightarrow t_{1}}\,,
\end{align}
and the solution is
\begin{align}
	Q_{4}(t_1)=-\frac{9}{4} e^{-i t_1} t_1^2+\frac{27}{2} i e^{-i t_1} t_1+\frac{207}{16} e^{-i t_1}+c_2 e^{i t_1}\,.
\end{align}
As before, the free parameter is fixed by time-translation invariance, and we find 
\begin{align}\label{G2-g2}
	G_{2}^{(2)}(t_{1},t_{2})=\frac{3}{16}e^{-3it_{1}+3it_{2}}-\frac{9}{16}e^{-it_{1}+it_{2}}\(4(t_{1}-t_{2})^{2}-24i(t_{1}-t_{2})-23\)\,.
\end{align}
The result agrees with \eqref{two-point-hamiltonian} at second order. 
We will not consider $G_{4}^{(2)}$ because that requires the knowledge of $U_{4}^{(1)}|_{g^{1}}$ and $U_{4}^{(2)}|_{g^{0}}$, which means that the calculation involves $G^{(1)}_{6}$ and $G^{(0)}_{8}$.

\subsubsection*{A simplified approach}

In the discussion above, we wrote $U_{n}^{(j)}$ in terms of higher-point functions and their derivatives, 
involving more time variables and thus more complicated functions that seem irrelevant. 
In a nonperturbative setting, the complexity of the Green's functions can grow much faster with the number of time variables than the perturbative case.
To avoid introducing additional time variables in the intermediate steps, 
we would like to express $U_{n}^{(j)}$ directly in terms of Green's functions with $n$ time variables and their derivatives.
In this approach, we can obtain $G_{n}$ without considering additional time variables.

To illustrate this approach, let us recall the example \eqref{J-calculation-example}. 
In fact, it can also be computed from
\begin{align}\label{simplified-example}
	\langle0|x(t_{1})x(t_{2})\ldots x(t_{n-1}) xp^{2}(t_{n})|0\rangle
	=\;&\frac{1}{6}\(\pa_{t_{n}}^{2}+3\)G_{n+2,t_{n+2},t_{n+1}\rightarrow t_{n}}
	\nn
	&+2g\sp G_{n+4,t_{n+4},t_{n+3},t_{n+2},t_{n+1}\rightarrow t_{n}}
	+i\pa_{t_{n}}G_{n}\,,
\end{align}
where we have used the DS equations at coincident times (or equivalently the canonical commutation relation).
The main difference is that we exchange the order of coincident time limits and time derivatives in the computation of $U_{n}^{(j)}$.
As we can see in \eqref{simplified-example}, we need to consider the Green's functions at coincident times.
They can be obtained by
\begin{align}
	\label{null-constraint-tk-1}
	&G_{n,t_{k+1},t_{k+2},\ldots,t_{n}\rightarrow t_{k}}+\frac{i}{n-k+1}\pa_{t_{k}}G_{n,t_{k+1},t_{k+2},\ldots,t_{n}\rightarrow t_{k}}
	\nn
	=\;&\(\frac{n-k}{2}G_{n-2}+gU_{n}^{(1)}+g^{2}U_{n}^{(2)}\)_{t_{k+1},t_{k+2},\ldots,t_{n}\rightarrow t_{k}}+O(g^{3})\,,
\end{align}
which is similar to \eqref{null-constraint-tk}, but we take the coincident time limit first and then take the time derivative.

As an example, let us solve for $G_{2}$ to order $g^{2}$ using only two time variables.
At zeroth order, the calculation of $G_{2}^{(0)}$ is the same as that in the example above.
Let us consider the perturbative corrections.
To obtain $G_{2}^{(1)}$, we need to compute $U_{2}^{(1)}|_{g^{0}}$, which is related to $G_{2}^{(0)}(t_1,t_2)$ and $G_{4}^{(0)}(t_{1},t_{2},t_{2},t_{2})$.
The null differential equation \eqref{null-constraint-tk-1} with $(n,k)=(4,2)$ reads
\begin{align}
	G_{4}^{(0)}(t_{1},t_{2},t_{2},t_{2})+\frac{i}{3}\pa_{t_{2}}G_{4}^{(0)}(t_{1},t_{2},t_{2},t_{2})=G_{2}^{(0)}(t_{1},t_{2})\,,
\end{align}
and the solution is
\begin{align}
	G_{4}^{(0)}(t_{1},t_{2},t_{2},t_{2})=\frac{3}{4}e^{-it_{1}+it_{2}}+e^{3it_{2}}Q_{5}(t_{1})\,.
\end{align}
Then we consider the $(n,k)=(4,1)$ null differential equation \eqref{null-constraint-tk-1}
\begin{align}
	\frac{3}{4}+\frac{1}{4}e^{3it_{1}}Q_{5}(t_{1})+\frac{1}{4}ie^{3it_{1}}\pa_{t_{1}}Q_{5}(t_{1})=\frac{3}{2}G_{2}^{(0)}(t_{1},t_{1})\,.
\end{align}
The solution is $Q_{5}(t_{1})=c_{4}e^{it_{1}}$, and we obtain the time-translation invariant solution
\begin{align}\label{G4-g0-reduced}
	G_{4}^{(0)}(t_{1},t_{2},t_{2},t_{2})=\frac{3}{4}e^{-it_{1}+it_{2}}\,,
\end{align}
which is \eqref{G4-g0} in the limit $t_{3},t_{4}\rightarrow t_{2}$.
Using the zeroth-order solutions \eqref{G2-g0} and \eqref{G4-g0}, we have
\begin{align}\label{U2-1-g0-2}
	U_{2}^{(1)}\big|_{g^{0}}=-\frac{3}{2}e^{-it_{1}+it_{2}}\,,
\end{align}
which is precisely \eqref{U2(1)-g0}. Then time-translation invariance and \eqref{U2-1-g0-2} allow us to determine $G_{2}^{(1)}$
\begin{align}\label{G2-g1-2}
	G_{2}^{(1)}(t_{1},t_{2})=\frac{3}{2} e^{-i t_1+i t_2} \( -1-it_{1}+it_{2} \)\,,
\end{align}
which is exactly \eqref{G2-g1}.

To obtain $G_{2}^{(2)}$, we need to consider $U_{2}^{(2)}|_{g^{0}}$ and $U_{2}^{(1)}|_{g^{1}}$, which are related to the following Green's functions:
\begin{align}
	&G_{2}^{(0)}(t_{1},t_{2})\,,\;
	G_{4}^{(0)}(t_{1},t_{2},t_{2},t_{2})\,,\;
	G_{6}^{(0)}(t_{1},t_{2},t_{2},t_{2},t_{2},t_{2})\,,
	\nn
	&G_{2}^{(1)}(t_{1},t_{2})\,,\;
	G_{4}^{(1)}(t_{1},t_{2},t_{2},t_{2})\,,
\end{align}
where $G_{6}^{(0)}(t_{1},t_{2},t_{2},t_{2},t_{2},t_{2})$ and $G_{4}^{(1)}(t_{1},t_{2},t_{2},t_{2})$ are unknown.
At zeroth order, the null differential equation \eqref{null-constraint-tk-1} with $(n,k)=(6,2)$ yields
\begin{align}
	G_{6}^{(0)}(t_{1},t_{2},t_{2},t_{2},t_{2},t_{2})=\frac{15}{8}e^{-it_{1}+it_{2}}+e^{5it_{2}}Q_{6}(t_{1})\,.
\end{align}
Then the $(n,k)=(6,1)$ null differential equation gives
\begin{align}
	Q_{6}(t_{1})=c_{5}e^{it_{1}}\,.
\end{align}
So we obtain the time-translation invariant solution
\begin{align}\label{G6-g0-complete-simplified}
	G_{6}^{(0)}(t_{1},t_{2},t_{2},t_{2},t_{2},t_{2})=\frac{15}{8}e^{-it_{1}+it_{2}}\,,
\end{align}
which is \eqref{G6-g0} in the limit $t_{3},t_{4},t_{5},t_{6}\rightarrow t_{2}$. 
Using \eqref{G2-g0}, \eqref{G4-g0-reduced}, and \eqref{G6-g0-complete-simplified}, we obtain
\begin{align}
	U^{(2)}_{2}\big|_{g^{0}}=\frac{81}{8}e^{-it_{1}+it_{2}}\,,
\end{align}
which is precisely \eqref{U2(2)-g0}.
Next we compute $G_{4}^{(1)}(t_{1},t_{2},t_{2},t_{2})$.
The $(n,k)=(4,2)$ null differential equation \eqref{null-constraint-tk-1} reads
\begin{align}
	G_{4}^{(1)}(t_{1},t_{2},t_{2},t_{2})+\frac{i}{3}\pa_{t_{2}}G_{4}^{(1)}(t_{1},t_{2},t_{2},t_{2})=G_{2}^{(1)}(t_{1},t_{2})+\(U_{4}^{(1)}\big|_{g^{0}}\)_{t_{3},t_{4}\rightarrow t_{2}}\,.
\end{align}
The second term on the right-hand side can be computed using the zeroth-order solutions \eqref{G2-g0}, \eqref{G4-g0} and \eqref{G6-g0}
\begin{align}
	\(U_{4}^{(1)}\big|_{g^{0}}\)_{t_{3},t_{4}\rightarrow t_{2}}=-3e^{-it_{1}+it_{2}}\,.
\end{align}
In the end, using the $(n,k)=(4,1)$ null differential equation \eqref{null-constraint-tk-1}, we obtain
\begin{align}\label{G4-g1-simplified}
	G_{4}^{(1)}(t_{1},t_{2},t_{2},t_{2})
	=\frac{3}{8} e^{ -3it_{1}+3it_{2} } + \frac{9}{8} e^{ -it_{1}+it_{2} } \( -5-2it_{1}+2it_{2} \)\,,
\end{align}
and this is \eqref{G4-g1} in the limit $t_{3},t_{4}\rightarrow t_{2}$. 
Together with \eqref{G2-g1}, we obtain
\begin{align}
	U_{2}^{(1)}\big|_{g^{1}}
	=-\frac{3}{8} e^{ -3it_{1}+3it_{2} } + \frac{9}{8} e^{ -it_{1}+it_{2} } \( 3+4it_{1}-4it_{2} \)\,.
\end{align}
which is the same as \eqref{U2(1)-g1}.
Assuming time-translation invariance, we can determine $G_{2}$ to order $g^{2}$ using only two time variables.
We have also verified that $G_{4}^{(1)}(t_{1},t_{2},t_{3},t_{4})$ can be determined using only four time variables. 

\subsection{Reduced procedure}
We have shown that the DS equations can be solved in the complete procedure, i.e., by using the null state condition from the exact ladder operators.
In the reduced procedure, the exact expressions of the ladder operators $L_{\pm k}$ are unknown, 
so the explicit form of the null state condition seems unclear.
However, we want to emphasize that the null state condition should be consistent with the DS equations \eqref{DS-equations-1}, 
which leads to strong constraints on the possible form of the null differential equations. 
This is parallel to the consistency of the Hamiltonian $H$ and the ladder operator $L_{-1}$ 
discussed in Sec. \ref{Anharmonic operator algebra}.\footnote{The stationary Schr\"{o}dinger equation cannot be encoded in an inconsistent null state condition as in \eqref{HLS}. }
In fact, the perturbative DS equations allow only two types of null state conditions, both of which can determine $G_{n}$ completely order by order in $g$.

Although we do not know the exact expressions of the ladder operators, we have some general idea about the form of the null state condition.
We assume that
\begin{align}
	L_{\pm1}(t)|0\rangle=0\,,
\end{align}
where $L_{\pm1}(t)$ has the small $g$ expansion
\begin{align}
	L_{\pm 1}(t)=L_{\pm1}^{(0)}(t)+gL_{\pm1}^{(1)}(t)+g^{2}L_{\pm1}^{(2)}(t)+O(g^{3})\,.
\end{align}
The $L_{\pm1}^{(j)}(t)$ are degree-$(2j+1)$ polynomials in $x(t)$ and $\dot x(t)=p(t)$.\footnote{In \cite{Li:2023nip}, the null operator is built from higher derivatives of $x(t)$:
	\begin{align}
		\mathcal O_{\text{null}}=\sum_m a_m\frac{\dd^m x(t)}{\dd t^m}\,,
	\end{align}
which should be equivalent to the ansatz here using both $x(t)$ and $\dot x(t)$. }
These are the minimal degrees for constructing the nontrivial part of the level-$1$ ladder operators.\footnote{ In Sec. \ref{Reduced procedure 1}, the null operators are associated with lower-degree polynomials in $x$ and $ip$, which helps to deduce considerably higher order results.
However, this is not be very useful for the reduced procedure here.
We can assume that $L_{\pm 1}(t)$ is of degree-$(1,3,3,\ldots)$ in $x(t)$ and $\dot x(t)$ at order $g^0,g^1,g^2,\ldots$ based on the $K=3$ results in \eqref{naive-K3-n1}.
Then in the examples below, we do not directly need $G_{6}^{(0)}$ to deduce \eqref{U-U}, 
which is the main ingredient in the derivation of $G^{(2)}_{2}$.
However, this simplification is insignificant, since \eqref{U-U} requires the knowledge of $G^{(1)}_{4}$, and 
the derivation of $G^{(1)}_{4}$ is based on the solution of $G_{6}^{(0)}$.
Nonetheless, the approach would be more useful if $G^{(1)}_{4}$ is already known and we do not need to consider the more complicated six-point function. }
We will see later that there are two choices $\pm1$ corresponding to the two types of null state conditions that the DS equations allow.

As indicated in \eqref{null-state-condition}, the null state condition can be rearranged into a more convenient form
\begin{align}\label{null-state-rearranged}
	\(x(t)+c_{\pm}i\frac{\dd}{\dd t}x(t)\)
	|0\rangle
	=-\(gL_{\pm1}^{(1)}(t)+g^{2}L_{\pm1}^{(2)}(t)+O(g^{3})\)
	|0\rangle\,,
\end{align}
where $c_{\pm}$ is the relative coefficient in the degree-1 polynomial.
Equation \eqref{null-state-rearranged} leads to the null differential equation for $G_{n}$
\begin{align}
	\label{null-condition-G-tn}
	G_{n}+ic_{\pm}\pa_{t_{n}}G_{n}=gU^{(1)}_{n,\pm}+g^{2}U^{(2)}_{n,\pm}+O(g^{3})\,,
\end{align}
where we have defined
\begin{align}
	U^{(j)}_{n,\pm}\equiv-\langle0|x(t_{1})x(t_{2})\ldots L_{\mp1}^{(j)}(t_{n})|0\rangle\,.
\end{align}
Here $c_{\pm}$ and $U_{\pm}^{(j)}$ are unknown, so the null differential equation \eqref{null-condition-G-tn} seems unclear, as opposed to \eqref{null-constraint-tn} in the complete procedure.
Nevertheless, let us write down the null differential equations with $t_{k}$ derivatives
\begin{align}
\label{null-diff}
	\(G_{n}+ic_{\pm}\pa_{t_{k}}G_{n}\)_{t_{k+1},t_{k+2},\ldots,t_{n}\rightarrow t_{k}}
	=\;&\[c_{\pm}\(n-k\)G_{n-2}+gU_{n,\pm}^{(1)}+g^{2}U_{n,\pm}^{(2)}\]_{t_{k+1},t_{k+2},\ldots,t_{n}\rightarrow t_{k}}
	+O(g^{3})\,.
\end{align}
Below, we will explain how to determine $c_{\pm}$ and $U_{\pm}^{(j)}$.
Then we can solve for $G_{n}$ using the null differential equations \eqref{null-diff} as in the complete procedure. 

For consistency, the solutions to the DS equations should also satisfy the null differential equations.
We will substitute the DS solutions into the null differential equation \eqref{null-condition-G-tn}, which yields strong constraints on $c_{\pm}$ and $U_{\pm}^{(j)}$.
There are two types of choices satisfying these constraints, corresponding to the null state condition for the raising and lowering operators.
More explicitly, at zeroth order, we consider the DS equation
\begin{align}
	\(\pa_{t_{n}^{2}}+1\)G_{n}^{(0)}(t_{1},t_{2},\ldots,t_{n})=0\,,
\end{align}
which has the solution
\begin{align}
	G_{n}^{(0)}(t_{1},t_{2},\ldots,t_{n})=e^{it_{n}}Q_{7}(t_{1},t_{2},\ldots,t_{n-1})+e^{-it_{n}}Q_{8}(t_{1},t_{2},\ldots,t_{n-1})\,.
\end{align}
We remind the reader that $Q_{i}$ represents the functional dependence that remains to be determined. 
The $k=n$ null differential equation \eqref{null-condition-G-tn} implies
\begin{align}\label{DS-soultion-tn}
	(1-c_{\pm})e^{it_{n}}Q_{7}(t_{1},t_{2},\ldots,t_{n-1})+(1+c_{\pm})e^{-it_{n}}Q_{8}(t_{1},t_{2},\ldots,t_{n-1})=0\,.
\end{align}
The nontrivial solutions are
\begin{align}\label{consistency-soultions}
	c_{+}=+1\,,\quad Q_{8}=0\,,\qquad
	\text{or}\qquad
	c_{-}=-1\,,\quad Q_{7}=0\,.
\end{align}
The $c_+$ case corresponds to the null state condition with the lowering operator $L_{-1}$, while the $c_-$ case is associated with the raising operator $L_{+1}$. 
In this way, we determine the zeroth-order null state condition.
We can solve for $G_{n}^{(0)}$ using time-translation invariance.

We repeat the procedure at first order in $g$, and $U_{n,\pm}^{(1)}$ is completely fixed by the consistency with the DS equations \eqref{DS-equations-1}, 
which will be explained more explicitly below.
Therefore, the null differential equations are determined to first order.
As in the complete procedure, together with time-translation invariance, we can solve the null differential equations to obtain $G_{n}^{(1)}$.
The procedure extends to higher orders, so we can solve for $G_{n}$ order by order in $g$ in the reduced procedure as well.\footnote{Again we assume that the parity symmetry is unbroken, so $G_{n}$ vanishes for odd $n$.}

\subsubsection*{Examples: $G_{2}$, $G_{4}$ and $G_{6}$}
As concrete examples, we consider $G_{2}$, $G_{4}$ and $G_{6}$.
Let us consider the more physical case $c_{+}=+1$.\footnote{Perturbatively, we can consider the case $c_{-}=-1$ in a similar fashion. } 
Below, we use the consistency with \eqref{DS-equations-1} to determine $U_{n,\pm}^{(j)}$ order by order, and we obtain $G_2$ to second order in $g$.

At first order, the $(n,k)=(2,2)$ null differential equation \eqref{null-diff} reads
\begin{align}
	\label{G2-g1-t2}
	G_{2}^{(1)}+i\pa_{t_{2}}G_{2}^{(1)}=U_{2,+}^{(1)}\big|_{g^{0}}\,.
\end{align}
On the other hand, we have the DS equation
\begin{align}\label{DS-G2-t2}
	\(\pa_{t_{2}}^{2}+1\)G_{2}^{(1)}(t_{1},t_{2})+4G_{4}^{(0)}(t_{1},t_{2},t_{2},t_{2})=0\,,
\end{align}
and the solution is
\begin{align}\label{DS-G2-t2-solution}
	G_{2}^{(1)}(t_1,t_2)=\frac{3}{4}e^{-it_1+it_2}+\frac{3i}{2}e^{-it_1+it_2}t_{2}+e^{it_{2}}Q_{9}(t_1)+e^{-it_{2}}Q_{10}(t_1)\,,
\end{align}
where the terms with unknown functions $Q_{9}(t_1)$ and $Q_{10}(t_1)$ correspond to the general solution of the associated homogeneous equation.
The solution \eqref{DS-G2-t2-solution} gives
\begin{align}\label{G2-g1-t2-2}
	G_{2}^{(1)}+i\pa_{t_{2}}G_{2}^{(1)}=-\frac{3}{2} e^{-i t_1+i t_2}+2e^{-i t_2} Q_{10}(t_1)\,.
\end{align}
Let us show that the consistency with \eqref{G2-g1-t2} implies $Q_{10}(t_{1})=0$.
The right-hand side $U_{2,+}^{(1)}|_{g^{0}}$ in \eqref{G2-g1-t2} is a linear combination of $G_{2}^{(0)}$, $G_{4}^{(0)}$ and their derivatives in the limit $t_{3},t_{4}\rightarrow t_{2}$.
Using the zeroth-order solutions \eqref{G2-g0} and \eqref{G4-g0}, we deduce that $U_{2,+}^{(1)}|_{g^{0}}$ cannot have the term with the factor $e^{-i t_2}$.
We have
\begin{align}\label{U2-1-g0}
	U_{2,+}^{(1)}\big|_{g^{0}}=-\frac{3}{2} e^{-i t_1+i t_2}\,, 
\end{align}
which is the same as \eqref{U2(1)-g0}.
Since the $(n,k)=(2,1)$ null differential equation \eqref{null-diff} has the same $U_{2,+}^{(1)}\big|_{g^{0}}$, it is also fixed at first order.
We can determine $G_{2}^{(1)}$ using time-translation invariance.
To calculate $G_{4}^{(1)}$, we consider the $(n,k)=(4,4)$ null differential equation \eqref{null-diff}
\begin{align}
	G_{4}^{(1)}+i\pa_{t_{4}}G_{4}^{(1)}=\;&U_{4,+}^{(1)}\big|_{g^{0}}\,.
\end{align}
The corresponding DS equation is
\begin{align}\label{DS-G4-t4}
	\(\pa_{t_{4}}^{2}+1\)G_{4}^{(1)}(t_{1},t_{2},t_{3},t_{4})+4G_{6}^{(0)}(t_{1},t_{2},t_{3},t_{4},t_{4},t_{4})=0\,,
\end{align}
which has the solution
\begin{align}\label{DS-G4-t4-solution}
	G_{4}^{(1)}(t_{1},t_{2},t_{3},t_{4})=\;&-\frac{3}{8}e^{-it_1+it_2-it_3+it_4}-\frac{3}{4}e^{-it_1-it_2+it_3+it_4}+\frac{3}{8}e^{-it_1-it_2-it_3+3it_4}
	\nn
	&+\frac{3i}{4}e^{-it_1+it_2-it_3+it_4}t_4+\frac{3i}{2}e^{-it_1-it_2+it_3+it_4}t_4
	\nn
	&+e^{it_4}Q_{11}(t_{1},t_{2},t_{3})
	+e^{-it_4}Q_{12}(t_{1},t_{2},t_{3})\,.
\end{align}
Using \eqref{DS-G4-t4-solution}, we find
\begin{align}
	G_{4}^{(1)}+i\pa_{t_{4}}G_{4}^{(1)}=\;&-\frac{3}{4} e^{-i t_1+i t_2-i t_3+i t_4}-\frac{3}{2} e^{-i t_1-i t_2+i t_3+i t_4}-\frac{3}{4} e^{-i t_1-i t_2-i t_3+3 i t_4}
	\nn
	&+2e^{-it_{4}}Q_{12}(t_{1},t_{2},t_{3})\,.
\end{align}
Note that $U_{4,+}^{(1)}|_{g^{0}}$ is a linear combination of $G_{4}^{(0)}$, $G_{6}^{(0)}$ and their derivatives in the limit $t_5,t_6\rightarrow t_4$. 
One can check that $U_{4,+}^{(1)}|_{g^{0}}$ cannot contain the term with the factor $e^{-it_{4}}$ using the zeroth-order solutions \eqref{G4-g0} and \eqref{G6-g0}.
Therefore, we have
\begin{align}
	U_{4,+}^{(1)}\big|_{g^{0}}=-\frac{3}{4} e^{-i t_1+i t_2-i t_3+i t_4}-\frac{3}{2} e^{-i t_1-i t_2+i t_3+i t_4}-\frac{3}{4} e^{-i t_1-i t_2-i t_3+3 i t_4}\,,
\end{align}
which agrees with \eqref{U4(1)-g0}. 
So the $n=4$ null differential equations for all $k$ are determined, and the time-translation invariant $G_{4}^{(1)}$ can be derived.

At second order, the $(n,k)=(2,2)$ null differential equation \eqref{null-diff} reads
\begin{align}\label{null-differential}
	G_{2}^{(2)}+i\pa_{t_{2}}G_{2}^{(2)}=U_{2,+}^{(1)}\big|_{g^{1}}+U_{2,+}^{(2)}\big|_{g^{0}}\,,
\end{align}
and we consider the DS equation
\begin{align}\label{DS-G2-g2-t2}
	\(\pa_{t_{2}}^{2}+1\)G_{2}^{(2)}(t_{1},t_{2})+4G_{4}^{(1)}(t_{1},t_{2},t_{2},t_{2})=0\,.
\end{align}
The solution to \eqref{DS-G2-g2-t2} is
\begin{align}\label{DS-G2-g2-t2-solution}
	G_{2}^{(2)}(t_{1},t_{2})=\;&\frac{27}{4}e^{-it_1+it_2}+\frac{3}{16}e^{-3it_1+3it_2}+\frac{9i}{4}e^{-it_1+it_2}t_1-\frac{27i}{2}e^{-it_1+it_2}t_2
	\nn
	&+\frac{9}{2}e^{-it_1+it_2}t_1 t_2-\frac{9}{4}e^{-it_1+it_2}t_2^2
	+e^{it_{2}}Q_{13}(t_1)+e^{-it_{2}}Q_{14}(t_1)\,,
\end{align}
which satisfies
\begin{align}\label{DS-gives-g2}
	G_{2}^{(2)}+i\pa_{t_{2}}G_{2}^{(2)}
	=-\frac{3}{8} e^{-3i t_1+3i t_2} + \frac{9}{2} e^{-i t_1+i t_2} \( 3+it_{1}-it_{2} \) + 2e^{-it_{2}}Q_{14}(t_1)\,.
\end{align}
Note that $U_{2,+}^{(1)}|_{g^{1}}+U_{2,+}^{(2)}|_{g^{0}}$ is a linear combination of the following Green's functions and their derivatives in the limit $t_3,t_4,t_5,t_6\rightarrow t_2$:
\begin{align}
	G_{2}^{(0)},\quad G_{4}^{(0)},\quad G_{6}^{(0)},\quad G_{2}^{(1)},\quad G_{4}^{(1)}\,.
\end{align}
Based on the zeroth- and first-order solutions for the Green's functions \eqref{G2-g0}, \eqref{G4-g0}, \eqref{G6-g0}, \eqref{G2-g1} and \eqref{G4-g1}, one can check that $U_{2,+}^{(1)}|_{g^{1}}+U_{2,+}^{(2)}|_{g^{0}}$ cannot have the term with factor $e^{-it_{2}}$.
So we have
\begin{align}\label{U-U}
	U_{2,+}^{(1)}\big|_{g^{1}}+U_{2,+}^{(2)}\big|_{g^{0}}=-\frac{3}{8} e^{-3i t_1+3i t_2} + \frac{9}{2} e^{-i t_1+i t_2} \( 3+it_{1}-it_{2} \)\,,
\end{align}
which is consistent with \eqref{U2(1)-g1} and \eqref{U2(2)-g0}. 
Then we can obtain the time-translation invariant $G_{2}^{(2)}$ using the null differential equations \eqref{null-differential} with $n=2$.

\subsubsection*{Remarks on $L_{-1}$}
One may wonder if we could determine $L_{-1}$ by considering more Green's functions and determining more $U_{n,+}^{(j)}$.
The answer is that we can determine $L_{-1}$ up to some free parameters, but they are not exactly the level-$1$ ladder operators in the complete procedure.
The reason is that we only impose that $L_{-1}$ annihilates $|0\rangle$, 
and this constraint here is weaker than the constraint that $L_{-1}$ is a lowering operator for all energy eigenstates. 

\subsubsection*{A simplified approach}
In parallel to the complete procedure, we can also determine the Green's functions without introducing more time variables.
We also consider the consistency between the DS equations and null differential equations to constrain $U_{n,\pm}^{(j)}$.
The difference is that here we exchange the order of time derivatives and coincident time limits, as shown in \eqref{simplified-example}.
In other words, we think of $U_{n,\pm}^{(j)}$ as a linear combination of Green's functions with $n$ time variables and their derivatives.
Once $U_{n,\pm}^{(j)}$ is fixed by the consistency condition, the corresponding Green's function will be determined by the null differential equations as above.

As an example, we carry out the calculation of $G_{2}^{(2)}(t_1,t_2)$.
We need to solve for the following Green's functions:
\begin{align}\label{G-list}
	&G_{2}^{(0)}(t_{1},t_{2})\,,\;
	G_{4}^{(0)}(t_{1},t_{2},t_{2},t_{2})\,,\;
	G_{6}^{(0)}(t_{1},t_{2},t_{2},t_{2},t_{2},t_{2})\,,
	\nn
	&G_{2}^{(1)}(t_{1},t_{2})\,,\;
	G_{4}^{(1)}(t_{1},t_{2},t_{2},t_{2})\,.
\end{align}
Below, we will restrict the discussion to the case $c_+=+1$ and determine these functions.
Since the ladder operator is fixed at zeroth order, we can determine the zeroth-order functions in \eqref{G-list}.

For the calculation of $G_{2}^{(1)}(t_1,t_2)$, we have the consistency condition 
\begin{align}\label{consistency-condition}
	U_{2,+}^{(1)}\big|_{g^{0}}
	=-\frac{3}{2} e^{-i t_1+i t_2}+2e^{-i t_2} Q_{10}(t_1)\,,
\end{align}
which follows from \eqref{G2-g1-t2} and \eqref{G2-g1-t2-2}.
Note that $U_{2,+}^{(1)}|_{g^{0}}$ is a linear combination of $G_{2}^{(0)}(t_{1},t_{2})$, $G_{4}^{(0)}(t_{1},t_{2},t_{2},t_{2})$ and their derivatives.
Using \eqref{G2-g0} and \eqref{G4-g0-reduced}, we deduce that the term with the factor $e^{-i t_2}$ cannot appear in $U_{2,+}^{(1)}|_{g^{0}}$.
Then \eqref{consistency-condition} implies
\begin{align}
	U_{2,+}^{(1)}\big|_{g^{0}}=-\frac{3}{2} e^{-i t_1+i t_2}\,,
\end{align}
which is the same as \eqref{U2-1-g0}.
Therefore, we can determine $G_{2}^{(1)}(t_1,t_2)$ using the null differential equations with $n=2$ and time-translation invariance.

The next Green's function in the list \eqref{G-list} is $G_{4}^{(1)}(t_{1},t_{2},t_{2},t_{2})$.
In our discussion, it is more convenient to consider $G_{4}^{(1)}(t_{1},t_{1},t_{1},t_{2})$, which is the same as $G_{4}^{(1)}(t_{1},t_{2},t_{2},t_{2})$ due to time-reversal symmetry. 
To derive the consistency condition, we consider the DS equation
\begin{align}\label{DS-G4-g1-t1-simplified}
	\(\pa_{t_{2}}^{2}+1\)G_{4}^{(1)}(t_{1},t_{1},t_{1},t_{2})+4G_{6}^{(0)}(t_{1},t_{1},t_{1},t_{2},t_{2},t_{2})=0\,,
\end{align}
where we use\footnote{We derive $G_{6}^{(0)}(t_{1},t_{1},t_{1},t_{2},t_{2},t_{2})$ using the null differential equation \eqref{null-constraint-tk-1} at zeroth order.
The case $n=6$ and $k=4$ reads
\begin{align}\label{G6-simplified-reduced-1}
	G^{(0)}_{6,t_{2},t_{3}\rightarrow t_{1};t_{5},t_{6}\rightarrow t_{4}}+\frac{i}{6-4+1}\pa_{t_{4}}G^{(0)}_{6,t_{2},t_{3}\rightarrow t_{1};t_{5},t_{6}\rightarrow t_{4}}
	=\frac{6-4}{2}G^{(0)}_{4,t_{2},t_{3}\rightarrow t_{1}}\,,
\end{align}
where we have also taken the limit $t_{2},t_{3}\rightarrow t_{1}$.
On the right-hand side, $G^{(0)}_{4}(t_1,t_1,t_1,t_4)=G^{(0)}_{4}(t_1,t_4,t_4,t_4)$ due to time-reversal symmetry.
Then we consider \eqref{null-constraint-tk-1} with $n=6$ and $k=1$
\begin{align}\label{G6-simplified-reduced-2}
	G^{(0)}_{6,t_{2},t_{3},t_{4},t_{5},t_{6}\rightarrow t_{1}}+\frac{i}{6-1+1}\pa_{t_{1}}G^{(0)}_{6,t_{2},t_{3},t_{4},t_{5},t_{6}\rightarrow t_{1}}
	=\frac{6-1}{2}G^{(0)}_{4,t_{2},t_{3},t_{4}\rightarrow t_{1}}\,.
\end{align}
The null differential equations \eqref{G6-simplified-reduced-1}, \eqref{G6-simplified-reduced-2} and time-translation invariance give \eqref{G6-g0-reduced}.}
\begin{align}\label{G6-g0-reduced}
	G_{6}^{(0)}(t_{1},t_{1},t_{1},t_{2},t_{2},t_{2})=\frac{9}{8}e^{-it_{1}+it_{2}}+\frac{3}{4}e^{-3it_{1}+3it_{2}}\,.
\end{align}
The solution to \eqref{DS-G4-g1-t1-simplified} is
\begin{align}
	G_{4}^{(1)}(t_{1},t_{1},t_{1},t_{2})=\;&-\frac{9}{8}e^{-it_1+it_2}+\frac{3}{8}e^{-3it_1+3it_2}+\frac{9i}{4}e^{-it_1+it_2}t_2
	\nn
	&+e^{it_2}Q_{15}(t_1)+e^{-it_2}Q_{16}(t_1)\,,
\end{align}
and it satisfies
\begin{align}\label{DS-G4-reduced}
	G_{4}^{(1)}(t_{1},t_{1},t_{1},t_{4})+i\pa_{t_{4}}G_{4}^{(1)}(t_{1},t_{1},t_{1},t_{4})=-\frac{9}{4} e^{-i t_1+i t_4}-\frac{3}{4} e^{-3 i t_1+3 i t_4}+2e^{-i t_4}Q_{16}(t_1)\,,
\end{align}
where we use $t_4$ instead of $t_2$ for convenience.
On the other hand, the corresponding null differential equation is
\begin{align}\label{null-differential-reduced-g1}
	G_{4}^{(1)}(t_{1},t_{1},t_{1},t_{4})+i\pa_{t_{4}}G_{4}^{(1)}(t_{1},t_{1},t_{1},t_{4})=\(U_{4,+}^{(1)}\big|_{g^{0}}\)_{t_{2},t_{3}\rightarrow t_{1}}\,,
\end{align}
which is \eqref{null-condition-G-tn} with $n=4$ at zeroth order in the limit $t_{2},t_{3}\rightarrow t_{1}$.
Since $\(U_{4,+}^{(1)}\big|_{g^{0}}\)_{t_{2},t_{3}\rightarrow t_{1}}$ is a linear combination of $G_{4}^{(0)}(t_{1},t_{1},t_{1},t_{2})=G_{4}^{(0)}(t_{1},t_{2},t_{2},t_{2})$, $G_{6}^{(0)}(t_{1},t_{1},t_{1},t_{2},t_{2},t_{2})$ and their derivatives, one can check that $\(U_{4,+}^{(1)}\big|_{g^{0}}\)_{t_{2},t_{3}\rightarrow t_{1}}$ cannot contain the term with the factor $e^{-it_{4}}$ using the zeroth-order solutions \eqref{G4-g0-reduced} and \eqref{G6-g0-reduced}.
Combining \eqref{DS-G4-reduced} and \eqref{null-differential-reduced-g1}, we have
\begin{align}\label{U-reduced-simplified-g1}
	\(U_{4,+}^{(1)}\big|_{g^{0}}\)_{t_{2},t_{3}\rightarrow t_{1}}=-\frac{9}{4} e^{-i t_1+i t_2}-\frac{3}{4} e^{-3 i t_1+3 i t_2}\,.
\end{align}
This is \eqref{U4(1)-g0} in the limit $t_{2},t_{3}\rightarrow t_{1}$. 
So we can determine $G_{4}^{(1)}(t_{1},t_{1},t_{1},t_{4})$ using $n=4$ null differential equations and time-translation invariance.

Now all the functions in \eqref{G-list} are obtained.
We consider the consistency condition
\begin{align}\label{consistency-G2-g2-simplified}
	U_{2,+}^{(1)}\big|_{g^{1}}+U_{2,+}^{(2)}\big|_{g^{0}}=-\frac{3}{8} e^{-3i t_1+3i t_2} + \frac{9}{2} e^{-i t_1+i t_2} \( 3+it_{1}-it_{2} \) + 2e^{-it_{2}}Q_{14}(t_1)\,,
\end{align}
which follows from \eqref{null-differential} and \eqref{DS-gives-g2}.
The left-hand side is a linear combination of the Green's functions in \eqref{G-list} and their derivatives.
Using \eqref{G2-g0}, \eqref{G4-g0-reduced}, \eqref{G2-g1-2}, \eqref{G4-g1-simplified} and \eqref{G6-g0-reduced}, one can check that the term with the factor $e^{-it_{2}}$ is cannot be present in $U_{2,+}^{(1)}\big|_{g^{1}}+U_{2,+}^{(2)}\big|_{g^{0}}$.
So \eqref{consistency-G2-g2-simplified} implies
\begin{align}
	U_{2,+}^{(1)}\big|_{g^{1}}+U_{2,+}^{(2)}\big|_{g^{0}}=-\frac{3}{8} e^{-3i t_1+3i t_2} + \frac{9}{2} e^{-i t_1+i t_2} \( 3+it_{1}-it_{2} \)\,,
\end{align}
which is the same as \eqref{U-U}. 
This allows us to determine the time-translation invariant $G_{2}^{(2)}(t_{1},t_{2})$ using the null differential equations with $n=2$.
We have also verified that $G_{4}^{(1)}(t_{1},t_{2},t_{3},t_{4})$ can be determined using only four time variables in the reduced procedure.
In conclusion, we can carry out the reduced procedure without introducing additional intermediate time variables. 

\section{Discussion}
\label{Discussion}
In this work, we have used the null bootstrap to investigate the quartic anharmonic oscillator in perturbation theory. 
Although the spacing of energy levels is not constant,
there exist anharmonic ladder operators that can generate the full energy spectrum from a given energy eigenstate. 
In the complete procedure, we determined the energy spectrum \eqref{En} and the matrix elements 
from the analytic solutions for the ladder operators.
To order $g^3$, the explicit expressions of the level-$1$ ladder operators are given in \eqref{ladder-nontrivial-g0}, \eqref{ladder-nontrivial-g1}, \eqref{ladder-nontrivial-g2} and \eqref{ladder-nontrivial-g3}.
In the reduced procedure, we derived the low energy eigenvalues and matrix elements 
without using the exact ladder operators.
We described three versions of the perturbative reduced procedure and compared the results to those from the nonperturbative method in the Hamiltonian formalism.
We obtained the results for the low energy levels to higher orders in the coupling constant $g$.
The results are presented in \eqref{E0-K=1}, \eqref{E0-K=2}, \eqref{E1-K=2}, and \eqref{E0-K=3}--\eqref{E3-K=4}. 
Moreover, we discussed some properties of the anharmonic operator algebra.  
The anharmonic number operator and the level-$1$ ladder operators form a closed algebra \eqref{L-1L+1} and \eqref{commutators-N-L}, as in the case of the harmonic oscillator. 
In \eqref{Hamiltonian-in-terms-of-N}, the Hamiltonian is written as a nonlinear function in the anharmonic number operator.
Furthermore, we showed that the dynamical Schr\"{o}dinger equation is encoded in the null state condition generated by the lowering operator, according to \eqref{HLS}.

Besides the Hamiltonian formalism, we have also studied the Dyson-Schwinger equations in the Lagrangian formalism.
We showed that the underdetermined system of a finite set of the DS equations can be solved by 
imposing the null state condition.
In the complete procedure, the null state conditions are deduced from the exact expression of the lowering operator and the Green's functions can be determined order by order from the null differential equation \eqref{null-constraint-tk}. 
In the reduced procedure, the exact expression for the lowering operator is not needed.
Using the null differential equation \eqref{null-constraint-tk-1} and the consistency with the DS equations, 
the $n$-point Green's functions can be computed order by order as well.
In the explicit examples, the numbers of points in the Green's functions range from $n=2$ to $n=6$. 
We also presented simplified methods that do not introduce additional time variables in the intermediate steps, 
which significantly reduces the complexity of the functions in the computation. 

We would like to extend these perturbative results to genuine quantum field theory with at least two spacetime dimensions. It would be interesting to examine if the standard issues of divergences and the renormalization procedure are simplified in the bootstrap approach. 
Some insights from our perturbative analysis should also extend to the nonperturbative bootstrap approach. 

Another interesting direction is to revisit the multiplet recombination method for conformal field theory \cite{Rychkov:2015naa}, 
which is closely related to the Dyson-Schwinger equations \cite{Nii:2016lpa}. 
The null state condition may be crucial to the derivation of higher order corrections in the $\e$ expansion.\footnote{Constraints on the subleading corrections can be derived from the null state condition \cite{unpublished}. }
In addition, the conformal field theory classification program at higher dimensions should share some features with the two-dimensional minimal models, in which the null state condition plays a central role. 
This may also shed light on the more ambitious goal of classifying more generic quantum field theories by the principle of nullness. 

\section*{Acknowledgments}
We would like to thank the referee for suggesting the comparison to the non-perturbative results and other valuable comments. 
This work was supported by the 100 Talents Program of Sun Yat-sen University,  
the Natural Science Foundation of China (Grant No. 12205386) and the Guangzhou Municipal
Science and Technology Project (Grant No. 2023A04J0006).

\appendix
\section{Traditional perturbation theory}
\label{The traditional method}
For comparison, we review the traditional perturbation theory for the quartic anharmonic oscillator in this appendix.
We will first solve the Schr\"{o}dinger equation to find the eigenenergies and then obtain the perturbed ladder operators. 

The stationary Schr\"{o}dinger equation reads
\begin{align}
	\label{schrodinger-equation}
	H|n\rangle=E_{n}|n\rangle\,.
\end{align}
The Hamiltonian is $H=H_{\text{harmonic}}+gH'$, where $g$ is a small parameter and $H'=x^{4}$.  
The quartic term leads to perturbative corrections to the harmonic oscillator eigenstates and energy eigenvalues.
They take the form of power series in $g$
\begin{align}
	E_{n}&=E_{n}^{(0)}+gE_{n}^{(1)}+g^{2}E_{n}^{(2)}+g^{3}E_{n}^{(3)}+\ldots\,,\\
	|n\rangle&=|n^{(0)}\rangle+g|n^{(1)}\rangle+g^{2}|n^{(2)}\rangle+g^{3}|n^{(3)}\rangle+\ldots\,.
\end{align}
To order $g^{3}$, Eq. \eqref{schrodinger-equation} implies
\begin{align}
	\label{schrodinger-equation-g1}
	H_{\text{harmonic}}|n^{(1)}\rangle
	+H'|n^{(0)}\rangle
	&=E_{n}^{(0)}|n^{(1)}\rangle
	+E_{n}^{(1)}|n^{(0)}\rangle\,,
	\\
	\label{schrodinger-equation-g2}
	H_{\text{harmonic}}|n^{(2)}\rangle
	+H'|n^{(1)}\rangle
	&=E_{n}^{(0)}|n^{(2)}\rangle
	+E_{n}^{(1)}|n^{(1)}\rangle
	+E_{n}^{(2)}|n^{(0)}\rangle\,,
	\\
	\label{schrodinger-equation-g3}
	H_{\text{harmonic}}|n^{(3)}\rangle
	+H'|n^{(2)}\rangle
	&=E_{n}^{(0)}|n^{(3)}\rangle
	+E_{n}^{(1)}|n^{(2)}\rangle
	+E_{n}^{(2)}|n^{(1)}\rangle
	+E_{n}^{(3)}|n^{(0)}\rangle\,.
\end{align}
The zeroth-order equation is the same as the Schr\"{o}dinger equation of the harmonic oscillator.
The first-order corrections $E_{n}^{(1)}$ and $|n^{(1)}\rangle$ are obtained by projecting \eqref{schrodinger-equation-g1} onto $|m^{(0)}\rangle$
\begin{align}
	E_{m}^{(0)}\langle m^{(0)}|n^{(1)}\rangle
	+\langle m^{(0)}|H'|n^{(0)}\rangle
	=E_{n}^{(0)}\langle m^{(0)}|n^{(1)}\rangle
	+E_{n}^{(1)}\langle m^{(0)}|n^{(0)}\rangle\,.
\end{align}
When $m=n$, we obtain the first-order correction to the energy
\begin{align}
	E_{n}^{(1)}
	=\langle m^{(0)}|H'|n^{(0)}\rangle
	=\frac{3}{4}\pt{1+2n+2n^{2}}\,.
\end{align}
When $m\neq n$, we have
\begin{align}
	\langle m^{(0)}|n^{(1)}\rangle
	=\frac
	{
		\left\langle m^{(0)} \left|
		H'
		\right|n^{(0)}\right\rangle
	}
	{E_{n}^{(0)}-E_{m}^{(0)}}\,,
\end{align}
which are the expansion coefficients of $|n^{(1)}\rangle$ in $|m^{(0)}\rangle$.
However, the part parallel to $|n^{(0)}\rangle$ remains arbitrary.
The expansion coefficient $\langle n^{(0)}|n^{(1)}\rangle$ is assumed to be real and chosen such that the norm $\langle n|n \rangle$ is independent of $g$, or in other words $\langle n|n \rangle=1+O(g^{4})$.
We have
\begin{align}
	|n^{(1)}\rangle
	=\pt{
		\frac{3}{4}a^{2}-\frac{3}{4}\ad{2}+\frac{1}{16}a^{4}+\frac{1}{2}\adl a^{3}-\frac{1}{2}\ad{3}a-\frac{1}{16}\ad{4}
	}|n^{(0)}\rangle
	\equiv f^{(1)}|n^{(0)}\rangle\,.
\end{align}
We have written the terms involving the occupation number $n$ in terms of Dirac's ladder operators, 
which is more natural for the operator algebra perspective 
and more convenient for the discussion of the anharmonic ladder operators below.

For the second-order corrections, we consider \eqref{schrodinger-equation-g2}.
Following the same procedure of projection, we obtain
\begin{align}
	E_{m}^{(0)}\langle m^{(0)}|n^{(2)}\rangle
	+\langle m^{(0)}|H'|n^{(1)}\rangle
	=E_{n}^{(0)}\langle m^{(0)}|n^{(2)}\rangle
	+E_{n}^{(1)}\langle m^{(0)}|n^{(1)}\rangle
	+E_{n}^{(2)}\langle m^{(0)}|n^{(0)}\rangle\,.
\end{align}
The case of $m=n$ gives
\begin{align}
	E_{n}^{(2)}
	=\langle n^{(0)}|H'|n^{(1)}\rangle
	=-\frac{1}{8}
	\pt{
		21+59n+51n^{2}+34n^{3}
	}\,.
\end{align}
For $m\neq n$, we have 
\begin{align}
	\langle m^{(0)}|n^{(2)}\rangle
	=\frac
	{
		\left\langle m^{(0)} \left|
		\pt{ H'-E_{n}^{(1)} } f^{(1)} 
		\right|n^{(0)}\right\rangle
	}
	{E_{n}^{(0)}-E_{m}^{(0)}}\,.
\end{align}
As before, we assume the expansion coefficient $\langle n^{(0)}|n^{(2)}\rangle$ is real, and the part parallel to $|n^{(0)}\rangle$ is fixed by the condition that $\langle n|n \rangle$ is independent of $g$
\begin{align}
	|n^{(2)}\rangle
	=\;&\left(
		\frac{\pt{a^\dagger}^8}{512}+\frac{\pt{a^\dagger}^7 a}{32}+\frac{\pt{a^\dagger}^6 a^2}{8}+\frac{17 \pt{a^\dagger}^6}{192}-\frac{\pt{a^\dagger}^5
		a^3}{32}+\frac{15 \pt{a^\dagger}^5 a}{16}-\frac{65 \pt{a^\dagger}^4 a^4}{256}
	\right.
	\nn
		&+\frac{117 \pt{a^\dagger}^4 a^2}{64}+\frac{9\pt{a^\dagger}^4}{8}-\frac{\pt{a^\dagger}^3 a^5}{32}-\frac{65 \pt{a^\dagger}^3 a^3}{32}+\frac{15 \pt{a^\dagger}^3 a}{2}+\frac{\pt{a^\dagger}^2
		a^6}{8}
	\nn
		&-\frac{147 \pt{a^\dagger}^2 a^4}{64}-\frac{333 \pt{a^\dagger}^2 a^2}{64}+\frac{75 \pt{a^\dagger}^2}{16}+\frac{a^\dagger
		a^7}{32}+\frac{9 a^\dagger a^5}{16}-9 a^\dagger a^3-\frac{69 a^\dagger
		a}{16}
	\nn
		&\left.+\frac{a^8}{512}+\frac{25 a^6}{192}+\frac{3 a^4}{16}-6 a^2-\frac{39}{64}\right)|n^{(0)}\rangle
	\nn
	\equiv\;&f^{(2)}|n^{(0)}\rangle\,.
\end{align}

Finally, for the third-order corrections, Eq. \eqref{schrodinger-equation-g3} yields
\begin{align}
	&E_{m}^{(0)}\langle m^{(0)}|n^{(3)}\rangle
	+\langle m^{(0)}|H'|n^{(2)}\rangle \nn
	=\;&E_{n}^{(0)}\langle m^{(0)}|n^{(3)}\rangle
	+E_{n}^{(1)}\langle m^{(0)}|n^{(2)}\rangle
	+E_{n}^{(2)}\langle m^{(0)}|n^{(1)}\rangle
	+E_{n}^{(3)}\langle m^{(0)}|n^{(0)}\rangle\,.
\end{align}
When $m=n$, we obtain
\begin{align}
	E_{n}^{(3)}
	&=\langle n^{(0)}|H'|n^{(2)}\rangle
	-E_{n}^{(1)}\langle n^{(0)}|n^{(2)}\rangle
	\nn
	&=\frac{3}{16}\pt{
		111+347n+472n^{2}+250n^{3}+125n^{4}
	}\,.
\end{align}
When $m\neq n$, we have
\begin{align}
	\langle m^{(0)}|n^{(3)}\rangle
	=\frac
	{
		\left\langle m^{(0)} \left|
		\pt{ H'-E_{n}^{(1)} } f^{(2)} 
		-E_{n}^{(2)} f^{(1)}
		\right|n^{(0)}\right\rangle
	}
	{E_{n}^{(0)}-E_{m}^{(0)}}\,,
\end{align}
The third-order corrections to the eigenstates are
\begin{align}
	|n^{(3)}\rangle
	=\;& \left( -\frac{(a^\dagger)^{12}}{24576}-\frac{(a^\dagger)^{11} a}{1024}-\frac{(a^\dagger)^{10} a^2}{128}-\frac{25
	(a^\dagger)^{10}}{6144}-\frac{61 (a^\dagger)^9 a^3}{3072}-\frac{61 (a^\dagger)^9 a}{768}+\frac{129 (a^\dagger)^8
	a^4}{8192}\right.
	\nn
	&-\frac{1005 (a^\dagger)^8 a^2}{2048}-\frac{37 (a^\dagger)^8}{256}+\frac{33 (a^\dagger)^7 a^5}{512}-\frac{1213
	(a^\dagger)^7 a^3}{1536}-\frac{145 (a^\dagger)^7 a}{64}+\frac{4145 (a^\dagger)^6 a^4}{3072}
	\nn
	&-\frac{10123 (a^\dagger)^6
		a^2}{1024}-\frac{587 (a^\dagger)^6}{256}-\frac{33 (a^\dagger)^5 a^7}{512}+\frac{267 (a^\dagger)^5 a^5}{128}-\frac{7
	(a^\dagger)^5 a^3}{2}-\frac{6903 (a^\dagger)^5 a}{256}
	\nn
	&-\frac{129 (a^\dagger)^4 a^8}{8192}-\frac{2785 (a^\dagger)^4
		a^6}{3072}+\frac{6675 (a^\dagger)^4 a^4}{256}-\frac{17097 (a^\dagger)^4 a^2}{256}-\frac{16709
	(a^\dagger)^4}{1024}
	\nn
	&+\frac{61 (a^\dagger)^3 a^9}{3072}-\frac{1987 (a^\dagger)^3 a^7}{1536}+\frac{141 (a^\dagger)^3
		a^5}{16}+\frac{3531 (a^\dagger)^3 a^3}{32}-\frac{16703 (a^\dagger)^3 a}{128}
	\nn
	&+\frac{(a^\dagger)^2
		a^{10}}{128}+\frac{93 (a^\dagger)^2 a^8}{2048}-\frac{12277 (a^\dagger)^2 a^6}{1024}+\frac{5967 (a^\dagger)^2
		a^4}{64}+\frac{5877 (a^\dagger)^2 a^2}{32}
	\nn
	&-\frac{12099 (a^\dagger)^2}{256}+\frac{a^\dagger a^{11}}{1024}+\frac{59
		a^\dagger a^9}{768}-\frac{83 a^\dagger a^7}{64}-\frac{7425 a^\dagger a^5}{256}+\frac{23445 a^\dagger
		a^3}{128}+\frac{3141 a^\dagger a}{32}
	\nn
	&\left.+\frac{a^{12}}{24576}+\frac{41 a^{10}}{6144}+\frac{17 a^8}{128}-\frac{125 a^6}{32}-\frac{14571 a^4}{1024}+\frac{19485 a^2}{256}+\frac{279}{32} \right)|n^{(0)}\rangle
	\nn
	\equiv\;& f^{(3)}|n^{(0)}\rangle\,,
\end{align}
where the expansion coefficient $\langle n^{(0)}|n^{(3)}\rangle$ is assumed to be real and then fixed by the condition that $\langle n|n \rangle$ is independent of $g$.

We can also study the corrections to Dirac's ladder operators.
We define the raising and the lowering operators by their action on a generic energy eigenstate
\begin{align}
	\label{ladder-definition}
	L_{\pm1}|n\rangle=C_{\pm}|n\pm 1\rangle\,,
\end{align}
where $C_{\pm}$ are real numbers.
The ladder operators $L_{\pm1}$ and $C_{\pm}$ have the perturbative expansion
\begin{align}
	L_{\pm1}&=L^{(0)}_{\pm1}+gL^{(1)}_{\pm1}+g^{2}L^{(2)}_{\pm1}+g^{3}L^{(3)}_{\pm1}+\ldots\,, \\
	C_{\pm}&=C_{\pm}^{(0)}+gC_{\pm}^{(1)}+g^{2}C_{\pm}^{(2)}+g^{3}C_{\pm}^{(3)}+\ldots\,.
\end{align}
To be consistent with the discussion in Sec. \ref{The null bootstrap in perturbation theory}, we adopt the normalization
\begin{align}
	\qquad C^{(0)}_{-}=\sqrt{n}\,,\qquad C^{(0)}_{+}=\sqrt{n+1}\,,\qquad C^{(1)}_{\pm}=C^{(2)}_{\pm}=\ldots=0\,.
\end{align}
The zeroth-order solution to \eqref{ladder-definition} is the Dirac's ladder operators: $L_{-1}^{(0)}\big|_{\text{n.t.}}=a$, $L_{+1}^{(0)}\big|_{\text{n.t.}}=a^{\dagger}$.
At higher orders, let us first focus on the lowering operator.
Equation \eqref{ladder-definition} implies
\begin{align}
	L_{-1}^{(0)}|n^{(1)}\rangle
	+L^{(1)}_{-1}|n^{(0)}\rangle
	&=C_{-}^{(0)}|(n-1)^{(1)}\rangle\,,
	\\
	L_{-1}^{(0)}|n^{(2)}\rangle
	+L^{(1)}_{-1}|n^{(1)}\rangle
	+L^{(2)}_{-1}|n^{(0)}\rangle
	&=C_{-}^{(0)}|(n-1)^{(2)}\rangle\,,
	\\
	L_{-1}^{(0)}|n^{(3)}\rangle
	+L^{(1)}_{-1}|n^{(2)}\rangle
	+L^{(2)}_{-1}|n^{(1)}\rangle
	+L^{(3)}_{-1}|n^{(0)}\rangle
	&=C_{-}^{(0)}|(n-1)^{(3)}\rangle\,.
\end{align}
To solve these equations, we write all the states in terms of $|n^{(0)}\rangle$
\begin{align}
	\pt{
		L_{-1}^{(0)}f^{(1)}+L^{(1)}_{-1}-f^{(1)}L_{-1}^{(0)}
	}
	|n^{(0)}\rangle
	&=0\,,
	\\
	\pt{
		L_{-1}^{(0)}f^{(2)}+L^{(1)}_{-1}f^{(1)}+L^{(2)}_{-1}-f^{(2)}L_{-1}^{(0)}
	}
	|n^{(0)}\rangle
	&=0\,,
	\\
	\pt{
		L_{-1}^{(0)}f^{(3)}+L^{(1)}_{-1}f^{(2)}+L^{(2)}_{-1}f^{(1)}+L^{(3)}_{-1}-f^{(3)}L_{-1}^{(0)}
	}
	|n^{(0)}\rangle
	&=0\,.
\end{align}
At first order, the solution reads
\begin{align}
	L^{(1)}_{-1}\big|_{\text{n.t.}}=\left[f^{(1)},L_{-1}^{(0)}\big|_{\text{n.t.}}\right]
	=\frac{3}{2}\adl-\frac{1}{2}a^{3}+\frac{3}{2}(a^{\dagger})^{2}a+\frac{1}{4}(a^{\dagger})^{3}\,.
\end{align}
At second order, we obtain
\begin{align}
	L^{(2)}_{-1}\big|_{\text{n.t.}}=\;&\left[f^{(2)},L_{-1}^{(0)}\big|_{\text{n.t.}}\right]-L^{(1)}_{-1}\big|_{\text{n.t.}}f^{(1)}
	\nn
	=\;&\frac{9}{16}a-\frac{81}{8}a^{\dagger}+9a^{3}+\frac{81}{32}a^{\dagger}a^{2}-\frac{189}{8}(a^{\dagger})^{2}a-\frac{9}{8}(a^{\dagger})^{3}+\frac{9}{16}a^{5}+\frac{9}{2}a^{\dagger}a^{4}
	\nn
	&+\frac{27}{32}(a^{\dagger})^{2}a^{3}-\frac{63}{8}(a^{\dagger})^{3}a^{2}-\frac{9}{16}(a^{\dagger})^{4}a+\frac{1}{8}(a^{\dagger})^{5}\,,
\end{align}
At third order, we have
\begin{align}
	L^{(3)}_{-1}\big|_{\text{n.t.}}=\;&\left[f^{(3)},L_{-1}^{(0)}\big|_{\text{n.t.}}\right]-L^{(1)}_{-1}\big|_{\text{n.t.}}f^{(2)}-L^{(2)}_{-1}\big|_{\text{n.t.}}f^{(1)} \nn
	=\;&-\frac{11 (a^\dagger)^7}{64}-\frac{103 (a^\dagger)^6 a}{32}-\frac{285 (a^\dagger)^5 a^2}{128}-\frac{309 (a^\dagger)^5}{32}+\frac{4159
	(a^\dagger)^4 a^3}{64}-\frac{1425 (a^\dagger)^4 a}{128}
	\nn
	&-\frac{351 (a^\dagger)^3 a^4}{64}+\frac{12477 (a^\dagger)^3
	a^2}{32}-\frac{243 (a^\dagger)^3}{128}-\frac{2673 (a^\dagger)^2 a^5}{64}-\frac{1053 (a^\dagger)^2 a^3}{32}
	\nn
	&+\frac{33519
		(a^\dagger)^2 a}{64}-\frac{315 a^\dagger a^6}{64}-\frac{13365 a^\dagger a^4}{64}-\frac{729 a^\dagger
	a^2}{32}+\frac{8565 a^\dagger}{64}+\frac{7 a^7}{32}-\frac{945 a^5}{64}
	\nn
	&-\frac{5985 a^3}{32}+\frac{81 a}{8}\,.
\end{align}
For the raising operator, the solutions are
\begin{align}
	L_{+1}^{(1)}\big|_{\text{n.t.}}=\;&\left[f^{(1)},L_{+1}^{(0)}\big|_{\text{n.t.}}\right] 
	=\frac{3}{2}a+\frac{1}{4}a^{3}+\frac{3}{2}\adl a^{2}-\frac{1}{2}\ad{3}\,,
	\\
	L_{+1}^{(2)}\big|_{\text{n.t.}}=\;&\left[f^{(2)},L_{+1}^{(0)}\big|_{\text{n.t.}}\right]-L_{+1}^{(1)}\big|_{\text{n.t.}}f^{(1)} \nn
	=\;&\frac{9 (a^\dagger)^5}{16}+\frac{9 (a^\dagger)^4 a}{2}+\frac{27 (a^\dagger)^3 a^2}{32}+9 (a^\dagger)^3-\frac{63 (a^\dagger)^2 a^3}{8}+\frac{81 (a^\dagger)^2 a}{32}-\frac{9 a^\dagger a^4}{16}
	\nn
	&-\frac{189 a^\dagger a^2}{8}+\frac{9 a^\dagger}{16}+\frac{a^5}{8}-\frac{9 a^3}{8}-\frac{81 a}{8}\,,
	\\
	L^{(3)}_{+1}\big|_{\text{n.t.}}=\;&\left[f^{(3)},L_{+1}^{(0)}\big|_{\text{n.t.}}\right]-L_{+1}^{(1)}\big|_{\text{n.t.}}f^{(2)}-L_{+1}^{(2)}\big|_{\text{n.t.}}f^{(1)} \nn
	=\;&\frac{7 (a^\dagger)^7}{32}-\frac{315 (a^\dagger)^6 a}{64}-\frac{2673 (a^\dagger)^5 a^2}{64}-\frac{945 (a^\dagger)^5}{64}-\frac{351
	(a^\dagger)^4 a^3}{64}-\frac{13365 (a^\dagger)^4 a}{64}
	\nn
	&+\frac{4159 (a^\dagger)^3 a^4}{64}-\frac{1053 (a^\dagger)^3
	a^2}{32}-\frac{5985 (a^\dagger)^3}{32}-\frac{285 (a^\dagger)^2 a^5}{128}+\frac{12477 (a^\dagger)^2 a^3}{32}
	\nn
	&-\frac{729
		(a^\dagger)^2 a}{32}-\frac{103 a^\dagger a^6}{32}-\frac{1425 a^\dagger a^4}{128}+\frac{33519 a^\dagger
	a^2}{64}+\frac{81 a^\dagger}{8}-\frac{11 a^7}{64}-\frac{309 a^5}{32}
	\nn
	&-\frac{243 a^3}{128}+\frac{8565 a}{64}\,.
\end{align}
Using the explicit expression of Dirac's ladder operators \eqref{a-ad-in-terms-of-xp}, 
one can check that the results agree exactly with those from the null bootstrap in the main text.


\begin{thebibliography}{99}

\bibitem{Anderson:2016rcw}
P.~D.~Anderson and M.~Kruczenski,
``Loop Equations and bootstrap methods in the lattice,''
Nucl. Phys. B \textbf{921} (2017), 702-726
[arXiv:1612.08140 [hep-th]].

\bibitem{Lin:2020mme}
H.~W.~Lin,
``Bootstraps to strings: solving random matrix models with positivity,''
JHEP \textbf{06} (2020), 090
[arXiv:2002.08387 [hep-th]].

\bibitem{Han:2020bkb}
X.~Han, S.~A.~Hartnoll and J.~Kruthoff,
``Bootstrapping Matrix Quantum Mechanics,''
Phys. Rev. Lett. \textbf{125} (2020) no.4, 041601
[arXiv:2004.10212 [hep-th]].

\bibitem{Han}
X. Han,
``Quantum Many-body Bootstrap,''
[arXiv:2006.06002[cond-mat]].

\bibitem{Hessam:2021byc}
H.~Hessam, M.~Khalkhali and N.~Pagliaroli,
``Bootstrapping Dirac ensembles,''
J. Phys. A \textbf{55}, no.33, 335204 (2022)
[arXiv:2107.10333 [hep-th]].

\bibitem{Kazakov:2021lel}
V.~Kazakov and Z.~Zheng,
``Analytic and numerical bootstrap for one-matrix model and \textquotedblleft{}unsolvable\textquotedblright{} two-matrix model,''
JHEP \textbf{06} (2022), 030
[arXiv:2108.04830 [hep-th]].

\bibitem{Berenstein:2021dyf}
D.~Berenstein and G.~Hulsey,
``Bootstrapping Simple QM Systems,''
[arXiv:2108.08757 [hep-th]].


\bibitem{Bhattacharya:2021btd}
J.~Bhattacharya, D.~Das, S.~K.~Das, A.~K.~Jha and M.~Kundu,
``Numerical bootstrap in quantum mechanics,''
Phys. Lett. B \textbf{823} (2021), 136785
[arXiv:2108.11416 [hep-th]].

\bibitem{Aikawa:2021eai}
Y.~Aikawa, T.~Morita and K.~Yoshimura,
``Application of bootstrap to a \ensuremath{\theta} term,''
Phys. Rev. D \textbf{105} (2022) no.8, 085017
[arXiv:2109.02701 [hep-th]].

\bibitem{Berenstein:2021loy}
D.~Berenstein and G.~Hulsey,
``Bootstrapping more QM systems,''
J. Phys. A \textbf{55} (2022) no.27, 275304
[arXiv:2109.06251 [hep-th]].

\bibitem{Tchoumakov:2021mnh}
S.~Tchoumakov and S.~Florens,
``Bootstrapping Bloch bands,''
J. Phys. A \textbf{55} (2022) no.1, 015203
[arXiv:2109.06600 [cond-mat.mes-hall]].


\bibitem{Aikawa:2021qbl}
Y.~Aikawa, T.~Morita and K.~Yoshimura,
``Bootstrap method in harmonic oscillator,''
Phys. Lett. B \textbf{833} (2022), 137305
[arXiv:2109.08033 [hep-th]].

\bibitem{Du:2021hfw}
B.~n.~Du, M.~x.~Huang and P.~x.~Zeng,
``Bootstrapping Calabi\textendash{}Yau quantum mechanics,''
Commun. Theor. Phys. \textbf{74} (2022) no.9, 095801
[arXiv:2111.08442 [hep-th]].

\bibitem{Lawrence:2021msm}
S.~Lawrence,
``Bootstrapping Lattice Vacua,''
[arXiv:2111.13007 [hep-lat]].

\bibitem{Bai:2022yfv}
D.~Bai,
``Bootstrapping the deuteron,''
[arXiv:2201.00551 [nucl-th]].

\bibitem{Nakayama:2022ahr}
Y.~Nakayama,
``Bootstrapping microcanonical ensemble in classical system,''
Mod. Phys. Lett. A \textbf{37} (2022) no.09, 2250054
[arXiv:2201.04316 [hep-th]].

\bibitem{Khan:2022uyz}
S.~Khan, Y.~Agarwal, D.~Tripathy and S.~Jain,
``Bootstrapping PT symmetric quantum mechanics,''
Phys. Lett. B \textbf{834} (2022), 137445
[arXiv:2202.05351 [quant-ph]].

\bibitem{Kazakov:2022xuh}
V.~Kazakov and Z.~Zheng,
``Bootstrap for lattice Yang-Mills theory,''
Phys. Rev. D \textbf{107} (2023) no.5, L051501
[arXiv:2203.11360 [hep-th]].

\bibitem{Berenstein:2022ygg}
D.~Berenstein and G.~Hulsey,
``Anomalous bootstrap on the half-line,''
Phys. Rev. D \textbf{106}, no.4, 045029 (2022)
[arXiv:2206.01765 [hep-th]].

\bibitem{Cho:2022lcj}
M.~Cho, B.~Gabai, Y.~H.~Lin, V.~A.~Rodriguez, J.~Sandor and X.~Yin,
``Bootstrapping the Ising Model on the Lattice,''
[arXiv:2206.12538 [hep-th]].

\bibitem{Morita:2022zuy}
T.~Morita,
``Universal bounds on quantum mechanics through energy conservation and the bootstrap method,''
PTEP \textbf{2023} (2023) no.2, 023A01
[arXiv:2208.09370 [hep-th]].

\bibitem{Blacker:2022szo}
M.~J.~Blacker, A.~Bhattacharyya and A.~Banerjee,
``Bootstrapping the Kronig-Penney model,''
Phys. Rev. D \textbf{106} (2022) no.11, 11
[arXiv:2209.09919 [quant-ph]].

\bibitem{Berenstein:2022unr}
D.~Berenstein and G.~Hulsey,
``Semidefinite programming algorithm for the quantum mechanical bootstrap,''
Phys. Rev. E \textbf{107} (2023) no.5, L053301
[arXiv:2209.14332 [hep-th]].

\bibitem{Nancarrow:2022wdr}
C.~O.~Nancarrow and Y.~Xin,
``Bootstrapping the gap in quantum spin systems,''
[arXiv:2211.03819 [hep-th]].

\bibitem{Lawrence:2022vsb}
S.~Lawrence,
``Semidefinite programs at finite fermion density,''
Phys. Rev. D \textbf{107} (2023) no.9, 094511
[arXiv:2211.08874 [hep-lat]].

\bibitem{Lin:2023owt}
H.~W.~Lin,
``Bootstrap bounds on D0-brane quantum mechanics,''
JHEP \textbf{06} (2023), 038
[arXiv:2302.04416 [hep-th]].

\bibitem{Li:2022prn}
W.~Li,
``Null bootstrap for non-Hermitian Hamiltonians,''
Phys. Rev. D \textbf{106} (2022) no.12, 125021
[arXiv:2202.04334 [hep-th]].

\bibitem{John:2023him}
R.~R.~John and K.~P.~R,
``Anharmonic oscillators and the null bootstrap,''
[arXiv:2309.06381 [quant-ph]].

\bibitem{Belavin:1984vu}
A.~A.~Belavin, A.~M.~Polyakov and A.~B.~Zamolodchikov,
``Infinite Conformal Symmetry in Two-Dimensional Quantum Field Theory,''
Nucl. Phys. B \textbf{241} (1984), 333-380

\bibitem{DiFrancesco:1997nk} 
P.~Di Francesco, P.~Mathieu and D.~Senechal,
``Conformal Field Theory,''
Graduate Texts in Contemporary Physics,
Springer-Verlag, New York (1997).

\bibitem{Cardy:1985yy}
J.~L.~Cardy,
``Conformal Invariance and the Yang-lee Edge Singularity in Two-dimensions,''
Phys. Rev. Lett. \textbf{54} (1985), 1354-1356

\bibitem{Cardy:1989fw}
J.~L.~Cardy and G.~Mussardo,
``S Matrix of the Yang-Lee Edge Singularity in Two-Dimensions,''
Phys. Lett. B \textbf{225} (1989), 275-278

\bibitem{Yang:1952be}
C.~N.~Yang and T.~D.~Lee,
``Statistical theory of equations of state and phase transitions. 1. Theory of condensation,''
Phys. Rev. \textbf{87}, 404-409 (1952)

\bibitem{Lee:1952ig}
T.~D.~Lee and C.~N.~Yang,
``Statistical theory of equations of state and phase transitions. 2. Lattice gas and Ising model,''
Phys. Rev. \textbf{87}, 410-419 (1952)

\bibitem{Kortman:1971zz}
P.~J.~Kortman and R.~B.~Griffiths,
``Density of Zeros on the Lee-Yang Circle for Two Ising Ferromagnets,''
Phys. Rev. Lett. \textbf{27} (1971), 1439-1442

\bibitem{Fisher:1978pf}
M.~E.~Fisher,
``Yang-Lee Edge Singularity and phi**3 Field Theory,''
Phys. Rev. Lett. \textbf{40}, 1610-1613 (1978)

\bibitem{Ferrara:1973yt}
S.~Ferrara, A.~F.~Grillo and R.~Gatto,
``Tensor representations of conformal algebra and conformally covariant operator product expansion,''
Annals Phys. \textbf{76} (1973), 161-188

\bibitem{Polyakov:1974gs}
A.~M.~Polyakov,
``Nonhamiltonian approach to conformal quantum field theory,''
Zh. Eksp. Teor. Fiz. \textbf{66} (1974), 23-42

\bibitem{Rattazzi:2008pe}
R.~Rattazzi, V.~S.~Rychkov, E.~Tonni and A.~Vichi,
``Bounding scalar operator dimensions in 4D CFT,''
JHEP \textbf{12} (2008), 031
[arXiv:0807.0004 [hep-th]].

\bibitem{Poland:2018epd}
D.~Poland, S.~Rychkov and A.~Vichi,
``The Conformal Bootstrap: Theory, Numerical Techniques, and Applications,''
Rev. Mod. Phys. \textbf{91}, 015002 (2019)
[arXiv:1805.04405 [hep-th]].

\bibitem{GN}
I. M. Gelfand, M. A. Naimark,
``On the imbedding of normed rings into the ring of operators on a Hilbert space,''
Matematicheskii Sbornik 12 (2) (1943), 197–217.

\bibitem{S}
I. E. Segal,
``Irreducible representations of operator algebras,''
Bull. Amer. Math. Soc. 53 (1) (1947), 73–88.

\bibitem{Dong}
S.~H.~Dong, ``Factorization method in quantum mechanics,'' Springer Science \& Business Media (2007).

\bibitem{Bender:1969si}
C.~M.~Bender and T.~T.~Wu,
``Anharmonic oscillator,''
Phys. Rev. \textbf{184}, 1231-1260 (1969)
doi:10.1103/PhysRev.184.1231

\bibitem{Dyson:1949ha}
F.~J.~Dyson,
``The S matrix in quantum electrodynamics,''
Phys. Rev. \textbf{75} (1949), 1736-1755

\bibitem{Schwinger:1951ex}
J.~S.~Schwinger,
``On the Green's functions of quantized fields. 1.,''
Proc. Nat. Acad. Sci. \textbf{37} (1951), 452-455

\bibitem{Schwinger:1951hq}
J.~S.~Schwinger,
``On the Green's functions of quantized fields. 2.,''
Proc. Nat. Acad. Sci. \textbf{37} (1951), 455-459

\bibitem{Bender:1988bp}
C.~M.~Bender, F.~Cooper and L.~M.~Simmons,
``Nonunique Solution to the Schwinger-dyson Equations,''
Phys. Rev. D \textbf{39} (1989), 2343-2349

\bibitem{Bender:2022eze}
C.~M.~Bender, C.~Karapoulitidis and S.~P.~Klevansky,
``Underdetermined Dyson-Schwinger Equations,''
Phys. Rev. Lett. \textbf{130} (2023) no.10, 101602
[arXiv:2211.13026 [math-ph]].

\bibitem{Bender:2023ttu}
C.~M.~Bender, C.~Karapoulitidis and S.~P.~Klevansky,
``Dyson-Schwinger equations in zero dimensions and polynomial approximations,''
[arXiv:2307.01008 [math-ph]].

\bibitem{Li:2023nip}
W.~Li,
``Taming Dyson-Schwinger Equations with Null States,''
Phys. Rev. Lett. \textbf{131} (2023) no.3, 031603
[arXiv:2303.10978 [hep-th]].

\bibitem{Li:2023ewe}
W.~Li,
``Principle of minimal singularity for Green's functions,''
[arXiv:2309.02201 [hep-th]].

\bibitem{Rychkov:2015naa}
S.~Rychkov and Z.~M.~Tan,
``The $\epsilon$-expansion from conformal field theory,''
J. Phys. A \textbf{48} (2015) no.29, 29FT01
[arXiv:1505.00963 [hep-th]].

\bibitem{Bender:1973rz}
C.~M.~Bender and T.~T.~Wu,
``Anharmonic oscillator. 2: A Study of perturbation theory in large order,''
Phys. Rev. D \textbf{7} (1973), 1620-1636

\bibitem{Sulejmanpasic:2016fwr}
T.~Sulejmanpasic and M.~\"Unsal,
``Aspects of perturbation theory in quantum mechanics: The BenderWu Mathematica \textregistered{}  package,''
Comput. Phys. Commun. \textbf{228} (2018), 273-289
[arXiv:1608.08256 [hep-th]].

\bibitem{Nii:2016lpa}
K.~Nii,
``Classical equation of motion and Anomalous dimensions at leading order,''
JHEP \textbf{07}, 107 (2016)
[arXiv:1605.08868 [hep-th]].

\bibitem{unpublished}
W.~Li, work in progress. 


\end{thebibliography}
\end{document}